\documentclass[prb,aps,twocolumn,superscriptaddress,showpacs]{revtex4-2}
\usepackage[colorlinks=true,linkcolor=black, citecolor=blue, urlcolor=blue, 
    unicode=true]{hyperref}

\usepackage{graphicx,mathtools}% Include figure files

\begin{document}

%\title{Correlated magnetic and ferroelectric relaxor response in PbFe$_{1/2}$Nb$_{1/2}$O$_{3}$}

\title{Fast broadband ``cluster" spin-glass dynamics in Pb(Fe$_{1/2}$Nb$_{1/2}$)O$_{3}$}

\author{C. Stock}
\affiliation{School of Physics and Astronomy, University of Edinburgh, Edinburgh EH9 3JZ, UK}
\author{B. Roessli}
\affiliation{Laboratory for Neutron Scattering and Imaging (LNS), Paul Scherrer Institut (PSI), 5232 Villigen PSI, Switzerland}
\author{P. M. Gehring}
\affiliation{NIST Center for Neutron Research, National Institute of Standards  and Technology, Gaithersburg, Maryland 20899-6100, USA}
\author{J. A. Rodriguez-Rivera}
\affiliation{NIST Center for Neutron Research, National Institute of Standards  and Technology, Gaithersburg, Maryland 20899-6100, USA}
\affiliation{Department of Materials Science, University of Maryland, College Park, MD  20742}
\author{N. Giles-Donovan}
\affiliation{Centre for Medical and Industrial Ultrasonics, James Watt School of Engineering, University of Glasgow, Glasgow G12 8QQ, United Kingdom}
\author{S. Cochran}
\affiliation{Centre for Medical and Industrial Ultrasonics, James Watt School of Engineering, University of Glasgow, Glasgow G12 8QQ, United Kingdom}
\author{G. Xu}
\affiliation{NIST Center for Neutron Research, National Institute of Standards  and Technology, Gaithersburg, Maryland 20899-6100, USA}
\author{P. Manuel}
\affiliation{ISIS Neutron and Muon Source, Rutherford Appleton Laboratory, Chilton, Didcot OX11 0QX, United Kingdom}
\author{M. J. Gutmann}
\affiliation{ISIS Neutron and Muon Source, Rutherford Appleton Laboratory, Chilton, Didcot OX11 0QX, United Kingdom}
\author{W. D. Ratcliff}
\affiliation{NIST Center for Neutron Research, National Institute of Standards  and Technology, Gaithersburg, Maryland 20899-6100, USA}
\author{T. Fennell}
\affiliation{Laboratory for Neutron Scattering and Imaging (LNS), Paul Scherrer Institut (PSI), 5232 Villigen PSI, Switzerland}
\affiliation{Institut Laue-Langevin,  71 avenue des Martyrs, CS 20156, 38042 Grenoble Cedex 9, France}
\author{Y. Su}
\affiliation{J\"ulich Centre for Neutron Science (JCNS) at Heinz Maier-Leibnitz Zentrum (MLZ), Forschungszentrum J\"ulich, Lichtenbergstrasse 1, D-85747 Garching, Germany}
\author{X. Li}
\affiliation{Shanghai Institute of Ceramics, Chinese Academy of Sciences, Shanghai 201800, China}
\author{H. Luo}
\affiliation{Shanghai Institute of Ceramics, Chinese Academy of Sciences, Shanghai 201800, China}

\date{\today}

\begin{abstract}

PbFe$_{1/2}$Nb$_{1/2}$O$_{3}$ (PFN) is a relaxor ferroelectric (T$_{c}$ $\sim$ 400 K) consisting of magnetic Fe$^{3+}$ (S=${5\over2}$, L$\approx$0) ions disordered throughout the lattice and hosts a spin-glass phase at low temperatures built from spatially isolated clusters of Fe$^{3+}$ ions, termed a ``cluster glass" (W. Kleemann $\textit{et al.}$ Phys. Rev. Lett. ${\bf{105}}$, 257202 (2010)).  We apply neutron scattering to investigate the magnetism and dynamics of this phase in a large single crystal which displays a low temperature spin glass transition ($T_{g} \sim$ 15 K, found with magnetization), but no observable macroscopic and spatially long-range antiferromagnetic order.  The static response in the low temperature cluster glass phase (sampled on the timescale set by our resolution) is found to be characterized by an average magnetic spin direction that lacks any preferred direction.  The dynamics that drive this phase are defined by a magnetic correlation length that gradually increases with decreasing temperature.  However, below $\sim$ 50 K the opposite is found with spatial correlations gradually becoming more short range indicative of increasing disorder on cooling, thereby unravelling magnetism, until the low temperature glass phase sets in at T$_{g}$ $\sim$ 15 K. Neutron spectroscopy is used to characterize the spin fluctuations in the cluster glass phase and are found to be defined by a broadband of frequencies on the scale of $\sim$ THz, termed here ``fast" fluctuations.  The frequency bandwidth driving the magnetic fluctuations mimics the correlation length and decreases until $\sim$ 50 K, and then increases again until the glass transition.  Through investigating the low-energy acoustic phonons we find evidence of multiple distinct structural regions which form the basis of the clusters, generating a significant amount of local disorder.  We suggest that random molecular fields originating from conflicting interactions between clusters is important for the destruction of magnetic order and the eventual formation of the cluster glass in PFN.

\end{abstract}


\maketitle

\section{Introduction}

Spin glasses are an example of a non-equilibrium phenomena and occur in magnets which have a large ground state spin degeneracy originating from conflicting interactions.~\cite{Binder86:58}  Spin glasses have been heavily studied in disordered metallic compounds with the magnetic susceptibility predominately characterized by a range of slow frequencies.  Relaxor ferroelectrics display some analogous traits to that of spin glasses with polar correlations displaying disorder and the dielectric properties being defined by a range of frequencies.  The magnetic relaxor PbFe$_{1/2}$Nb$_{1/2}$O$_{3}$ (PFN) is a material were both properties exist with relaxor ferroelectricity characterized by short-range polar correlations and also a low temperature spin-glass phase based upon clustering of magnetic Fe$^{3+}$ (S=${5\over2}$, L$\approx$0) ions.  In this paper we apply neutron scattering to study the static and dynamic magnetic properties in PFN.  We will show that the magnetism is driven by a broad frequency band of fluctuations on the $\sim$ THz frequency scale.  The results demonstrate the dynamic magnetic response in an insulating spin-glass built from clusters.

%Coupling magnetic and ferroelectric order parameters in bulk materials could provide the framework for new applications which require switching under applied electric and magnetic fields.~\cite{Gajek07:296,Chu08:7,Kimura03:426}  However, finding real stochiometric materials that exhibit such coupling has been challenging.~\cite{Eerenstein06:442,Cheong07:6,Bochenek08:114}  One possible series of candidates are the disordered lead-based relaxor ferroelectrics that contain a magnetic ion and display frequency and temperature broadened peaks in the dielectric response.  Indeed, motivated by this the relaxor PbFe$_{1/2}$Nb$_{1/2}$O$_{2}$ (PFN) has recently been found to display exceptionally large magnetoelectric coefficients at room temperature~\cite{Laguta16:51} and domain switching~\cite{Laguta17:95}, illustrating coupling between magnetic and polar spatial correlations.  We apply neutron scattering to study of PbFe$_{1/2}$Nb$_{1/2}$O$_{3}$ (PFN) to characterize the magnetic static and dynamic correlations that underpin this.

Relaxor ferroelectrics are materials that display exceptionally large piezoelectric coefficients~\cite{Park97:82}, particularly in single crystal form, making them of interest for applications.~\cite{Ye04:155,Ye09:34,Bokov06:41,Ye04:302,Ye96:184}  PbMg$_{1/3}$Nb$_{2/3}$O$_{3}$ (PMN) and PbZn$_{1/3}$Nb$_{2/3}$O$_{3}$ (PZN) are prototypical relaxors~\cite{Cowley11:60} that display a temperature broadened and frequency dependent dielectric response~\cite{Viehland91:43,Viehand92:46,Pirc07:76,Pirc01:63}.  This behavior contrasts with conventional ferroelectrics, such as PbTiO$_{3}$~\cite{Shirane51:6,Shirane70:2,Kempa06:41,Tomeno06:73,Hlinka06:73} (PT) where the dielectric constant is measured to have a well defined, and frequency independent peak, as a function of temperature indicative of ferroelectric order.  Nonmagnetic relaxors PMN and PZN display two distinct temperature scales.~\cite{Gehring09:79}  First, at a high temperature defined by the Burns temperature~\cite{Burns83:48}, an energy broadened~\cite{Gehring01:63,Hlinka03:91} zone center transverse optic mode softens to a minimum in energy.~\cite{Nab99:11,Gehring00:84,Gehring01:87,Waki02:65,Cao08:78,Vak02:66,Kamba05:17,Nuzhnyy17:96,Bishop10:81}  This softening is concomitant with the development of spatially short range~\cite{Vak89:90,You97:79,Dkhil01:65,Hirota02:65,Xu04:69,Mats06:74} and dynamic~\cite{Stock10:81,Xu12:86,Vak05:7} structural correlations characterized by momentum broadened diffuse scattering in the neutron and x-ray cross sections~\cite{Vak98:40,Vak95:37,Gvas03:63}.  This contrasts to expectations for a bulk ferroelectric transition, defined by a resolution limited Bragg peak as found in PbTiO$_{3}$~\cite{Shirane51:6}.  A second lower temperature scale is defined when long-range ferroelectric correlations can be stabilized under an applied electric field~\cite{Ye93:145,Stock07:76} and where local polar nanoregions become static.

These two temperature scales characterizing the dielectric properties of PMN and PZN can be understood in terms of random fields from the quenched dipolar disorder, originating from the valence difference on the $B$ site of the A$B$X$_{3}$ perovskite lattice.  These disorder induced random dipolar fields~\cite{Westphal92:68,Fisch03:67,Stock04:69} have been suggested to be a unique experimental realization in condensed matter physics of a continuous phase transition in the presence of random fields~\cite{Cowley09:378}.  The case of discrete, or Ising, symmetries in a random field have been extensively studied in magnetic systems where the application of a magnetic field can induce random fields~\cite{Imry75:35} in the magnetic Hamiltonian through the ``Aharony-trick".~\cite{Fishman79:12}  Given that applied magnetic fields also break the rotational symmetry, such an experimental technique cannot be applied to continuous phase transitions and hence has been confined to the study of Ising magnets~\cite{Birgeneau83:28}.  Random fields in continuous symmetries have been investigated in soft matter systems, such as liquid crystals~\cite{Bellini01:294,Park02:65}, or quantum liquids (such as liquid helium)~\cite{Reppy92:87} in aerogel or aerosil frameworks.  

PbFe$_{1/2}$Nb$_{1/2}$O$_{3}$ (PFN) is a relaxor ferroelectric with similar dielectric properties to that discussed for PMN and PZN discussed above~\cite{Gridnev12:54,Brze18:737} undergoing a ferroelectric transition at $\sim$ 400 K and displaying spatially short-range polar correlations similar to the polar nanoregions in PMN and PZN~\cite{Burns83:48}.  Interest in PFN originates from efforts in multiferroics ~\cite{Cheong07:6,Kimura03:426,Santos02:122} where strong coupling between ferromagnetism and ferroelectricity is sought affording external control of either magnetic or ferroelectric domains.  In this context, PFN has shown large magnetoelectric coupling and domain switching at room temperature~\cite{Laguta16:51,Laguta17:95} illustrating that magnetic relaxors maybe a promising avenue to study for coupling between spatial magnetic and polar correlations.  Evidence for coupling between dielectric and magnetic properties has been further reported in Ref. \onlinecite{Yang04:70,Correa11:83,Garcia11:23,Turik13:88} based on dielectric and magnetic susceptibility.  

PFN belongs to a series of compounds which include  PbFe$_{1/2}$Ta$_{1/2}$O$_{3}$ (PFT)~\cite{Chillal14:89} and PbFe$_{2/3}$W$_{1/3}$O$_{3}$ (PFW)~\cite{Chen16:6} that contain a magnetic Fe$^{3+}$ (S=5/2, L$\approx$0)  ion~\cite{Pavlenko12:38}.  Depending on the subtle change in the iron concentration in PFN (see for example Ref. \onlinecite{Raevski19:542}), it can display an antiferromagnetic transition at T$_{N}$ $\sim$ 150 K or a complete paramagnetic response at all temperatures.  Studies of the bulk ordered magnetic moment in samples where a magnetic Bragg peak have been observed have reported heavily reduced magnetic moments from the predicted $gS=5$ $\mu_{B}$ for Fe$^{3+}$ ions with an example of only 2.55 $\mu_{B}$ reported in Ref. \onlinecite{Pietrzak81:65}.  This supports the notion that PFN is based on spatially separated clusters.  At lower temperatures defined by T$_{g}$ $\sim$ 15 K, PFN undergoes a glass transition characterized by a differing zero/field cooled susceptibilities.  This spin-glass phase has been found to be characterized by clusters of Fe$^{3+}$ rich regions and hence termed a ``cluster glass".~\cite{Kleeman10:105}

In this paper, we investigate the magnetic and structural properties in single crystalline PFN that define the cluster glass phase.  This manuscript is divided into four parts including this introduction.  We first outline the various neutron instruments used to investigate the structural and magnetic response.  We then discuss the structural properties by studying the nuclear Bragg peaks, momentum broadened diffuse scattering, and long wavelength acoustic phonons.  This is followed by an investigation of the static and dynamic magnetic properties using both polarized and unpolarized diffractometers and triple-axis spectrometers.  We show in this section a crossover to where the magnetism is increasingly disordered as the temperature is decreased.  We then finish the paper with a discussion of the crossover from paramagnetism to glassy dynamics in PFN on cooling and its origin.

\section{Experiment}

Neutron scattering experiments were performed on different instruments to compare the structural and magnetic responses in PFN.  This section outlines details on sample characterization, different neutron experiment setups, and then the formalism used to describe the data through the remainder of the paper.

\subsection{PFN sample and characterization}

The single crystal was grown using the modified Bridgeman technique (outlined in Ref. \onlinecite{Luo00:39}) following the procedure in Ref. \onlinecite{Kozlenko14:89}.  The structural properties of our single crystal have previously been studied with neutron spectroscopy~\cite{Stock13:88} which observed a zone center transverse optic phonon softening at $\sim$ 400 K consistent with the formation of ferroelectric order.  Raman scattering work~\cite{Wilfong16:47} on a piece of our large single crystal previously reported that this corresponded to two ferroelectric transitions from a cubic to tetragonal ($P4mm$) then rhombohedral ($R3m$) unit cells (on cooling) which is consistent with the structural properties reported using x-ray diffraction~\cite{Bonny97:102}.  This transition has further been confirmed by us using high resolution neutron powder diffraction (HRPD,ISIS) on a piece of our single crystal confirming the structural transitions reported in Refs. \onlinecite{Bonny97:102,Wilfong16:47}. We note that due to the comparatively poorer momentum resolution afforded by neutrons and also the formation of domains which preserve the bulk high temperature structural symmetry, we index the reciprocal lattice using an average cubic unit cell at all temperatures in this paper.

We note that our sample does not show a sharp peak in the magnetic susceptibility~\cite{Stock13:88} characteristic of spatially long-range magnetic order which is highly sensitive to Fe$^{3+}$ ordering and relative Fe$^{3+}$ and Nb$^{5+}$ concentrations.  While single crystal neutron diffraction on our sample finds equal Nb:Fe concentrations within a few percent, small changes in this relative concentration have been shown to tune the presence of a spatially long-range antiferromagnetic component in samples~\cite{Zakorodniy18:2}, with even the complete absence of long-range magnetic order reported in some samples~\cite{Bochenek18:11}.  We emphasize that single crystal samples that have displayed this peak in the susceptibility, and that have been analyzed with neutron diffraction, have shown both spatially short-range and long-range components.  As far as we are aware however, all reported samples show ferroelectric transitions at $\sim$400 K and a glass transition below $\sim$ 15 K, both of which are displayed in our sample studied here.  The presence of a spatially long-ranged magnetic component is not required for the formation of the low temperature spin-glass phase which we study in this paper.

\subsection{Instruments}

\textit{Structural diffuse scattering with unpolarized neutrons (TASP and SPINS) :}  The use of cold neutrons has been shown to be versatile in studying structural properties, and in particular the momentum broadened diffuse scattering cross section, in conventional nonmagnetic relaxors like PMN.~\cite{Hiraka04:70}   Motivated by this previous work, we investigated the temperature dependent structural response, both spatially long- and short- range, using cold triple-axis spectrometers TASP (PSI) and SPINS (NIST).  On SPINS, the collimation sequences was set to $open$-80$'$-$S$-80$'$-$open$ with the incident and final energies fixed to E$_{i}$=5.0 meV using a PG(002) vertically focussed monochromator and flat PG(002) analyzer crystals.   Cooled Beryllium filters were placed before and after the sample to remove higher order contamination of the neutron beam.  On TASP, the same collimation sequence was used as on SPINS with E$_{i}$=4.7 meV and a Beryllium filter was used on the scattered side.

\textit{Magnetic diffuse scattering with polarized (DNS and D7) and unpolarized (TASP and PRISMA) neutrons:}  The magnetic diffuse scattering in PFN, indicative of spatially short-range correlations, was investigated using several instruments.   Unpolarized neutron scattering was performed using TASP (with the same configuration stated above) and the time of flight instrument PRISMA (ISIS, UK).  PRISMA was used in diffraction mode, integrating over energy transfers and using time-of-flight for the kinematics as described in Ref. \onlinecite{Zinkin97:56}. 

To confirm the magnetic nature of the scattering, characteristic of the underlying average magnetic structure of the Fe$^{3+}$ ions,  and also to separate nuclear and magnetic contributions to the neutron cross section, we initially used the DNS~\cite{Schwieka01:297} diffractometer (MLZ, Garching, Germany).  The incident energy was fixed at E$_{i}$=3.7 meV with no energy analysis on the scattered side (two-axis mode) therefore approximately measuring the energy integrated structure factor $S(\vec{Q})\equiv \int dE S(\vec{Q}, E)$.   The incident beam was monochromated using a horizontally and vertically focused PG002 monochromator and polarized with $m=3$ Sch\"{a}rpf supermirrors.  The direction of polarization at the sample was defined using an $XYZ$ geometry described in the Appendix. The flipping ratio was 20 $\pm$ 1.  Experiments on D7~\cite{Stewart09:42} (ILL) involved a similar setup to that of DNS (MLZ) but used an incident energy of E$_{i}$=3.6 meV.  

\textit{Magnetic and structural dynamics measured with a cold triple-axis (TASP and SPINS):}  Cold triple-axis measurements were done on SPINS at NIST (Gaithersburg, USA) and TASP (PSI, Switzerland) to investigate low-energy magnetic dynamics and acoustic phonons.  On SPINS (NIST), the collimation sequence was set to $open$-80$'$-$S$-80$'$-$open$ with a cooled Beryllium filter on the scattered side to remove higher order contamination of the neutron beam.  The incident beam energy was selected with a vertically focused PG(002) graphite monochromator and the final energy was fixed at E$_{f}$=5.0 meV using a flat PG(002) analyzer crystal.  On TASP (PSI), the collimation was set to $open$-$open$-$S$-$open$-$open$ with E$_{f}$=4.7 meV with a curved analyzer.  A cooled Beryllium filter was used on the scattered side.  

\textit{Structural dynamics measured with a thermal triple-axis (BT4):}  To probe the structural dynamics and in particular the zone center transverse optic mode, we used the BT4 thermal triple-axis measurements at NIST (Gaithersburg, USA).   The collimation was set to $open$-80$'$-$S$-80$'$-$open$ with a graphite filter placed on the scattered side of the sample to remove higher order contamination of the neutron beam.  The incident energy was selected with a vertically focused PG(002) graphite monochromator and the final energy was fixed to E$_{f}$=14.7 meV using a flat PG(002) analyzer crystal.

\subsection{Neutron Cross Section}

The measured neutron scattering intensity at a particular energy transfer defined $\hbar\omega \equiv E\equiv E_{i}-E_{f}$ on a reactor based instrument, with a monitor detector before the sample, is directly proportional to the structure factor $S(\vec{Q},E)$.  This in turn is related to the imaginary part of the susceptibility $\chi''(Q,\omega)$ by,

\begin{equation}\label{eq:def}
	I(\vec{Q},E) \propto S(\vec{Q},E) \equiv {1 \over \pi} [n(E)+1] \chi''(\vec{Q},E),
\end{equation} 

\noindent where $[n(E)+1]$ is the Bose thermal population factor.  For the purposes of this paper, we divide the susceptibility up into two terms depending on momentum and energy transfer as follows,

\begin{equation}\label{eq:suscept}
	\chi''(\vec{Q}, E) = \chi(\vec{Q}) F(E),
\end{equation} 

\noindent with $\int_{-\infty}^{\infty} dE \ F(E)=1$.  We describe the particular form for both the momentum dependent $\chi(\vec{Q})$ and energy dependent $F(E)$ in the text below.

\section{Results}

\subsection{Structural properties:}

\begin{figure}
	\includegraphics[width=88mm,trim=0.25cm 0.2cm 1.5cm 1.0cm,clip=true]{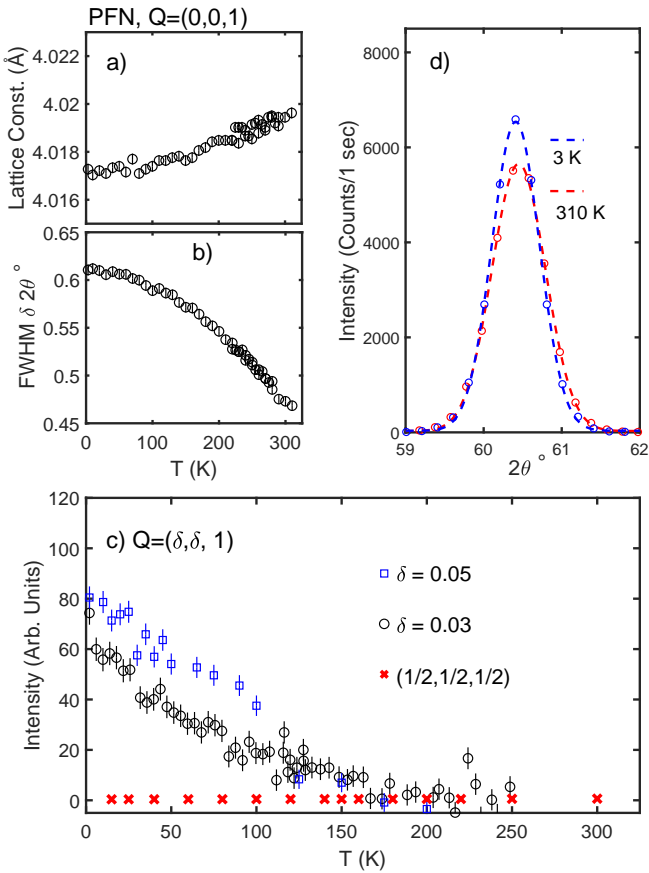}
	\caption{ \label{fig:structure_bragg} Temperature dependence of the structural properties of our 1 cm$^{3}$ crystal.  $(a)$ illustrates the lattice constant as a function of temperature and $(b)$ the full width of the $\vec{Q}$=(0,0,1) Bragg peak as a function of temperature.  $(c)$ compares the temperature dependent structural diffuse scattering near $\vec{Q}$=(0, 0, 1) with the structural intensity at $\vec{Q}$=(1/2, 1/2, 1/2) measured with polarized neutrons on DNS.  $(d)$ display representative scans through the $\vec{Q}$=(0,0,1) Bragg peak measured on SPINS (note the errorbars are approximately the size of the data points). }
\end{figure} 

\begin{figure}
	\includegraphics[width=88mm]{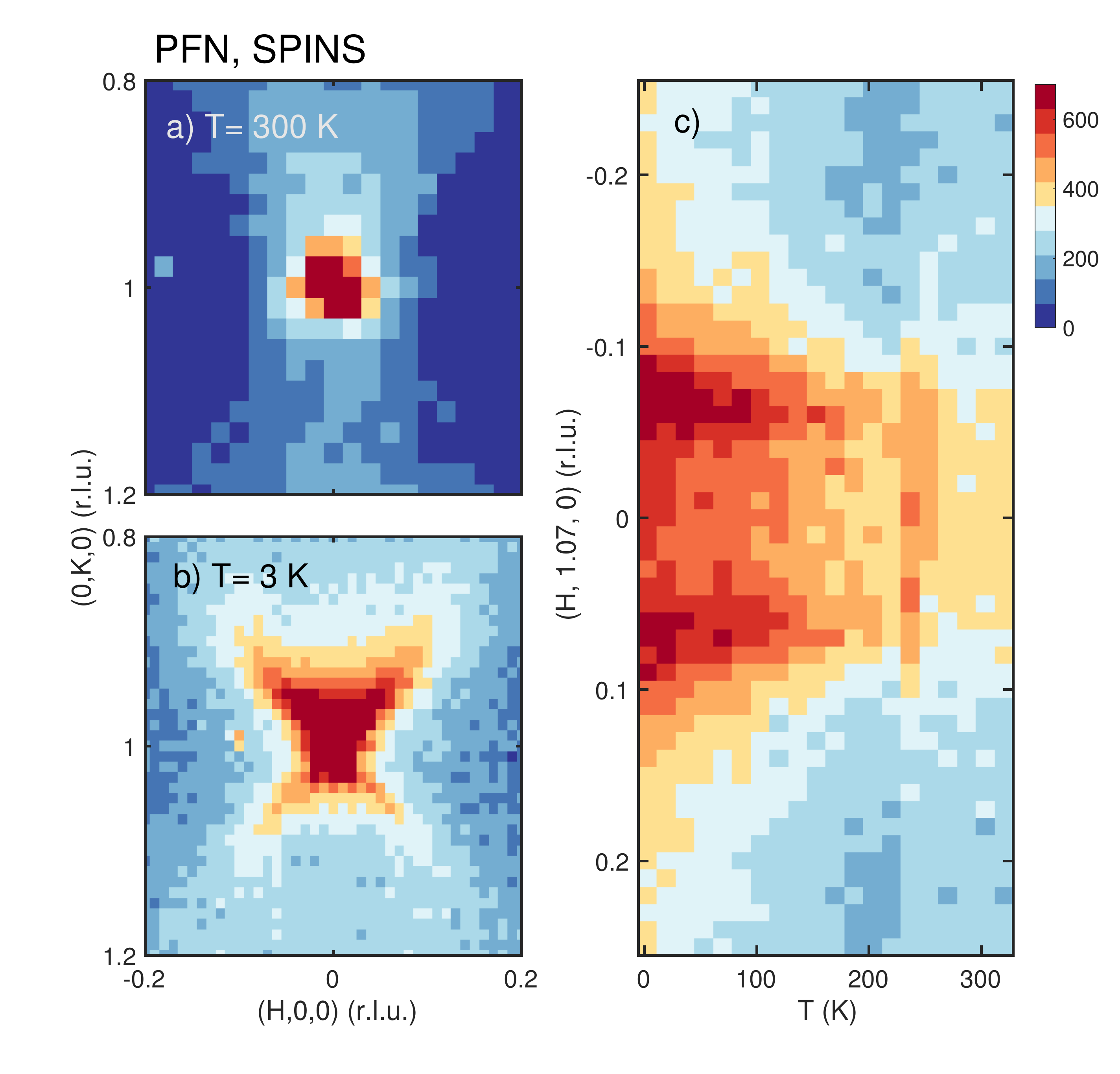}
	\caption{ \label{fig:structure_diffuse} The structural diffuse scattering near $\vec{Q}$=(0, 0, 1) at $(a)$ T=300 K and $(b)$ T=3 K measured on SPINS.  The temperature dependence of a line cut through the diffuse scattering at $\vec{Q}$=(H, 1.07, 0) (r.l.u.) is shown in $(c)$.}
\end{figure} 

The magnetic and structural energy scales are different in PFN.   PFN has a magnetic Curie-Weiss temperature of $\Theta_{CW}$ $\sim$ - 250 K (Figure 1 of Ref. \onlinecite{Stock13:88}) while the zone center soft transverse optic mode reaches an energy minimum at $\sim$ 400 K~\cite{Stock13:88,Wilfong15:70} defining the ferroelectric transition.  We first present the structural properties below the ferroelectric transition temperature, but at temperatures relevant to the magnetic response which we discuss later in this paper.  

\subsubsection{Static structural properties:}

We first experimentally investigate the structural properties of our PFN crystal with neutron scattering and are summarized in Fig. \ref{fig:structure_bragg} applying the cold neutron spectrometer SPINS to investigate the elastic scattering near the $\vec{Q}$=(0,0,1) Bragg peak.   Fig. \ref{fig:structure_bragg} $(a-b)$ plot the measured lattice constant $a$ and also the longitudinal full-width at half maximum $\delta 2\theta$ as a function of temperature below 300 K (note this is below the ferroelectric transition at $\sim$ 400 K).  Representative $\theta-2\theta$ scans are illustrated in Fig. \ref{fig:structure_bragg} $(d)$.    The lattice constant in panel Fig. \ref{fig:structure_bragg} $(a)$ shows a monotonic decrease as the temperature is lowered with no observable discontinuity that would indicate a bulk structural transition and a similar featureless trend is seen in the longitudinal linewidth Fig. \ref{fig:structure_bragg} $(b)$ which only shows a very small (on the scale of the resolution) and monotonic change with temperature.  

%If a structural transition was present, a splitting of the $\vec{Q}$=(0,0,1) Bragg peak or at least a discontinuous longitudinal broadening at a particular temperature would be expected.  Examples of such a response is illustrated in previous works on neutron scattering in bulk PMN-60PT~\cite{Stock06:73} or the skin region of PZN~\cite{Xu04:70} with x-rays. There is no such observable structural transition or discontinuity in the lattice constant, within the resolution of our experiments.  We note that this is consistent with the ``Invar-like" response reported in Ref. \onlinecite{Pavlenko12:57}.

In Fig. \ref{fig:structure_bragg} $(c)$, we plot the temperature dependence of the momentum broadened structural diffuse scattering near $\vec{Q}$=(0,0,1) which displays an increase in intensity on cooling.  This scattering is outside the momentum resolution of the spectrometer and indicative of spatially short range correlations. The data at $\vec{Q}$=(${1\over 2}$,${1\over 2}$,${1\over 2}$) are taken at a position sensitive to the magnetic scattering discussed below, however applying polarized neutrons to extract the nuclear component as discussed in the Appendix.  There is no observable change in the nuclear component of the scattering at $\vec{Q}$=(${1\over 2}$,${1\over 2}$,${1\over 2}$).  In terms of the nuclear Brillouin zone this is the $R$ point zone boundary (taking the space group as $Pm\overline{3}m$) and temperature dependent structural diffuse scattering has been reported at this point in PMN.~\cite{Swainson09:79}  Such zone boundary scattering is not clearly evident in our PFN crystal and is suggestive that the Fe$^{3+}$ and Nb$^{5+}$ ions, though in approximately equal 50 \% stochiometric amounts, are disordered in our single crystal, consistent with previous powder diffraction~\cite{Lampis99:11} and dielectric~\cite{Maj06:99} studies.  Temperature dependent nuclear diffuse scattering in PFN exists near the nuclear Bragg peaks, indicative of spatially short range correlations, but not at the zone boundaries.

Mesh scans near $\vec{Q}$=(0,1,0) (Fig. \ref{fig:structure_diffuse} $(a,b)$) illustrate the presence of both resolution limited Bragg peaks (discussed above) and momentum broadened ridges of scattering.  The diffuse ridges of scattering extend along the [1$\overline{1}$0] directions and are qualitatively consistent with the diffuse scattering reported in non magnetic relaxors PMN~\cite{You97:79,Hirota02:65,Xu04:69} and PZN~\cite{Xu04:70,Stock04:69}.  The temperature dependence is plotted in Fig. \ref{fig:structure_bragg} $(c)$ (discussed above) and is further illustrated in Fig. \ref{fig:structure_diffuse} $(c)$ through a contour plot of the elastic momentum line scans along the (H, 1.07, 0) direction.  The cut extends through the two ridges of the ``butterfly'' nuclear diffuse scattering shown in panel Fig. \ref{fig:structure_diffuse} $(a, b)$ and further shows a smooth increase in intensity as the temperature is lowered.  The presence of diffuse scattering which grows with decreasing temperature is indicative of localized polar correlations analogous to prototypical relaxors PMN and PZN.  There presence indicates localized structural disorder as characterized extensively in prototypical relaxors.

\subsubsection{Low-energy lattice vibrations and phonons:}

\begin{figure}
	\includegraphics[width=95mm]{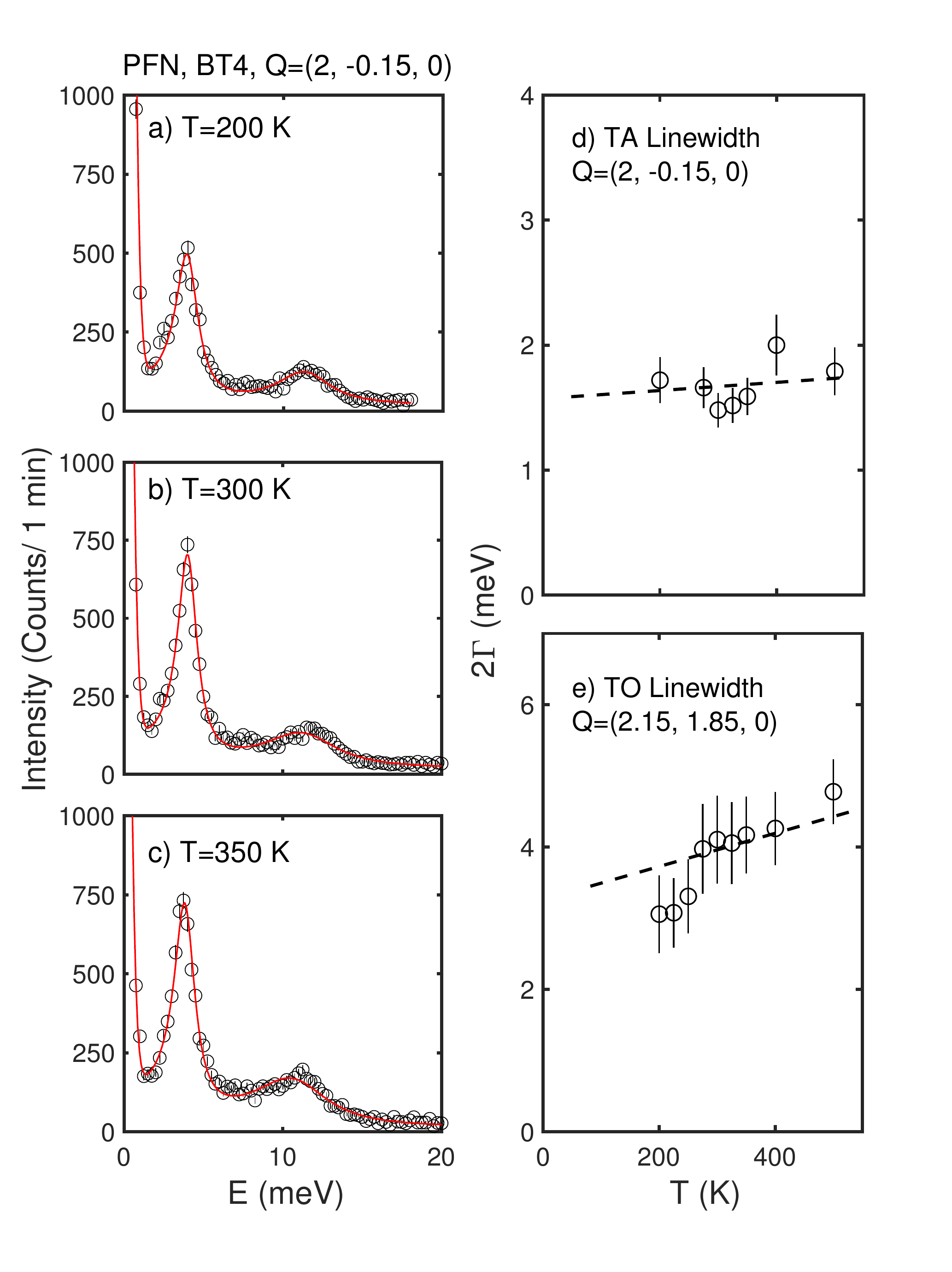}
	\caption{ \label{fig:BT4_summary} The temperature dependence of the optic and acoustic phonons near $\vec{Q}$=(2, 0, 0).  $(a-c)$ illustrates constant momentum cuts at $\vec{Q}$=(2, -0.15, 0).  The temperature dependence of the linewidth $2\Gamma$ of the transverse $(d)$ acoustic and  $(e)$ optic phonons.}
\end{figure} 

Having discussed the continuous static response at low temperatures where magnetism is relevant, we now discuss the phonons.  Fig. \ref{fig:BT4_summary} $(a-c)$ displays a series of constant momentum scans at $\vec{Q}$=(2, -0.15, 0) finding two peaks characterizing the lower energy transverse acoustic and a higher energy transverse optic modes.  The data displays phonons with the solid curve a fit to the sum of two simple harmonic oscillators with each oscillator being defined by antisymmetric lorentzians,

\begin{equation}\label{eq:SHO}
	F (E) \propto \left({1 \over {1+\left({{\omega-\Omega_{0}} \over \Gamma}\right)^{2}}}- {1 \over {1+\left({{\omega+\Omega_{0}} \over \Gamma}\right)^{2}}}\right) 
\end{equation}

\noindent with $\hbar \Gamma$ the energy linewidth, inversely proportional to the lifetime $\tau$, and with an energy position of $\hbar\Omega_{0}$.  The lorentzian at $+\hbar\Omega$ in Eqn. \ref{eq:SHO} accounts for the cross section for neutron energy gain (energy transfer negative $E < 0$).  This is required for the cross section to follow detailed balance which implies the imaginary part of the susceptibility $\chi''(\vec{Q},E)$ is an odd function in energy.  

\begin{figure}
	\includegraphics[width=80mm]{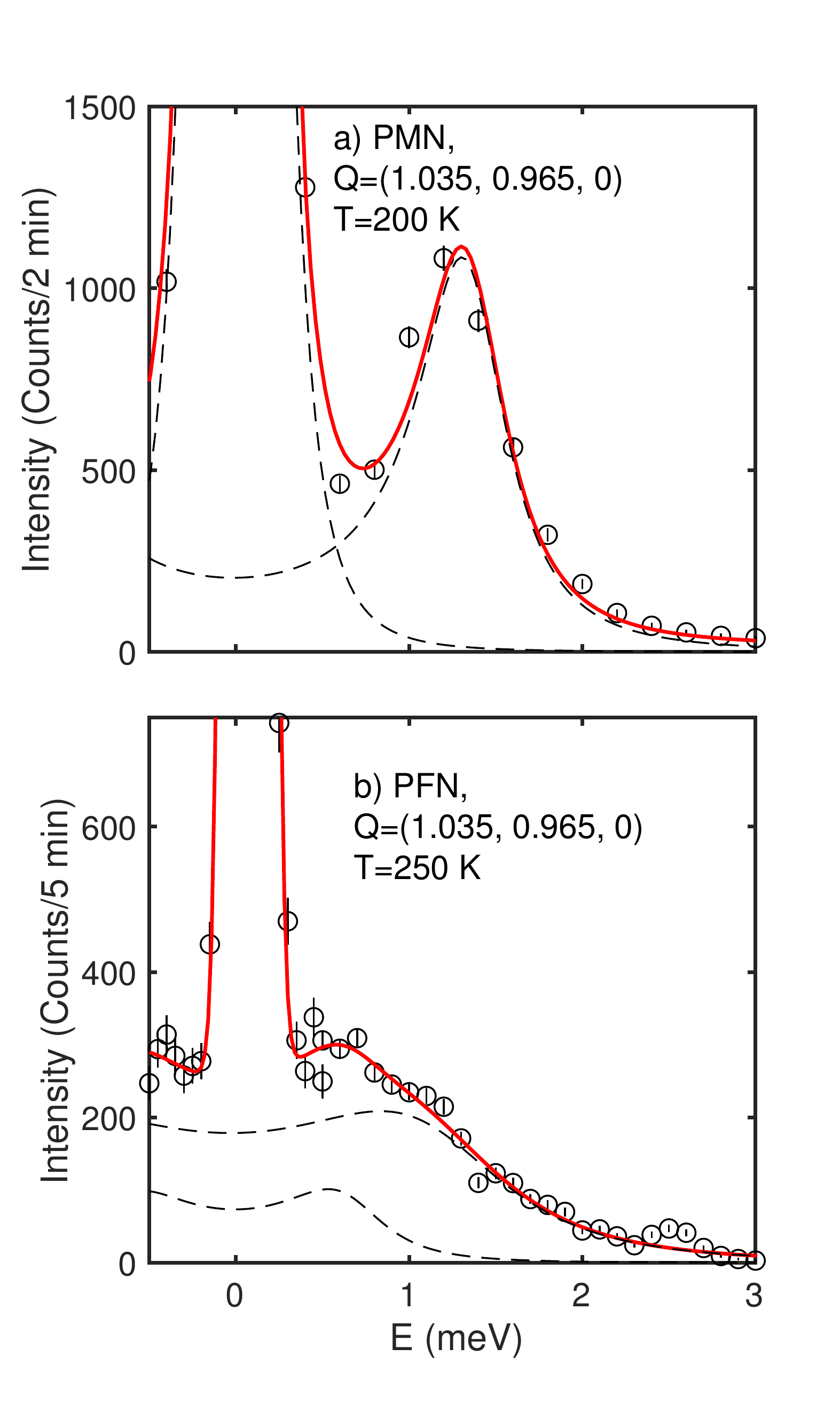}
	\caption{ \label{fig:compare_PMN} A comparison of the low-energy T$_{2}$ acoustic phonons near $\vec{Q}$=(1, 1, 0) measured on the non magnetic relaxor $(a)$  PMN and the magnetic relaxor $(b)$ PFN with identical spectrometer configurations on SPINS. }
\end{figure}

\begin{figure}
	\includegraphics[width=95mm]{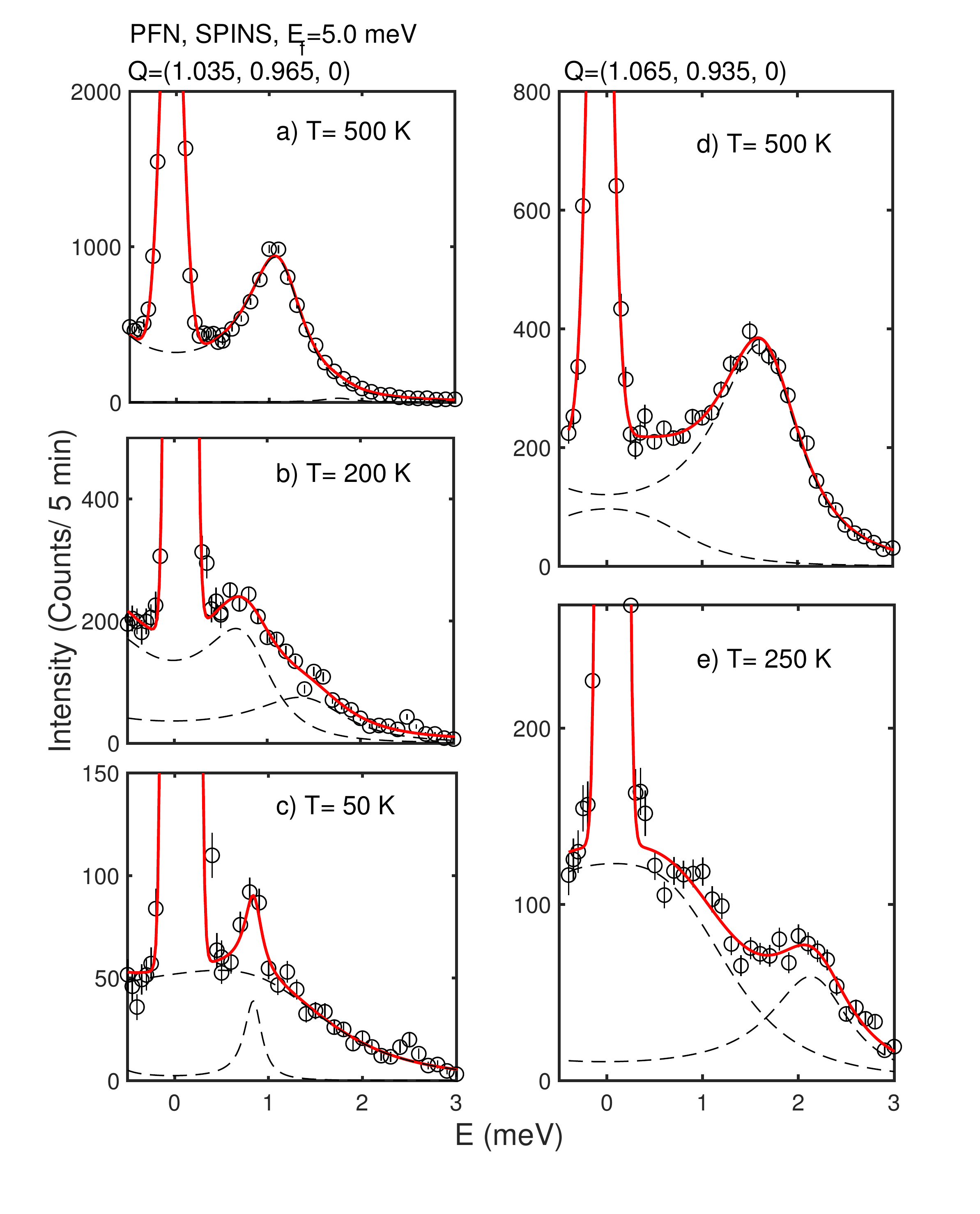}
	\caption{ \label{fig:SPINS_T} The temperature dependence of the T$_{2}$ acoustic phonons in PFN measured on SPINS.  $(a-c)$ show constant momentum cuts $\vec{Q}$=(1.035, 0.965, 0) and $(d-e)$ illustrate scans taken at $\vec{Q}$=(1.065, 0.935, 0).}
\end{figure} 

%\begin{figure}
%	\includegraphics[width=95mm]{SPINS_QT_depend.pdf}
%	\caption{ \label{fig:SPINS_QT} The momentum dependence of the T$_{2}$ acoustic phonons in PFN at $(a-b)$ T=100 K and $(c-e)$ T= 500 K measured on SPINS.}
%\end{figure} 

Both acoustic and optic modes at $\vec{Q}$=(2,-0.15,0) are underdamped and well defined in energy and this is confirmed in Fig. \ref{fig:BT4_summary} $(d,e)$ which do not show any large anomaly over the temperature range studied.  This response is in agreement with non magnetic relaxors PMN and PZN measured at similar positions in momentum away from the nuclear zone center where the waterfall effect~\cite{Gehring00:84,Gehring01:63,Stock18:2} is present.  However, this is in contrast to the phonon response near $\vec{Q}$=(1,1,0). Figure \ref{fig:compare_PMN} illustrates a comparison between constant momentum cuts in non magnetic PMN (taken from Ref. \onlinecite{Stock12:86}) compared to PFN.  Both scans are performed at the same position in momentum using the same instrumental setup on SPINS.   These scans are taken along the transverse direction near the $\vec{Q}$=(1,1,0) Bragg position and are therefore sensitive to the $T_{2}$ acoustic phonons which propagate along [1$\overline{1}$0] and are polarized along [110].  The PMN data displays a resolution limited acoustic phonon described by the lineshape above, however the PFN data is indicative of a broader energy lineshape.  Given the lack of any other model, we have chosen to fit the data to two damped harmonic oscillators as illustrated by the curves in Fig. \ref{fig:compare_PMN} to provide a description of this multi component lineshape.

As noted in Refs. \onlinecite{Stock05:74,Stock12:86,Hiraka04:70}, the dynamic lineshape near $\vec{Q}$=(1,1,0) is complicated over measurements near $\vec{Q}$=(2,0,0) by the presence of a strong elastic diffuse scattering cross section which contributes to a low-energy quasielastic-like~\cite{Gvas05:17,Gvas04:69,Gvas04:49} response.  This has been interpreted as either a coupling between harmonic phonons with relaxational dynamics~\cite{Stock05:74,Stock12:86} or in terms of quasielastic scattering indicative of non harmonic relaxation.  However, as displayed in Fig. \ref{fig:SPINS_T}, the lineshape near $\vec{Q}$=(1,1,0) is much more complex than reported previously in PMN.  Fig. \ref{fig:SPINS_T} $(a)$ shows that at T=500 K the linshape is dominated by a single underdamped phonon mode.  At T=200 K (Fig. \ref{fig:SPINS_T} $(b)$) the lineshape becomes more intricate and is well described by two temporally damped harmonic oscillators.  At low temperature of T=50 K (Fig. \ref{fig:SPINS_T} $(c)$), the lineshape is described by two distinct components with a sharp underdamped phonon mode and a much broader in energy overdamped mode.  The development of a multi component lineshape is also found at larger momentum transfers as illustrated in Fig. \ref{fig:SPINS_T} $(d,e)$ which display constant momentum scans at T=500 K and 250 K.  The multiple components are particularly clear at 250 K in panel $(e)$.

%The momentum dependence is shown in Fig. \ref{fig:SPINS_QT} at T=100 K (panels $(a,b)$) and 500 K (panels $(c-e)$). At T=100 K, the several components of the lineshape become distinct at larger momentum transfers and at least two peaks are apparent in Fig. \ref{fig:SPINS_QT} $(b)$ at $\vec{Q}$=(1.05, 0.95,0).  At 500 K shown in Fig. \ref{fig:SPINS_QT} $(c-e)$, as the momentum transfer is increased the phonon lineshape broadens in energy indicative of a shorter lifetime for the long-wavelength acoustic phonons, however the lineshape is dominated by a single damped harmonic oscillator.    

\begin{figure*}
 \includegraphics[trim=0.0cm 0.0cm 0.0cm 0.0cm,width=150mm]{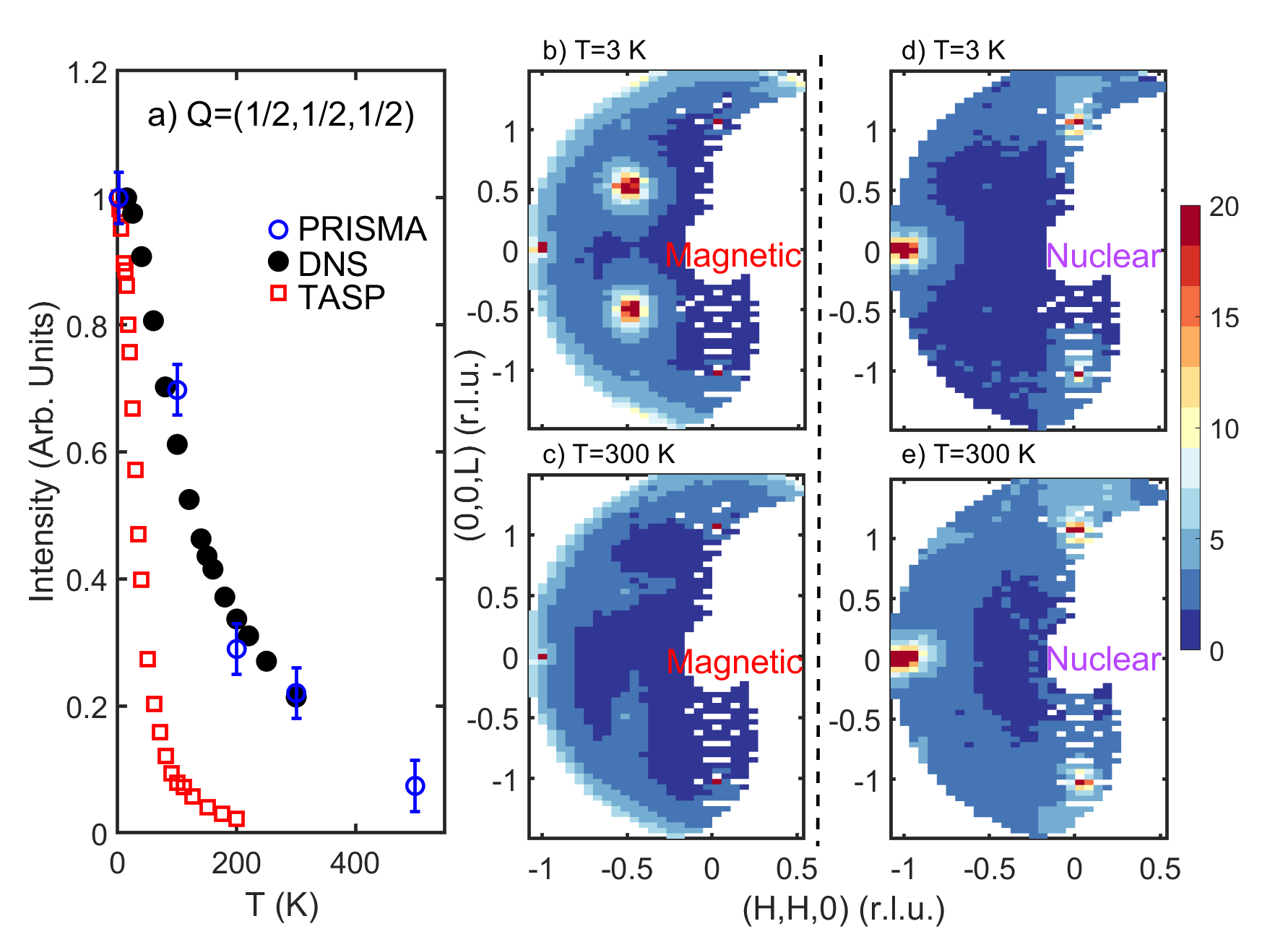}
 \caption{ \label{fig:mag_T} The magnetic diffuse scattering in PFN.  $(a)$ The temperature dependence of the magnetic intensity at $\vec{Q}$=(1/2, 1/2, 1/2) measured on energy integrating instruments DNS (MLZ) and PRISMA (ISIS).  These are compared against measurements on TASP (PSI) with an energy resolution of $2\delta$E=0.1 meV.  Note that all measurements have been normalized to the intensity at the base temperature.  $(b-c)$ illustrate the purely magnetic diffuse scattering taken at T=3 and 300 K, respectively, measured using polarized neutrons at DNS (MLZ).  We note that the intense scattering at the Bragg position of $\vec{Q}$=(-1,-1,0) is due to incomplete polarization of the neutron beam discussed in the appendix.   This is compared to the extracted nuclear cross section in panels $(d-e)$.}
\end{figure*} 

While non magnetic PMN displays a broadening of the acoustic phonons near $\vec{Q}$=(1,1,0)~\cite{Stock12:86,Stock05:74,Hiraka04:70}, the lineshapes shown above in PFN are more complicated and require at least two components to parameterize.  We will discuss this below in the context of the magnetic properties, however these indicate a strong amount of localized structural disorder in PFN not observed in prototypical relaxors PMN and PZN.  We note that this breakdown of well defined phonons is not due to any distinct low temperature distortion given the continuous response observed in the statics discussed in the previous section.

\subsection{Magnetism in PFN:}

We now discuss the magnetic properties originating from the Fe$^{3+}$ ($S$=5/2, $L$ $\approx$ 0) moments.  First, we show energy integrated data which is an approximate measure of $S(\vec{Q})$ that defines the static magnetic response.  Second, we show triple-axis measurements which are energy resolved to discuss the static (on the timescale of our resolution) and dynamic magnetism.  We note that, owing to resolution, neutrons are sensitive to dynamics on $\sim$ THz frequency scale in comparison to the slower dynamics $\sim$ MHz probed with, for example, muons.

\subsubsection{Energy integrated magnetic diffuse scattering:}

\begin{figure}
 \includegraphics[trim=0.6cm 2.2cm 0.5cm 0.0cm,width=85mm]{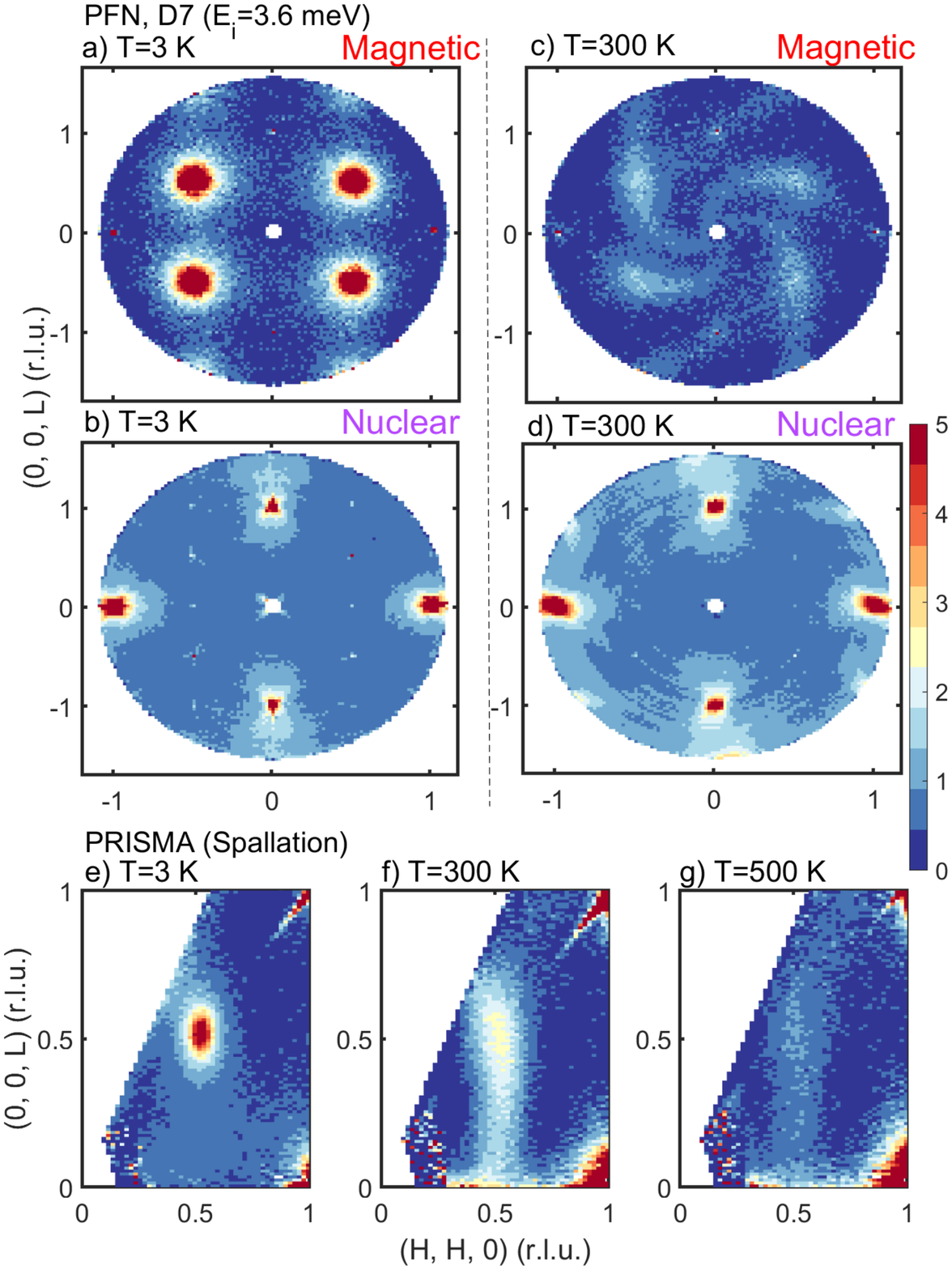}
 \caption{ \label{fig:mag_diffuse} The magnetic diffuse scattering measured with energy integrating neutron instruments.  Polarized neutrons are used from D7 (ILL) to compare magnetic and nuclear cross sections at T=3 K and 300 K in $(a-d)$.  Unpolarized measurements from PRISMA (ISIS) are showin in $(e-g)$.}
\end{figure} 

Figure \ref{fig:mag_T} illustrates a summary of the elastic magnetic properties as a function of temperature using both polarized (DNS) and unpolarized neutrons (PRISMA and TASP).  Fig. \ref{fig:mag_T} $(a)$ plots the temperature dependence of the momentum integrated magnetic scattering at $\vec{Q}$=(${1\over 2}$,${1\over 2}$,${1\over 2}$) corresponding to antiferromagnetically correlated spins along the three crystallographic directions.   Three different datasets from PRISMA, DNS, and TASP are shown normalized to base temperature.  Differences are seen in the temperature dependence of TASP data in comparison with PRISMA and DNS.  While the PRISMA and DNS data illustrate a consistent temperature dependence with a gradual and continuous increase of intensity with decreasing temperature, TASP displays more of a well defined intensity increase at low temperatures.  We note that none of these data sets is representative of a sharp increase of intensity, indicative of a clear phase transition defined by a critical temperature where a spatially long-range magnetic order parameter appears that subsequently saturates at low temperatures.

The difference between the temperature dependences from the three different data sets can be understood by the differing energy integration of the corresponding instruments resulting from the differing energy resolutions.~\cite{Murani78:41}  Both DNS and PRISMA are two-axis diffractometers which integrate the dynamics owing to the lack of any energy analysis.  TASP is a triple-axis with an analyzer PG(002) crystal that selects a particular final energy.  The energy resolution on TASP for the data discussed here is $2\delta E$ =0.1 meV  where $\delta E$ is the half-width at half maximum in energy.  We therefore conclude the presence of a significant amount of dynamic spectral weight which gradually slows and enters the elastic resolution with decreasing temperature which causes a gradual increase in intensity for diffractometers that integrate over the dynamics.  It also results in the lower onset temperature on TASP as the dynamics which slow will enter the time window on TASP at a lower temperature.  

Figs. \ref{fig:mag_T} $(b-e)$ illustrate the polarization analysis described in the Appendix separating nuclear and magnetic contributions.   Figs.  \ref{fig:mag_T} $(b-c)$ illustrate the purely magnetic contribution at 3 K and 300 K showing the presence of magnetic correlations which develop at $\vec{Q}$=(${1\over 2}$,${1\over 2}$,${1\over 2}$).  We note that this is not a Bragg peak as the magnetic correlations are extended in momentum indicative of spatially short-range correlations.   Figs. \ref{fig:mag_T} $(d,e)$ display the nuclear cross section which shows the absence of observable nuclear scattering at the $\vec{Q}$=(${1\over 2}$,${1\over 2}$,${1\over 2}$) antiferromagnetic position, however does illustrate the diffuse nuclear scattering near $\vec{Q}$=(0,0,$\pm$1) discussed above.

A more detailed momentum dependence is displayed in Fig. \ref{fig:mag_diffuse} taken on the D7 (ILL) polarized diffractometer in panels $(a-d)$ and unpolarized data from PRISMA in panels $(e-g)$.  The polarized data in panels $(a-d)$ are separated into magnetic and nuclear components for datasets taken at T=3 K and T=300 K. While the low-temperature polarized data show symmetric momentum broadened peaks at $\vec{Q}$=(${1\over 2}$,${1\over 2}$,${1\over 2}$), the high temperature T=300 K maps shown in panels Figs. \ref{fig:mag_diffuse} $(c)$ and Fig. \ref{fig:mag_T} $(c)$ illustrate an unusual lineshape in momentum.  In particular, the T=300 K magnetic cross section (Fig. \ref{fig:mag_diffuse} $(c)$) displays a ``wheel-like" response which is unphysical given the average cubic nature of the crystal structure.  Such lineshapes have been reported in the study of energy integrated diffuse scattering and have been referred to as ``Catherine Wheels".~\cite{Welberry03:36,Hohlwein03:68,Zeiske97:241,Hohlwein97:234}  The extended and curved lineshape is further confirmed using PRISMA in panels Fig. \ref{fig:mag_diffuse} $(e-g)$ and is arguably more pronounced.  We note that PRISMA integrates over a broader energy range than DNS, owing to the fact it is on a spallation source, and therefore this is suggestive that the lack of energy analysis is the origin of this unphysical lineshape.  We discuss this in the Appendix and show that it is an instrument artifact resulting from the energy integration on D7, DNS, and PRISMA combined with the presence of thermally excited magnetic dynamics sampled through the energy gain neutron cross section.

\subsubsection{Static Magnetism}

\begin{figure}
	\includegraphics[trim=2.2cm 3.5cm 2.2cm 2.0cm,width=90mm]{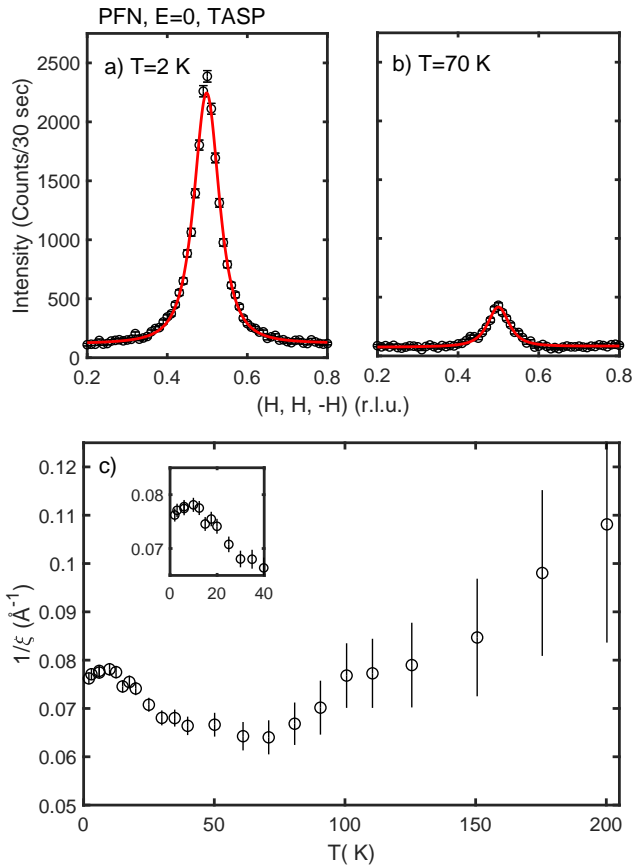}
	\caption{ \label{fig:mag_elastic} Correlated elastic magnetic scattering measured on TASP at $(a)$ T=2 K and $(b)$ 70 K.  The solid red lines are fits to the lineshape discussed in the text.  $(c)$ The inverse correlation length as a function of temperature.  Note the local maximum at $\sim$ 15 K highlighted by the inset.}
\end{figure} 

To isolate the static magnetic cross section from the dynamics, we use the energy analysis on the TASP triple-axis spectrometer.  Representative scans through the correlated magnetic scattering are shown in Fig. \ref{fig:mag_elastic} $(a,b)$ at T=2 K and 70 K. To model the elastic lineshape we have followed Ref. \onlinecite{Stock13:88} and fit a lorentzian squared with one direction integrated to account for the coarse vertical integration of the spectrometer, hence resulting in a power of ${3\over 2}$,

\begin{figure*}
	\includegraphics[width=165mm]{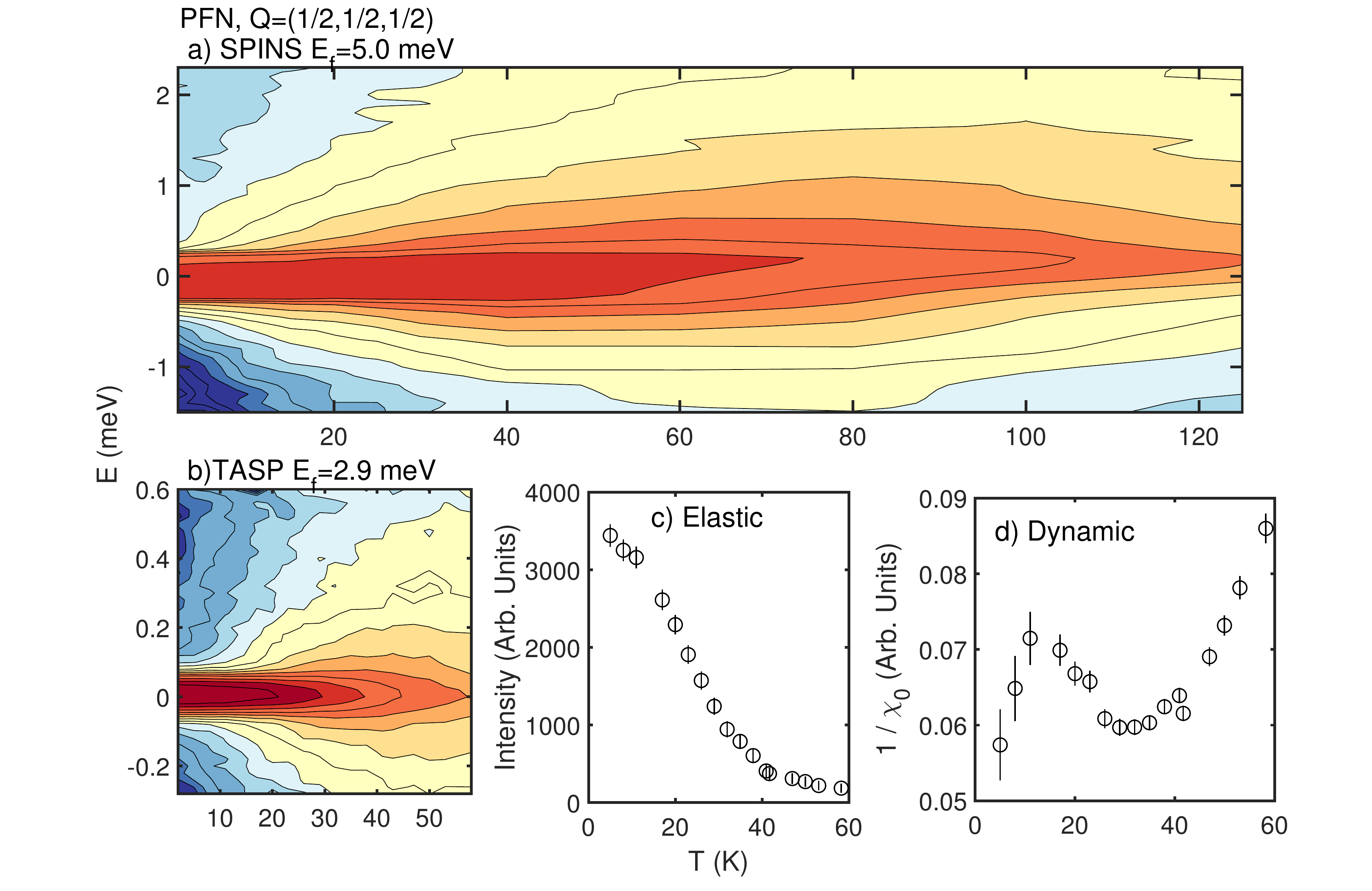}
	\caption{ \label{fig:magnetic_dynamics_T}  $(a,b)$ Contour plots of the magnetic scattering as a function of temperature measured on SPINS (E$_{f}$=5.0 meV, $2\delta E$=0.25 meV) and TASP (E$_{f}$=2.9 meV, $2\delta E$=0.1 meV).  $(c)$ The elastic magnetic intensity and $(d)$ dynamic component measured on TASP with E$_{f}$=2.9 meV.}
\end{figure*} 

\begin{equation}\label{eq:elastic}
	\chi(\vec{Q}) \propto {1 \over {\left[ 1+\left({|\vec{Q}-\vec{Q}_{0}|\xi}\right)^{2}\right]^{3\over2}}}.
\end{equation}

\noindent The inverse correlation length $1/\xi$ is plotted in Fig. \ref{fig:mag_elastic} $(c)$.  On cooling from room temperature, the spatial correlations increase gradually until $\sim$ 50 K.  On further cooling these shorten until $\sim$ 15 K where spatial correlations increase once again. This lower temperature coincides with anomalies in the temperature dependent magnetization and attributed to the formation of a glass phase~\cite{Falqui05:109,Kuma08:93} and the temperature dependence of the correlation length is similar to that reported in ferromagnetic spin-glasses~\cite{Aeppli84:55}.  The lengthscale studied here never gets particularly large and only reaches a maximum of $\sim$ 4 unit cells.  This behavior is different from typical critical scattering where the correlation length scale becomes large and diverges on cooling and neither inflection points in the correlation length can be considered as a phase transition in PFN.  As discussed in the Appendix in the context of our polarized neutron results, the magnetic structure that we measure with neutrons is isotropic on average.  However, given the issue of domains, it is not possible from our bulk experiments to determine if the structure is uniaxial as proposed in Ref. \onlinecite{Chillal13:87}. However, unlike Ref. \onlinecite{Chillal13:87}, we do not observe a magnetic Bragg peak indicative of a spatially long-range component.  
%We discuss this difference in the companion paper where we suggest it is the result of relative Fe$^{3+}$/Nb$^{5+}$ concentration.

\subsubsection{Magnetic dynamics:}

In Fig. \ref{fig:magnetic_dynamics_T} $(a)$ we show a contour plot of constant momentum scans at $\vec{Q}$=(${1\over 2}$,${1\over 2}$,${1\over 2}$) taken on SPINS (NIST) showing the temperature dependence of the low-energy magnetic dynamics.  The data indicate a gradual slowing of the dynamic magnetic response as the temperature is decreased and the fluctuations collapse to the elastic line which defines static magnetism on the timescale of the instrument resolution.  The same dynamics are studied with finer resolution using TASP (PSI) with E$_{f}$=2.9 meV in Fig. \ref{fig:magnetic_dynamics_T} $(b)$ and indicates a freezing of the dynamics on the timescale of TASP at a lower temperature of $\sim$ 20 K.  Similar to the comparison above (Fig. \ref{fig:mag_T}), the measured temperature where static magnetism is observable is determined by instrumental resolution with finer energy resolution illustrating lower temperature scales.

In Fig. \ref{fig:magnetic_dynamics_T} $(c,d)$ we plot the parameters of a fit to the TASP data to the following lineshape, convolved with the spectrometer resolution, indicative of a static (on the timescale of the spectrometer resolution) and dynamic magnetism defined by a single energy scale,

\begin{equation}
F(E)=\chi_{0} [n(E)+1] {{\omega \Gamma} \over {\omega^{2} + \Gamma^{2}}}+B\delta(E),
\label{s_q_SPINS_TASP}
\end{equation}

\noindent where $\chi_{0}$ is the dynamic intensity and is related to the real part of the susceptibility by the Kramers Kronig relation and $\Gamma$ is the single dominant energy scale inversely proportional to the fluctuation timescale.  The amplitude $B$ is the component which is resolution limited and residing in the elastic line (denoted as a $\delta$-function in the lineshape above).  Fig. \ref{fig:magnetic_dynamics_T} $(c)$ plots the elastic component defined by the parameter $B$ as a function of temperature and shows a gradual increase in intensity with decreasing temperature starting $\sim$ 50 K.  As illustrated above in our discussion of energy integrated diffuse scattering, this temperature dependence is tied to the spectrometer resolution and not indicative of a phase transition to static magnetic order.  We note that muon spectroscopy studies magnetism on the $\sim$ MHz scale which is within the experimental resolution here characterized by the elastic $\delta$-function in Eqn. \ref{s_q_SPINS_TASP}.  In Fig. \ref{fig:magnetic_dynamics_T} $(d)$ we plot $1/\chi_{0}$ as a function of temperature which displays a peak below $\sim$ 15 K at a similar temperature where we see an anomaly in the correlation length above (Fig. \ref{fig:mag_elastic}) and magnetization~\cite{Falqui05:109,Kuma08:93}.  Based on the freezing of the dynamics, characterized by an elastic contribution to the magnetic neutron cross section, and also the peak in the susceptibility, we relate this low temperature to a glass transition in analogy to prototypical spin-glasses~\cite{Nagata79:19,Sherrington75:8}.

\begin{figure}
	\includegraphics[trim=0.6cm 3.0cm 1.5cm 2.8cm,width=80mm]{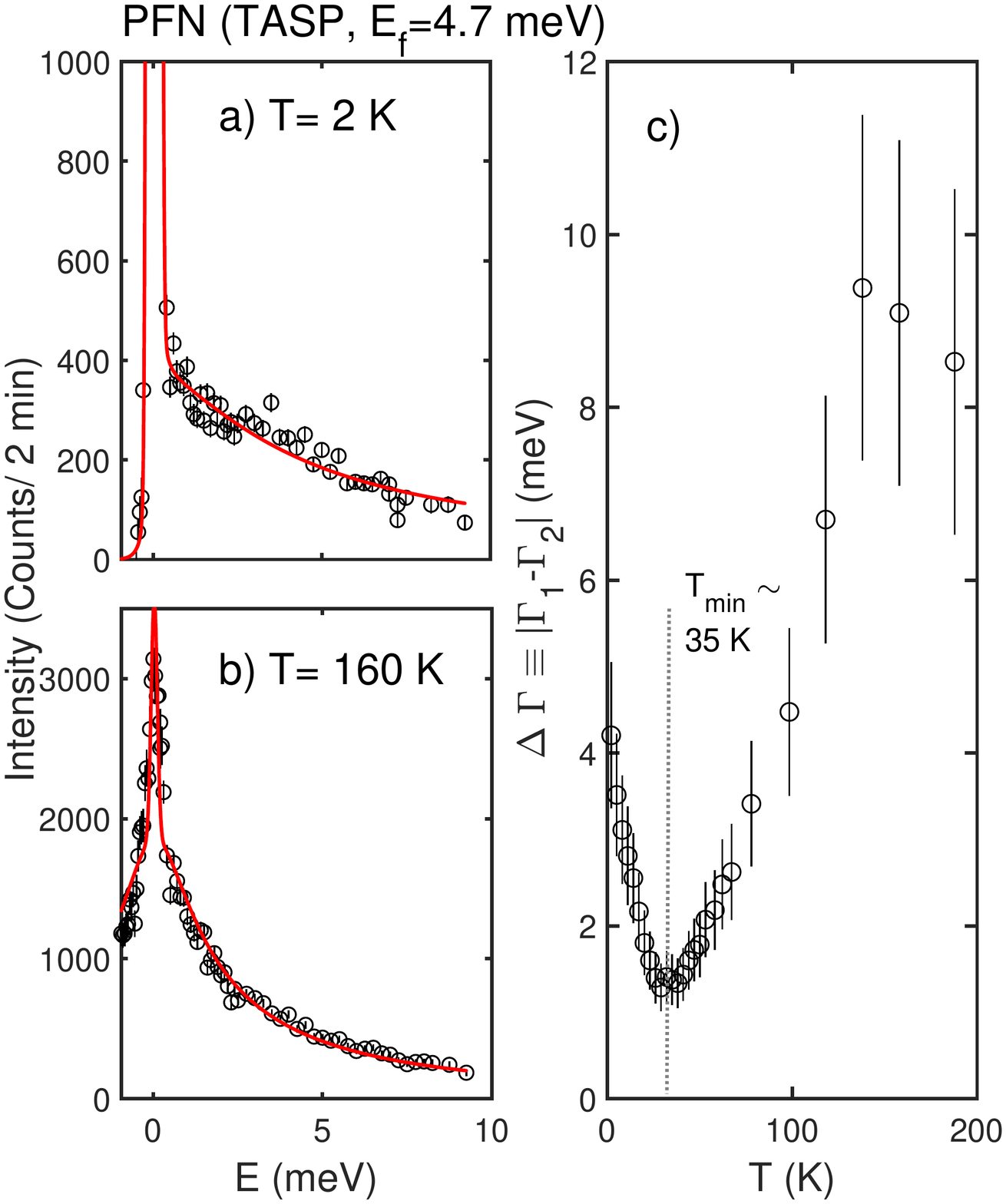}
	\caption{ \label{fig:bern_fit}  Fits to the bandwidth model described in the text at $(a)$ 2 K and $(b)$ 160 K.  $(c)$ illustrates the bandwidth as a function of temperature.  The data has been corrected for the background by measuring intensity at a reciprocal space position away from $\vec{Q}=({1\over 2},{1\over 2},{1\over 2})$ as a function of energy transfer.}
\end{figure}

We investigate the magnetic dynamics over a broader energy range in Fig. \ref{fig:bern_fit} using the TASP triple-axis which highlights a problem as the data cannot be well described by this single energy scale lineshape (Eqn. \ref{s_q_SPINS_TASP}).  The magnetic scattering displayed by PFN at low temperatures shows a continuous intensity distribution over all measured energy transfers.  This contrasts with typical critical scattering of a magnet where the cross section is peaked at a finite energy defining the characteristic timescale $\tau$ of the fluctuations which become static at the ordering temperature (examples shown in Refs. \onlinecite{Yamani15:91,Stock08:77}).  While the single energy scale is an effective approximation, describing data over a small energy range (like in Fig. \ref{fig:magnetic_dynamics_T}) or with coarser energy resolution (as in Ref. \onlinecite{Stock13:88}), the finer energy resolved experiments presented here illustrate the neutron cross section is not consistent with this analysis.  

Given that Fig. \ref{fig:bern_fit} $(a)$ displays a featureless cross section at all observable energy transfers with no observable peak in energy, we consider a model that describes a bandwidth of energy scales.  Instead of the approach applied above where the susceptibility is

\begin{equation}\label{eq:mod_lor}
	\chi(E) \propto {1 \over {\Gamma -i E}}
\end{equation}

\noindent we integrate the susceptibility over a range of frequencies,

\begin{equation}\label{eq:mod_lor}
	\chi(E) \propto \int_{\Gamma_1}^{\Gamma_2} d\gamma {1 \over {\gamma -i E}}.
\end{equation}

\noindent This analysis is motivated by work on non-Fermi liquids~\cite{Bernhoeft01:13} and has recently been applied to understand the unusual quasielastic lineshape in hybrid perovskites~\cite{Stock20:32}.  Performing the integral and taking the imaginary part, which is proportional to the neutron cross section, we find

\begin{equation}\label{eq:mod_lor}
	F(E) \propto {1\over {\Gamma_{2}-\Gamma_{1}}} \left[ \arctan \left({E \over \Gamma_{1}} \right) - \arctan \left({E \over \Gamma_{2}} \right) \right].
\end{equation}

\noindent A fit of this lineshape to the magnetic scattering in PFN is shown in Fig. \ref{fig:bern_fit} $(a-b)$ at T=2 K and 160 K and provides a good description of the continuous range of scattering over the entire dynamics range.  The fits have also included an elastic contribution at E=0 defined by the energy resolution which parameterizes scattering from localized magnetic ions that are static on the timescale of the resolution of TASP.  In comparison to the frequency range sampled by muons, we note that the frequency scale was on the order of $\sim$ MHz, which is within the elastic line measured here with neutrons. 

The bandwidth obtained from this fitting procedure, defined by $\Delta \Gamma \equiv |\Gamma_{1} - \Gamma_{2}|$, is plotted in Fig. \ref{fig:bern_fit} $(c)$.  It is worth noting that the energy range of the scans essentially fixes the maximum bandwith that can be derived.  In this experiment applying cold neutron triple-axis spectrometers, this energy range is confined to below $\Delta \Gamma_{max} \sim$ 10 meV and therefore the errorbars on the measurements at high temperatures are large when it is expected the bandwidth exceeds the dynamic range of the spectrometer.    

Fig. \ref{fig:bern_fit} $(c)$ shows a decrease in $\Delta \Gamma$ with decreasing temperature from 200 K until below $\sim$ 40 K the bandwidth is seen to increase.  This is a similar temperature scale to where the elastic, or static, component of the low-energy neutron cross section was found to increase in Fig. \ref{fig:magnetic_dynamics_T} above.  The range of frequencies contributing to the imaginary part of the susceptibility $\chi''(\omega)$ makes this response analogous to a glass response.~\cite{Sherrington19:52,Kumar08:93}

\begin{figure}
	\includegraphics[trim=0.5cm 1.5cm 0.5cm 0.7cm,width=85mm]{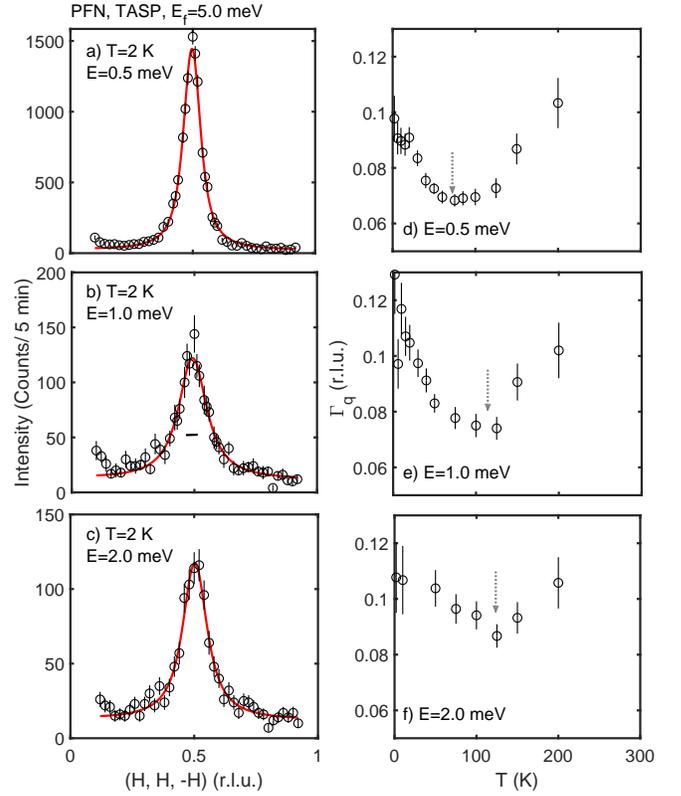}
	\caption{ \label{fig:constant_E}  $(a-c)$ Representative constant energy cuts at T=2 K with the horizontal bar a measure of the calculated momentum resolution. Plots of the momentum half-widths of the correlated magnetic scattering as a function of temperature at $(d)$ 0.5 meV, $(e)$ 1.0 meV, and $(f)$ 2.0 meV.}
\end{figure} 

In Fig. \ref{fig:constant_E} we investigate the momentum dependence of the dynamic correlated magnetic scattering with temperature.  Example constant energy scans are illustrated in Figs. \ref{fig:constant_E} $(a-c)$ and the peak half widths in momentum, at each energy transfer, is shown in Figs. \ref{fig:constant_E} $(d-f)$.  The lineshape used was defined as follows to fit the momentum dependence of the dynamic component of the scattering cross section,

\begin{equation}\label{eq:mod_lor}
	\chi(\vec{Q})\propto {1 \over {1+\left({{|\vec{Q}-\vec{Q}_{0}|}\over{\Gamma_{q}}}\right)^{2}}}.
\end{equation}

\noindent The plots of the $\Gamma_{q}$ in Fig. \ref{fig:constant_E} illustrates a qualitatively similar temperature dependence to the bandwidth $\Delta \Gamma$ discussed above.  On decreasing temperature from 300 K, the linewidth initially decreases, indicative of larger spatial dynamic correlations.  However, below some ``cross over" temperature (indicated by the arrow in Fig. \ref{fig:constant_E}) the linewidth increases indicative of spatially shorter dynamic correlations.   It is interesting to note that this cross over temperature occurs at progressively lower temperatures, when sampled at lower energies.  The measured unraveling of magnetism in PFN depends on the energy scale experimentally sampled.

Based on this study of the dynamics, we characterize the fast dynamics that drive the cluster glass which forms at low temperatures.  We find there is a temperature range $\sim$ 30-40 K where the frequency range over which the spins fluctuate reaches a minimum.  Below this temperature, PFN becomes more disordered spatially and temporally.  This temperature range also defines where the magnetic fluctuations become static (albeit spatially short-ranged) on the resolution of cold neutron spectroscopy.  However, below this a second temperature scale $\sim$ 15 K develops where the correlation lengths increase which is coincident with a peak in $1 / \chi_{0}$.  This coincides with previous reports of the formation of a cluster spin-glass phase.

\section{Discussion and conclusions}

The magnetic static and dynamics discussed above show a spin response characterized by a range of frequencies on the scale $\sim$ THz.  On decreasing temperature, we observe a development of correlations characterized by an increase in the magnetic fluctuation timescale and increasing correlation lengths.  However, over a particular crossover temperature range this trend towards spatially long range order is interrupted and the static and dynamic correlation lengths decrease and the frequency response is characterized by a broadband of fluctuations.  We note this magnetic response occurs on an underlying lattice with considerable disorder evidenced by our acoustic phonon investigation.

This magnetic response differs from disordered or frustrated magnets.  Frustrated magnets~\cite{Stock10:105} generally show a gradual slowing of dynamics with decreasing temperature and similar results are found in disordered two dimensional cuprates~\cite{Stock08:77,Yamani15:91}.   Qualitatively similar results are observed in disordered metallic compounds~\cite{Aeppli83:28} which enter into a spin-glass phase at low temperatures.  In such systems, there are two lengthscales that are important, first the cluster~\cite{Kleeman10:105} size consisting of a high concentration of magnetic ions, and second the lengthscale being the distance between the magnetic clusters.  The spin-glass phase is driven by frustrating interactions that become relevant once the correlation length becomes large enough. Unlike reports in conventional spin-glasses~\cite{Aeepli84:29} where the dynamics appear to be characterized by a single energy scale, the dynamics in the cluster glass discussed here are governed by a band of frequencies.  The effects of localized clusters has a strong impact on the spin fluctuations.  The presence of spatially localized regions, which presumably form the basis of the clusters, with differing chemical environments is corroborated by our phonon results illustrative of multiple components and hence regions in the crystal.

However, such a model does not explain the change in spin correlations in PFN at low temperatures where the spatial correlation length decreases on decreasing temperature.  Such a scenario based on two length scales with interactions becoming frustrated would be suggestive of a saturation of static and dynamic correlations.  It might be tempting to associate this with the onset of local polar correlations characterized by the temperature dependence of the diffuse scattering near the nuclear Bragg peaks described above.  Indeed such a scenario has been proposed in lightly doped SrTiO$_{3}$ where spin and polar freezing are coupled.~\cite{Shvartsman08:101}  However, the temperature scale of the inception of static polar correlations in PFN is much higher than where we observe an unfolding of magnetism when measured with comparable resolution. The temperature scale where we observe a decrease in magnetic spatial correlation lengths, characterizing increased disorder on cooling, is near where static magnetism is observed through neutron scattering intensity at $E=0$ and $\vec{Q}$=(${1\over 2}$,${1\over 2}$,${1 \over 2}$).

We therefore propose that the origin of the disordering of the magnetism in PFN on cooling is the result of random fields induced by the local molecular field from static magnetism.  Such a scenerio has been found to drive similar temperature responses in dilute ferromagnets.~\cite{Aeppli82:25}  We speculate and mention two different mechanisms for this based on competing interactions.  First, while PFN is not a ferromagnet, there has been the suggestion for the presence of competing antiferromagnetic and ferromagnetic exchange interactions.~\cite{Stephanovich16:18}  The exchange pathway in PFN is likely to be dominated by superexchange through the oxygen ion and this could be ferromagnetic if the pathway is 90$^{\circ}$ or antiferromagnetic if the pathway is 180$^{\circ}$~\cite{Kuzian14:89} resulting in differing localized molecular fields, based on the local environment which is randomized owing to the disorder on the Fe$^{3+}$/Nb$^{5+}$ site occupancy.  A second possibility is for the existence of iron rich spatially localized regions which are coupled in different manners depending on the environment.  Both mechanisms, result from conflicting interactions resulting in a local random field which inherently changes depending on the strength of the localized static magnetism.  With the presence of disorder and clustering of magnetic Fe$^{3+}$ spins, it is inevitable for the presence of localized random molecular fields.  We note that this induces a new random molecular field as static magnetic order gradually develops with decreasing temperature.  In turn, this would result in unravelled magnetism evidenced through shortened spatial correlation lengths.  We note that this mechanism depends on the correlation length increasing on decreasing temperature and this does not depend on the small changes of Fe$^{3+}$ concentration or ordering that may induce a magnetic Bragg peak.  Ultimately, the presence of a glass transition at lower temperatures where magnetism is unravelled is not dependent on the subtle details of Fe$^{3+}$ concentrations and may explain why this lower temperature anomaly is present in all reported samples.

The presence of significant disorder in the sample and also the correlation lengths of only a few unit cells amplifies the role that this field would impose on spins.  The presence of a random field contribution to the magnetic Hamiltonian would result in the lineshape describing the magnetic $S(\vec{Q},E)$ to have two components.~\cite{Birgeneau98:177}  The static component measured at the elastic line would be characterized by a lorentzian squared momentum dependence and the dynamics a lorentzian.  This is in agreement with the data analysis presented above.  While we observe two different temperature scales for the onset of static spatially short-range magnetism and also polar correlations, the onset of static diffuse scattering at higher temperatures likely enhances the disorder and random fields at lower temperatures.  %This maybe the origin of the connection between the first moment $\langle E \rangle \equiv \int_{-\infty}^{+\infty} E \ S(\vec{Q},E) dE$ investigated previously with neutron scattering.~\cite{Stock13:88}

We note, however, that other samples of PFN have reported a magnetic Bragg peak indicative of spatial long-range order set by the resolution of the diffractometer~\cite{Rotaru09:79} and supported by NMR~\cite{Blinc08:104}.  These have been suggested to be spatially distinct, hence the term ``cluster" in Ref. \cite{Kleeman10:105}.  We note Bragg peaks in the neutron response defining antiferromagnetic order have been observed to coexist with intense magnetic diffuse scattering.  It is not clear from the available published data how the integrated spectral weights, reflecting the relative volume fractions, compare.  Also, magnetic order has been reported to be sensitive to relative Fe/Nb concentrations~\cite{Bhat04:280,Bhat05:72}.  We note that similar issues have been discussed in the context of the unit cell shape in relaxors PMN and PZN are discussed elsewhere.~\cite{Xu03:67,Xu04:70,Conlon04:70,Kisi05:17,Xu06:79,Phelan15:88,Gehring04:16,Brown18:30}  In some respects, this is analogous to skin regions reported in random field Ising magnets.~\cite{Hill91:66}

Having discussed the disordering of the magnetic correlations in PFN, we now mention the low-temperature anomaly in Fig. \ref{fig:mag_elastic} $(c)$ (see inset) where an increase in spatial ordering at $\sim$ 15 K occurs evidenced by a subtle increase in the correlation length on cooling.  This low temperature scale is not reflected in the dynamics probed with triple-axis spectrometers discussed above, indicative that the fluctuations associated with this temperature are within the resolution of the neutron instruments applied here.   Higher resolution measurements such as with either spin-echo or muon spectroscopy are required to probe the dynamics in this temperature range.

In summary, we have reported a neutron study of the structural and magnetic correlations in PFN.  Based on the low-energy acoustic phonons, we find evidence that the sample consists of multiple chemically different regions presenting significant disorder in our single crystal.  The magnetic dynamics on cooling from room temperature initially show increased spatial correlation lengths yet at a crossover temperature the static and dynamic correlation lengths decrease.  This is followed by a transition that we suggest represents a glass transition comparing our neutron and susceptibility results.  The magnetic dynamics at low temperatures are well described by a broadband of contributing frequencies analogous to glasses.  We propose a model where the disordering of magnetic correlations at low temperatures is driven by random fields resulting from the local molecular field originating from the spatially short range Fe$^{3+}$ magnetic correlations.  These random fields eventually drive PFN into a magnetic low temperature glass ``cluster" phase with the dynamics described by a broadband of frequencies.

\section*{Acknowledgements}

We are grateful to the EPSRC and the STFC for funding.  N.G.-D. supported by EPSRC/Thales UK iCASE Award No. EP/P510506.

\section{Appendix A - Neutron polarization analysis}

 \begin{figure}
	\includegraphics[width=95mm,trim=4.5cm 6.5cm 3cm 7.5cm,clip=true]{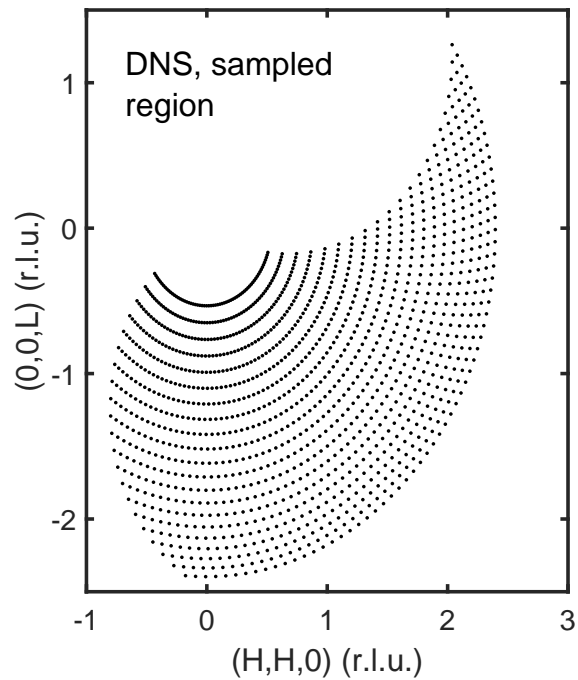}
	\caption{ \label{fig:detector_coverage} A visualization of the sampled region in momentum space on the DNS diffractometer by scanning the angle at the sample position.  Each point represents a detector and illustrates the detector density used to generate the plots for the DNS experiment.}
\end{figure} 

\begin{figure}
	\includegraphics[width=85mm]{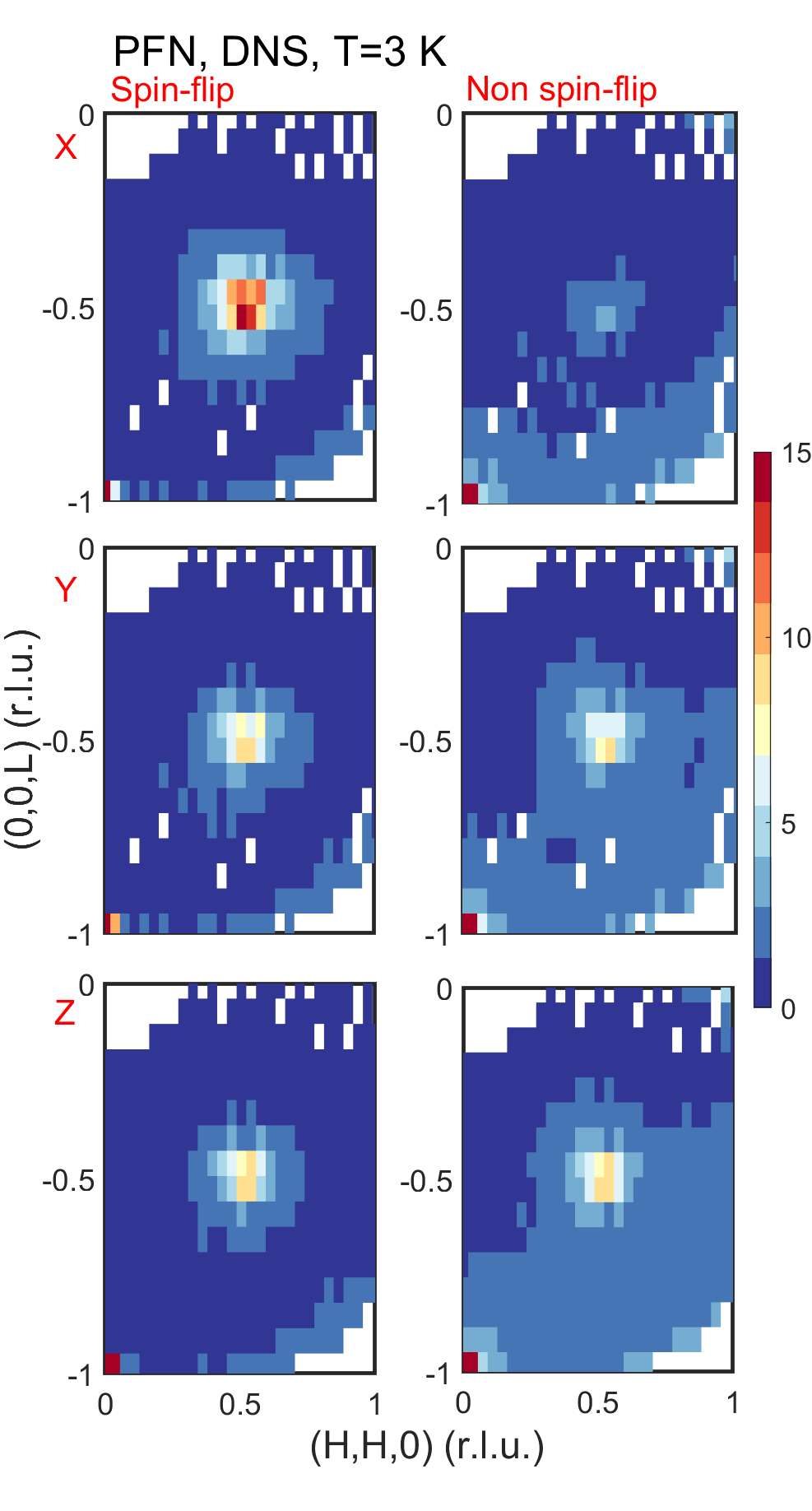}
	\caption{ \label{fig:pol_fig1} Colormaps illustrating the polarized neutron cross section at T=3 K for the different cross sections measured on DNS (MLZ).  The $X$ channel gives the total magnetic cross section in the spin-flip channel and the residual correlated scattering near $\vec{Q}$=(1/2, 1/2, 1/2) is the result of incomplete polarization of the neutron beam.}
\end{figure} 

\begin{figure}
	\includegraphics[width=85mm]{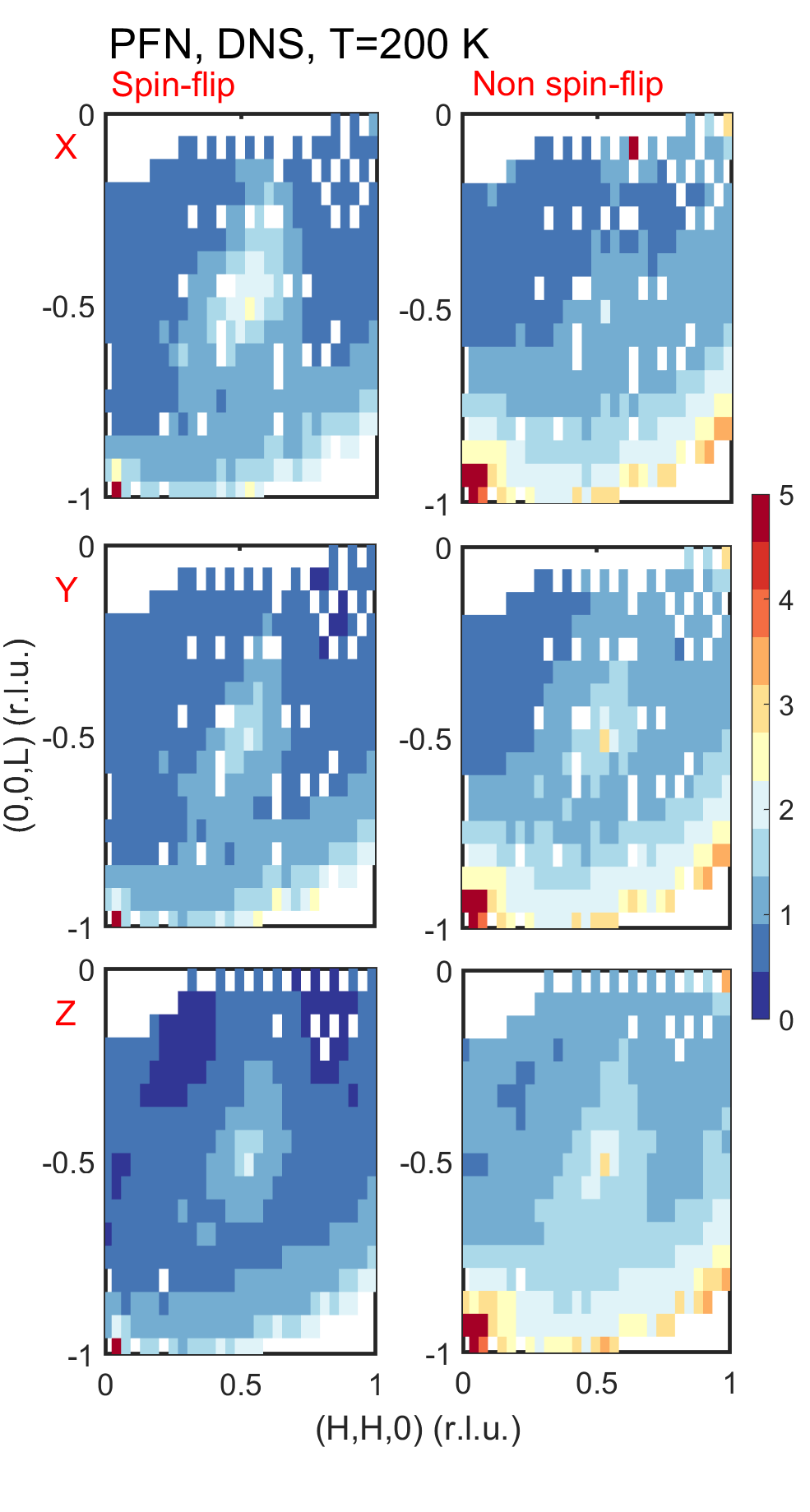}
	\caption{ \label{fig:pol_fig2} Colormaps illustrating the polarized neutron cross section at T=200 K for the different polarized cross sections measured on DNS (MLZ).}
\end{figure} 

In this appendix, we outline the polarization analysis used to separate the purely magnetic and nuclear cross sections discussed above.  For the discussion here, we focus our analysis using the DNS (MLZ) and the D7 (ILL) neutron diffractometers which both consist of arrays of neutron detectors confined to a horizontal scattering plane with the detectors equally spaced in angle.  For example,  the scattered beam at DNS was measured with 21 detectors equally spaced 5$^{\circ}$ apart in the horizontal scattering plane. These have been converted to momentum transfer in the main text of the paper and the conversion is illustrated in Fig. \ref{fig:detector_coverage}.  Both DNS and D7 operate with an $XYZ$ polarization analysis where the neutron spin direction can be tuned to be along three orthogonal axes.  The $Z$ axis defined as being vertical to the horizontal scattering plane and $X$ is defined as being parallel to the average momentum transfer $\vec{Q}$ of the scattered detector bank.  The $Y$ axis is defined as being perpendicular to the $X$ and $Z$ axes.   

The magnetic scattering cross section and how it is measured in neutron scattering has been outlined in several books~\cite{Shirane_book,Bacon_book} and review articles~\cite{Moon69:181,Stewart00:87}.  For completeness we summarize the important points here.  Unpolarized neutron scattering is sensitive to the component of the magnetic moment, perpendicular to the wavevector transfer defined by $\vec{Q}$.  This component of the magnetic moment is defined by $\vec{S}_{\perp}$. For the uniaxial polarization experiments used here, the components of $\vec{S}_{\perp}$ perpendicular to the neutron polarization will appear in the spin flip (SF) channel and the components of $\vec{S}_{\perp}$ parallel to the neutron polarization will be sampled in the non spin flip (NSF) channel.  Nuclear scattering will appear in the non spin flip channels.  These polarization dependent rules were used in this experiment to separate out purely magnetic and nuclear scattering cross sections.

An example plot of all six channels for an uniaxial polarization experiment is illustrated in Fig. \ref{fig:pol_fig1} for PFN aligned in the (HHL) scattering plane using the DNS diffractometer at T=3 K.  All of the magnetic scattering occurs in the SF channel with the polarization aligned along $X$.  The NSF channel for the neutron beam polarized along $X$ is a measure of the incomplete polarization of the neutron beam.  This originates predominately from several experimental aspects associated with polarized neutrons.  First, the flipping ratio, defined as the ratio of the intensities in the NSF and SF channels. On both DNS and D7 this was measured to be 20.  Second, we note that the $X$ direction is defined as parallel to the \textit{average} momentum transfer $\vec{Q}$ of the scattered detector bank.  This results in some feed through of intensity between the NSF and SF channels for the SF measurements in the $X$ channel as $\vec{Q}$ is not exactly along with the $X$ direction for all detectors.  This net feedthrough is evident in Fig. \ref{fig:pol_fig1} which shows some correlated scattering in the NSF channel with the neutron polarization along $X$.  The extra scattering which occurs at large momentum transfers in the NSF channels originates from nuclear scattering not sampled in the SF channel.

\begin{figure*}
	\includegraphics[width=170mm]{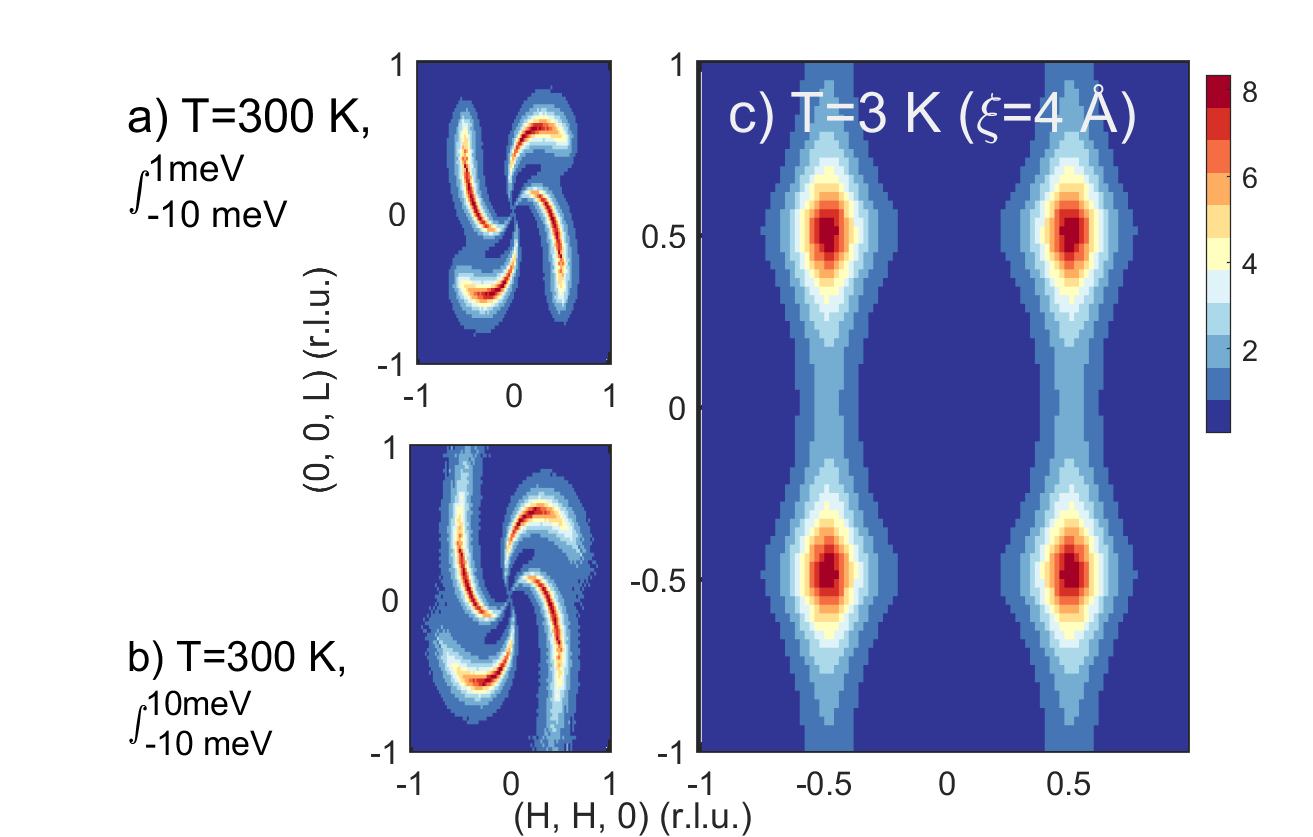}
	\caption{ \label{fig:mag_sim} Simulated magnetic diffuse scattering in PFN applying the model discussed in the main text.  $(a,b)$ Simulated magnetic diffuse scattering illustrating the effects of different integration ranges on the resulting pattern at T=300 K.  $(c)$ Further simulation of the magnetic diffuse scattering at T=3 K.}
\end{figure*} 

\begin{figure}
	\includegraphics[trim=1.5cm 2.5cm 0cm 0cm,width=110mm]{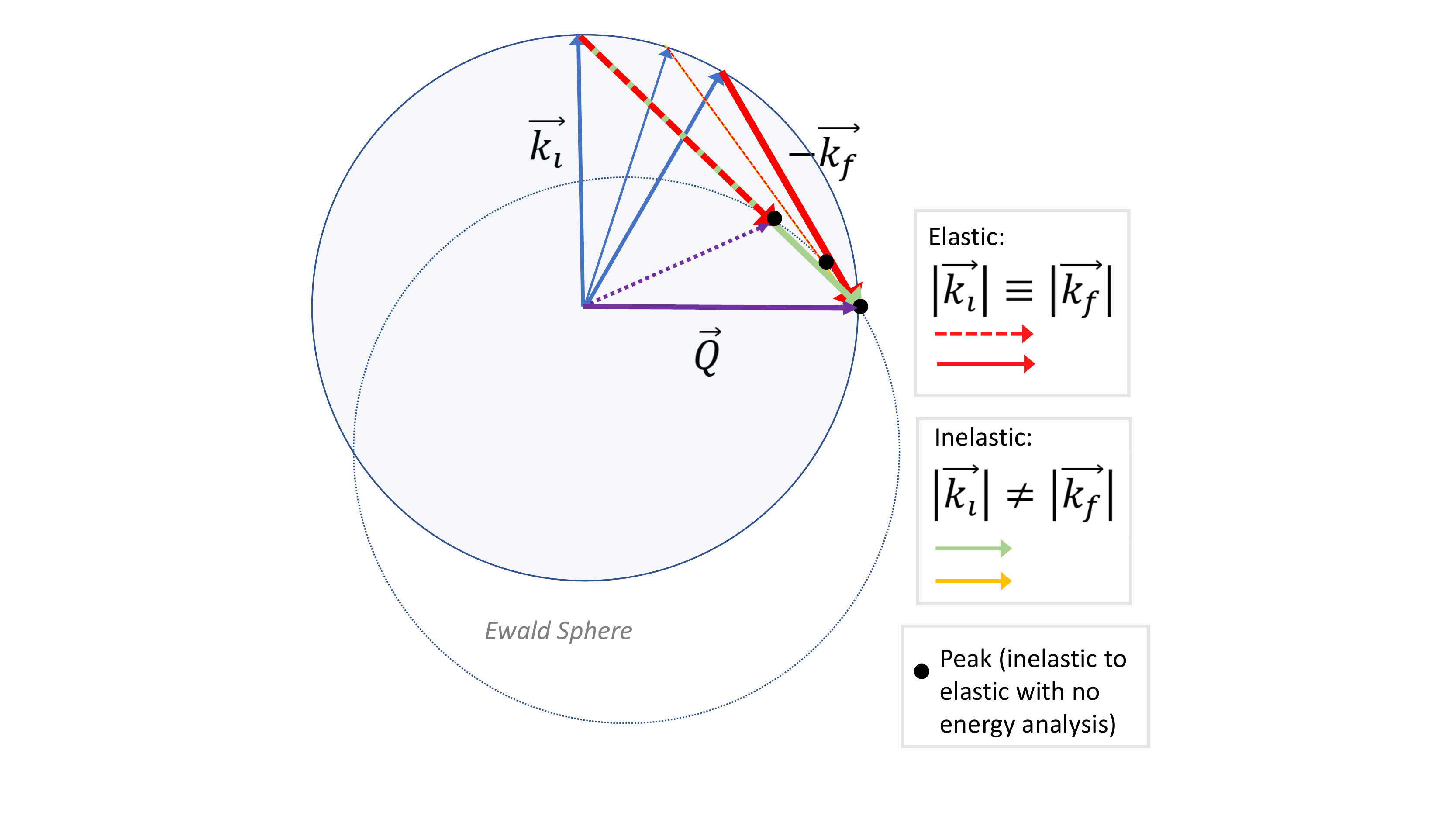}
	\caption{ \label{fig:diffuse_spurious} The kinematics for diffuse scattering measurements without energy analysis.  The peaks (or maximum intensity in the energy integrated neutron cross section $S(\vec{Q})$) is illustrated assuming a ridge of dynamical scattering centered around a single momentum position like that measured using triple-axis spectroscopy in PFN.}
\end{figure} 

When comparing the SF and NSF channels for the $Y$ and $Z$ polarizations (when the polarization is aligned perpendicular to the average momentum transfer) it can be seen that the correlated scattering around $\vec{Q}$ =$({1\over 2}, {1\over 2}, {1\over 2})$ in Fig. \ref{fig:pol_fig1} is isotropic and consistent with having the same intensity in all four of these channels.  The intensity in these four channels is also, within experimental error, ${1\over 2}$ of the intensity measured for the SF channel with the neutron polarization aligned along $X$ which samples the entire magnetic cross section.  This scattering resembles magnetic scattering with no preferred magnetic moment direction as is the case for isotropic magnetic scattering.  This isotropic scattering among these four channels is not indicative of spatially long-range magnetic ordering and is consistent with the short magnetic correlation lengths reported in this compound and the lack of any sharp, or well defined magnetic ordering in the data presented in the main text.  The magnetic scattering in PFN is isotropic with no preferred ordering direction.

This behavior across polarization channels is further consistent with the weaker magnetic response at T=200 K illustrated in Fig. \ref{fig:pol_fig2}.  In this case the overall magnetic cross section is reduced owing to the temperature, yet the the same pattern found above for T=3 K is repeated with generally isotropic scattering found in the SF and NSF channels for $Y$ and $Z$ polarizations.  

The properties of uniaxial polarization analysis of isotropic magnetic scattering was used to extract and disentangle the magnetic and nuclear cross sections in the main text.  On D7, the nuclear scattering was measured directly through the SF and NSF channels when the neutron was aligned along $X$.  For DNS, where the temperature dependence was studied in detail, we measured the SF and NSF with the neutron polarization along the $Z$ axis.  The isotropic scattering inherent to the underlying isotropic nature of the Fe$^{3+}$ moments implies that the magnetic scattering is the same in both channels.  However, the nuclear cross section is only present in the NSF channels.  By subtraction, this allowed us to extract the nuclear background and plot the purely magnetic and nuclear components shown in Figs. \ref{fig:mag_T} and \ref{fig:mag_diffuse} in the main text.

\section{Appendix B - Diffuse scattering and ``Catherine Wheels"}

In this appendix, we simulate the neutron diffuse scattering cross section measured in energy integrating two-axis mode in an attempt to understand the unusual momentum dependence discussed above in the magnetic channel at high temperatures.   Using a conventional triple-axis spectrometer with both a monochromator and analyzer crystal, the measured intensity provides a direct measure of $S(\vec{Q},\omega)$ with the energy transfer defined as $\hbar \omega=E_{i}-E_{f}$.  The diffractometers DNS and D7 operate in two-axis mode where the analyzer is removed which ideally provides a measure of $S(\vec{Q}) \equiv \int_{-\infty}^{\infty} d\omega S(\vec{Q},\omega)$.  DNS and D7 operate in a fixed E$_{i}$ mode with the incident energy defined by a monochromator.  PRISMA uses time of flight to assign an energy to every detected neutron under the assumption of elastic scattering where E$_{i}$$\equiv$E$_{f}$.  Given the constraints of the instrument, these only provide an approximate measure of the  $S(\vec{Q}) \equiv \int_{-\infty}^{\omega_{c}} d\omega  S(\vec{Q},\omega)$ with the cutoff energy $\hbar \omega_{c}$ being fixed by kinematics of the instrument used.  

To test whether the unusual lineshape measured on D7 (Fig. \ref{fig:mag_diffuse}) originates from dynamics, we have assumed, for simplicity, that the neutron scattering cross section can be approximated by a single relaxational lineshape multiplied by a lattice lorentzian accounting for exponentially decaying spatial correlations.  This is written with the measured neutron scattering cross section $I(\vec{Q},\omega)$, proportional to the structure factor $S(\vec{Q},\omega)$, which is in turn related to the imaginary part of the susceptibility $\chi '' (\vec{Q}, \omega)$ written above,

\begin{equation}\label{eq:mod_lor}
	\begin{aligned}	
		\chi'' (\vec{Q},\omega)\propto {{\omega \Gamma} \over {\omega^{2} +\Gamma^{2}}} \times {{\sinh(a/\xi)}\over {\cosh(a/\xi)-\cos(\vec{Q}\cdot\vec{a})}}.
	\end{aligned}
\end{equation}

\noindent $\Gamma$ is the single dominant relaxation rate proportional to the inverse of the lifetime $\tau$, $\xi$ is the spatial correlation length, and $\vec{a}$ is the lattice constant.  This form for the susceptibility was found to provide a good approximation of the scattering previously measured with a thermal triple-axis.~\cite{Stock13:88}  

To calculate an estimate of the diffuse scattering resulting from dynamic magnetism we have performed the following procedure.  1) We have taken the form of $S (\vec{Q},\omega)$ given by Eqn. \ref{eq:mod_lor} at each energy transfer as a function of scattering angles $A4=2\theta$ and sample rotation angle $\psi$. 2) We have then calculated the momentum transfer values based on these two angles assuming the scattering is entirely elastic ($\hbar \omega=0$).  3) We have then integrated the intensities over energy and imposed an energy cutoff which we assume is determined by the instrument configuration.  The results of this calculation are present in Fig. \ref{fig:mag_sim} integrating in energy from -10 meV to two cutoff energy values of 1 meV (panel $a$) and 10 meV (panel $b$).  At T=300 K, we have taken the measured value of $\Gamma \sim$ 10 meV found in Ref. \onlinecite{Stock13:88} and this is consistent with expectations based on the size of the Fe-O-Fe magnetic exchange~\cite{Benkhaled20:525}.   The value of $\Gamma$ restricts the cross section and implies that energies below -10 meV do not contribute significantly to the integral of $S(\vec{Q}$).  

We note that this calculation and the resulting diffuse scattering pattern in an energy integrated experiment like done on DNS or D7 depends on the kinematics of the experiment.  This is schematically shown in Fig. \ref{fig:diffuse_spurious} which outlines the scattering triangle for a particular geometry of $\vec{Q}=\vec{k}_{i}-\vec{k}_{f}$ and how the apparent ``elastic" cross section is influenced by the dynamics.  Three scattering triangles are illustrated with the solid line defining $\vec{Q}$ showing an elastic scattering process.  The dashed lines define inelastic (energy gain) scattering processes at $\vec{Q}=({1\over 2},{1\over 2},{1\over 2})$, but illustrate how they appear if assumed to be elastic with $|\vec{k}_{i}|\equiv |\vec{k}_{f}|$ (solid points).  It can be seen that such processes appear away from the $\vec{Q}=({1\over 2},{1\over 2},{1\over 2})$ position.  The position is defined by the Ewald Sphere with a radius defined by $|\vec{k}_{i}|$ in the case of DNS and D7.  We note that the situation on PRISMA is more complex given the range of incident and scattered wavelengths.  In such a case all scattering is defined as being elastic and fixed via geometry.~\cite{Liu21:54}

The scattering cross section from our simulation of the diffuse magnetic scattering in Figs. \ref{fig:mag_sim} $(a,b)$ are qualitatively in agreement with the unusual magnetic scattering in Fig. \ref{fig:mag_diffuse} $(d)$ which mimics ``Catherine Wheels".    Figs. \ref{fig:mag_sim} $(a,b)$ compare the effects on the scattering cross section with energy integration illustrating that the upper cuttoff on the integration only controls the extent of the scattering in momentum.  The main result of the scattering occurs due to the energy gain side of the integration.   This is further confirmed by the low temperature simulation at T=2 K in Fig. \ref{fig:mag_sim} $(c)$, which should be compared to Fig. \ref{fig:mag_sim} $(a)$, where it can be seen that the scattering cross section is described by momentum broadened points near $\vec{Q}$=(${1\over2},{1\over2},{1\over2}$).  We therefore conclude that the unusual diffuse scattering cross section experimentally measured in Fig. \ref{fig:mag_diffuse} $(c,f,g)$ originates from dynamics and is indicative of fluctuations of localized Fe$^{3+}$ moments and sampled through the thermally excited scattering on the energy gain side (negative energy transfers).


\begin{thebibliography}{136}%
	\makeatletter
	\providecommand \@ifxundefined [1]{%
		\@ifx{#1\undefined}
	}%
	\providecommand \@ifnum [1]{%
		\ifnum #1\expandafter \@firstoftwo
		\else \expandafter \@secondoftwo
		\fi
	}%
	\providecommand \@ifx [1]{%
		\ifx #1\expandafter \@firstoftwo
		\else \expandafter \@secondoftwo
		\fi
	}%
	\providecommand \natexlab [1]{#1}%
	\providecommand \enquote  [1]{``#1''}%
	\providecommand \bibnamefont  [1]{#1}%
	\providecommand \bibfnamefont [1]{#1}%
	\providecommand \citenamefont [1]{#1}%
	\providecommand \href@noop [0]{\@secondoftwo}%
	\providecommand \href [0]{\begingroup \@sanitize@url \@href}%
	\providecommand \@href[1]{\@@startlink{#1}\@@href}%
	\providecommand \@@href[1]{\endgroup#1\@@endlink}%
	\providecommand \@sanitize@url [0]{\catcode `\\12\catcode `\$12\catcode
		`\&12\catcode `\#12\catcode `\^12\catcode `\_12\catcode `\%12\relax}%
	\providecommand \@@startlink[1]{}%
	\providecommand \@@endlink[0]{}%
	\providecommand \url  [0]{\begingroup\@sanitize@url \@url }%
	\providecommand \@url [1]{\endgroup\@href {#1}{\urlprefix }}%
	\providecommand \urlprefix  [0]{URL }%
	\providecommand \Eprint [0]{\href }%
	\providecommand \doibase [0]{https://doi.org/}%
	\providecommand \selectlanguage [0]{\@gobble}%
	\providecommand \bibinfo  [0]{\@secondoftwo}%
	\providecommand \bibfield  [0]{\@secondoftwo}%
	\providecommand \translation [1]{[#1]}%
	\providecommand \BibitemOpen [0]{}%
	\providecommand \bibitemStop [0]{}%
	\providecommand \bibitemNoStop [0]{.\EOS\space}%
	\providecommand \EOS [0]{\spacefactor3000\relax}%
	\providecommand \BibitemShut  [1]{\csname bibitem#1\endcsname}%
	\let\auto@bib@innerbib\@empty
	%</preamble>
	\bibitem [{\citenamefont {Binder}\ and\ \citenamefont
		{Young}(1986)}]{Binder86:58}%
	\BibitemOpen
	\bibfield  {author} {\bibinfo {author} {\bibfnamefont {K.}~\bibnamefont
			{Binder}}\ and\ \bibinfo {author} {\bibfnamefont {A.~P.}\ \bibnamefont
			{Young}},\ }\href {https://doi.org/10.1103/RevModPhys.58.801} {\bibfield
		{journal} {\bibinfo  {journal} {Rev. Mod. Phys.}\ }\textbf {\bibinfo {volume}
			{58}},\ \bibinfo {pages} {801} (\bibinfo {year} {1986})}\BibitemShut
	{NoStop}%
	\bibitem [{\citenamefont {Park}\ and\ \citenamefont
		{Shrout}(1997)}]{Park97:82}%
	\BibitemOpen
	\bibfield  {author} {\bibinfo {author} {\bibfnamefont {S.~E.}\ \bibnamefont
			{Park}}\ and\ \bibinfo {author} {\bibfnamefont {T.~R.}\ \bibnamefont
			{Shrout}},\ }\href {https://doi.org/10.1063/1.365983} {\bibfield  {journal}
		{\bibinfo  {journal} {J. Appl. Phys.}\ }\textbf {\bibinfo {volume} {82}},\
		\bibinfo {pages} {1804} (\bibinfo {year} {1997})}\BibitemShut {NoStop}%
	\bibitem [{\citenamefont {Ye}(1998)}]{Ye04:155}%
	\BibitemOpen
	\bibfield  {author} {\bibinfo {author} {\bibfnamefont {Z.-G.}\ \bibnamefont
			{Ye}},\ }\href {https://doi.org/10.4028/www.scientific.net/KEM.155-156.81}
	{\bibfield  {journal} {\bibinfo  {journal} {Key Eng. Mater.}\ }\textbf
		{\bibinfo {volume} {155-156}},\ \bibinfo {pages} {81} (\bibinfo {year}
		{1998})}\BibitemShut {NoStop}%
	\bibitem [{\citenamefont {Ye}(2009)}]{Ye09:34}%
	\BibitemOpen
	\bibfield  {author} {\bibinfo {author} {\bibfnamefont {Z.-G.}\ \bibnamefont
			{Ye}},\ }\href {https://doi.org/10.1557/mrs2009.79} {\bibfield  {journal}
		{\bibinfo  {journal} {MRS Bull.}\ }\textbf {\bibinfo {volume} {34}},\
		\bibinfo {pages} {277} (\bibinfo {year} {2009})}\BibitemShut {NoStop}%
	\bibitem [{\citenamefont {Bokov}\ and\ \citenamefont {Ye}(2006)}]{Bokov06:41}%
	\BibitemOpen
	\bibfield  {author} {\bibinfo {author} {\bibfnamefont {A.~A.}\ \bibnamefont
			{Bokov}}\ and\ \bibinfo {author} {\bibfnamefont {Z.~G.}\ \bibnamefont {Ye}},\
	}\href {https://doi.org/10.1007/s10853-005-5915-7} {\bibfield  {journal}
		{\bibinfo  {journal} {J. Mater. Sci.}\ }\textbf {\bibinfo {volume} {41}},\
		\bibinfo {pages} {31} (\bibinfo {year} {2006})}\BibitemShut {NoStop}%
	\bibitem [{\citenamefont {Ye}\ and\ \citenamefont {Bokov}(2004)}]{Ye04:302}%
	\BibitemOpen
	\bibfield  {author} {\bibinfo {author} {\bibfnamefont {Z.~G.}\ \bibnamefont
			{Ye}}\ and\ \bibinfo {author} {\bibfnamefont {A.~A.}\ \bibnamefont {Bokov}},\
	}\href {https://doi.org/10.1080/00150190490455043} {\bibfield  {journal}
		{\bibinfo  {journal} {Ferroelectrics}\ }\textbf {\bibinfo {volume} {302}},\
		\bibinfo {pages} {473} (\bibinfo {year} {2004})}\BibitemShut {NoStop}%
	\bibitem [{\citenamefont {Ye}(1996)}]{Ye96:184}%
	\BibitemOpen
	\bibfield  {author} {\bibinfo {author} {\bibfnamefont {Z.~G.}\ \bibnamefont
			{Ye}},\ }\href {https://doi.org/10.1080/0015019960823026} {\bibfield
		{journal} {\bibinfo  {journal} {Ferroelectrics}\ }\textbf {\bibinfo {volume}
			{184}},\ \bibinfo {pages} {193} (\bibinfo {year} {1996})}\BibitemShut
	{NoStop}%
	\bibitem [{\citenamefont {Cowley}\ \emph {et~al.}(2011)\citenamefont {Cowley},
		\citenamefont {Gvasaliya}, \citenamefont {Lushnikov}, \citenamefont
		{Roessli},\ and\ \citenamefont {Rotaru}}]{Cowley11:60}%
	\BibitemOpen
	\bibfield  {author} {\bibinfo {author} {\bibfnamefont {R.~A.}\ \bibnamefont
			{Cowley}}, \bibinfo {author} {\bibfnamefont {S.~N.}\ \bibnamefont
			{Gvasaliya}}, \bibinfo {author} {\bibfnamefont {S.~G.}\ \bibnamefont
			{Lushnikov}}, \bibinfo {author} {\bibfnamefont {B.}~\bibnamefont {Roessli}},\
		and\ \bibinfo {author} {\bibfnamefont {G.~M.}\ \bibnamefont {Rotaru}},\
	}\href {https://doi.org/10.1080/00018732.2011.555385} {\bibfield  {journal}
		{\bibinfo  {journal} {Adv. Phys.}\ }\textbf {\bibinfo {volume} {60}},\
		\bibinfo {pages} {229} (\bibinfo {year} {2011})}\BibitemShut {NoStop}%
	\bibitem [{\citenamefont {Viehland}\ \emph {et~al.}(1991)\citenamefont
		{Viehland}, \citenamefont {Li}, \citenamefont {Jang}, \citenamefont {Cross},\
		and\ \citenamefont {Wuttig}}]{Viehland91:43}%
	\BibitemOpen
	\bibfield  {author} {\bibinfo {author} {\bibfnamefont {D.}~\bibnamefont
			{Viehland}}, \bibinfo {author} {\bibfnamefont {J.~F.}\ \bibnamefont {Li}},
		\bibinfo {author} {\bibfnamefont {S.~J.}\ \bibnamefont {Jang}}, \bibinfo
		{author} {\bibfnamefont {L.~E.}\ \bibnamefont {Cross}},\ and\ \bibinfo
		{author} {\bibfnamefont {M.}~\bibnamefont {Wuttig}},\ }\href
	{https://doi.org/10.1103/PhysRevB.43.8316} {\bibfield  {journal} {\bibinfo
			{journal} {Phys. Rev. B}\ }\textbf {\bibinfo {volume} {43}},\ \bibinfo
		{pages} {8316} (\bibinfo {year} {1991})}\BibitemShut {NoStop}%
	\bibitem [{\citenamefont {Viehland}\ \emph {et~al.}(1992)\citenamefont
		{Viehland}, \citenamefont {Li}, \citenamefont {Jang}, \citenamefont {Cross},\
		and\ \citenamefont {Wuttig}}]{Viehand92:46}%
	\BibitemOpen
	\bibfield  {author} {\bibinfo {author} {\bibfnamefont {D.}~\bibnamefont
			{Viehland}}, \bibinfo {author} {\bibfnamefont {J.~F.}\ \bibnamefont {Li}},
		\bibinfo {author} {\bibfnamefont {S.~J.}\ \bibnamefont {Jang}}, \bibinfo
		{author} {\bibfnamefont {L.~E.}\ \bibnamefont {Cross}},\ and\ \bibinfo
		{author} {\bibfnamefont {M.}~\bibnamefont {Wuttig}},\ }\href
	{https://doi.org/10.1103/PhysRevB.46.8013} {\bibfield  {journal} {\bibinfo
			{journal} {Phys. Rev. B}\ }\textbf {\bibinfo {volume} {46}},\ \bibinfo
		{pages} {8013} (\bibinfo {year} {1992})}\BibitemShut {NoStop}%
	\bibitem [{\citenamefont {Pirc}\ and\ \citenamefont {Blinc}(2007)}]{Pirc07:76}%
	\BibitemOpen
	\bibfield  {author} {\bibinfo {author} {\bibfnamefont {R.}~\bibnamefont
			{Pirc}}\ and\ \bibinfo {author} {\bibfnamefont {R.}~\bibnamefont {Blinc}},\
	}\href {https://doi.org/10.1103/PhysRevB.76.020101} {\bibfield  {journal}
		{\bibinfo  {journal} {Phys. Rev. B}\ }\textbf {\bibinfo {volume} {76}},\
		\bibinfo {pages} {020101(R)} (\bibinfo {year} {2007})}\BibitemShut {NoStop}%
	\bibitem [{\citenamefont {Pirc}\ \emph {et~al.}(2001)\citenamefont {Pirc},
		\citenamefont {Blinc},\ and\ \citenamefont {Bobnar}}]{Pirc01:63}%
	\BibitemOpen
	\bibfield  {author} {\bibinfo {author} {\bibfnamefont {R.}~\bibnamefont
			{Pirc}}, \bibinfo {author} {\bibfnamefont {R.}~\bibnamefont {Blinc}},\ and\
		\bibinfo {author} {\bibfnamefont {V.}~\bibnamefont {Bobnar}},\ }\href
	{https://doi.org/10.1103/PhysRevB.63.054203} {\bibfield  {journal} {\bibinfo
			{journal} {Phys. Rev. B}\ }\textbf {\bibinfo {volume} {63}},\ \bibinfo
		{pages} {054203} (\bibinfo {year} {2001})}\BibitemShut {NoStop}%
	\bibitem [{\citenamefont {Shirane}\ and\ \citenamefont
		{Hoshino}(1951)}]{Shirane51:6}%
	\BibitemOpen
	\bibfield  {author} {\bibinfo {author} {\bibfnamefont {G.}~\bibnamefont
			{Shirane}}\ and\ \bibinfo {author} {\bibfnamefont {S.}~\bibnamefont
			{Hoshino}},\ }\href {https://doi.org/10.1143/JPSJ.6.265} {\bibfield
		{journal} {\bibinfo  {journal} {J. Phys. Soc.}\ }\textbf {\bibinfo {volume}
			{6}},\ \bibinfo {pages} {265} (\bibinfo {year} {1951})}\BibitemShut {NoStop}%
	\bibitem [{\citenamefont {Shirane}\ \emph {et~al.}(1970)\citenamefont
		{Shirane}, \citenamefont {Axe}, \citenamefont {Harada},\ and\ \citenamefont
		{Remeika}}]{Shirane70:2}%
	\BibitemOpen
	\bibfield  {author} {\bibinfo {author} {\bibfnamefont {G.}~\bibnamefont
			{Shirane}}, \bibinfo {author} {\bibfnamefont {J.~D.}\ \bibnamefont {Axe}},
		\bibinfo {author} {\bibfnamefont {J.}~\bibnamefont {Harada}},\ and\ \bibinfo
		{author} {\bibfnamefont {J.~P.}\ \bibnamefont {Remeika}},\ }\href
	{https://doi.org/10.1103/PhysRevB.2.155} {\bibfield  {journal} {\bibinfo
			{journal} {Phys. Rev. B}\ }\textbf {\bibinfo {volume} {2}},\ \bibinfo {pages}
		{155} (\bibinfo {year} {1970})}\BibitemShut {NoStop}%
	\bibitem [{\citenamefont {Kempa}\ \emph {et~al.}(2006)\citenamefont {Kempa},
		\citenamefont {Hlinka}, \citenamefont {Kulda}, \citenamefont {Bourges},
		\citenamefont {Kania},\ and\ \citenamefont {Petzelt}}]{Kempa06:41}%
	\BibitemOpen
	\bibfield  {author} {\bibinfo {author} {\bibfnamefont {M.}~\bibnamefont
			{Kempa}}, \bibinfo {author} {\bibfnamefont {J.}~\bibnamefont {Hlinka}},
		\bibinfo {author} {\bibfnamefont {J.}~\bibnamefont {Kulda}}, \bibinfo
		{author} {\bibfnamefont {P.}~\bibnamefont {Bourges}}, \bibinfo {author}
		{\bibfnamefont {A.}~\bibnamefont {Kania}},\ and\ \bibinfo {author}
		{\bibfnamefont {J.}~\bibnamefont {Petzelt}},\ }\href
	{https://doi.org/10.1080/01411590600892021} {\bibfield  {journal} {\bibinfo
			{journal} {Phase Transitions}\ }\textbf {\bibinfo {volume} {79}},\ \bibinfo
		{pages} {351} (\bibinfo {year} {2006})}\BibitemShut {NoStop}%
	\bibitem [{\citenamefont {Tomeno}\ \emph {et~al.}(2006)\citenamefont {Tomeno},
		\citenamefont {Ishii}, \citenamefont {Tsunoda},\ and\ \citenamefont
		{Oka}}]{Tomeno06:73}%
	\BibitemOpen
	\bibfield  {author} {\bibinfo {author} {\bibfnamefont {I.}~\bibnamefont
			{Tomeno}}, \bibinfo {author} {\bibfnamefont {Y.}~\bibnamefont {Ishii}},
		\bibinfo {author} {\bibfnamefont {Y.}~\bibnamefont {Tsunoda}},\ and\ \bibinfo
		{author} {\bibfnamefont {K.}~\bibnamefont {Oka}},\ }\href
	{https://doi.org/10.1103/PhysRevB.73.064116} {\bibfield  {journal} {\bibinfo
			{journal} {Phys. Rev. B}\ }\textbf {\bibinfo {volume} {73}},\ \bibinfo
		{pages} {064116} (\bibinfo {year} {2006})}\BibitemShut {NoStop}%
	\bibitem [{\citenamefont {Hlinka}\ \emph {et~al.}(2006)\citenamefont {Hlinka},
		\citenamefont {Kempa}, \citenamefont {Kulda}, \citenamefont {Bourges},
		\citenamefont {Kania},\ and\ \citenamefont {Petzelt}}]{Hlinka06:73}%
	\BibitemOpen
	\bibfield  {author} {\bibinfo {author} {\bibfnamefont {J.}~\bibnamefont
			{Hlinka}}, \bibinfo {author} {\bibfnamefont {M.}~\bibnamefont {Kempa}},
		\bibinfo {author} {\bibfnamefont {J.}~\bibnamefont {Kulda}}, \bibinfo
		{author} {\bibfnamefont {P.}~\bibnamefont {Bourges}}, \bibinfo {author}
		{\bibfnamefont {A.}~\bibnamefont {Kania}},\ and\ \bibinfo {author}
		{\bibfnamefont {J.}~\bibnamefont {Petzelt}},\ }\href
	{https://doi.org/10.1103/PhysRevB.73.140101} {\bibfield  {journal} {\bibinfo
			{journal} {Phys. Rev. B}\ }\textbf {\bibinfo {volume} {73}},\ \bibinfo
		{pages} {140101(R)} (\bibinfo {year} {2006})}\BibitemShut {NoStop}%
	\bibitem [{\citenamefont {Gehring}\ \emph {et~al.}(2009)\citenamefont
		{Gehring}, \citenamefont {Hiraka}, \citenamefont {Stock}, \citenamefont
		{Lee}, \citenamefont {Chen}, \citenamefont {Ye}, \citenamefont {Vakhrushev},\
		and\ \citenamefont {Chowdhuri}}]{Gehring09:79}%
	\BibitemOpen
	\bibfield  {author} {\bibinfo {author} {\bibfnamefont {P.~M.}\ \bibnamefont
			{Gehring}}, \bibinfo {author} {\bibfnamefont {H.}~\bibnamefont {Hiraka}},
		\bibinfo {author} {\bibfnamefont {C.}~\bibnamefont {Stock}}, \bibinfo
		{author} {\bibfnamefont {S.-H.}\ \bibnamefont {Lee}}, \bibinfo {author}
		{\bibfnamefont {W.}~\bibnamefont {Chen}}, \bibinfo {author} {\bibfnamefont
			{Z.-G.}\ \bibnamefont {Ye}}, \bibinfo {author} {\bibfnamefont {S.~B.}\
			\bibnamefont {Vakhrushev}},\ and\ \bibinfo {author} {\bibfnamefont
			{Z.}~\bibnamefont {Chowdhuri}},\ }\href
	{https://doi.org/10.1103/PhysRevB.79.224109} {\bibfield  {journal} {\bibinfo
			{journal} {Phys. Rev. B}\ }\textbf {\bibinfo {volume} {79}},\ \bibinfo
		{pages} {224109} (\bibinfo {year} {2009})}\BibitemShut {NoStop}%
	\bibitem [{\citenamefont {Burns}\ and\ \citenamefont
		{Dacol}(1983)}]{Burns83:48}%
	\BibitemOpen
	\bibfield  {author} {\bibinfo {author} {\bibfnamefont {G.}~\bibnamefont
			{Burns}}\ and\ \bibinfo {author} {\bibfnamefont {F.~H.}\ \bibnamefont
			{Dacol}},\ }\href {https://doi.org/10.1016/0038-1098(83)90132-1} {\bibfield
		{journal} {\bibinfo  {journal} {Solid State Commun.}\ }\textbf {\bibinfo
			{volume} {48}},\ \bibinfo {pages} {853} (\bibinfo {year} {1983})}\BibitemShut
	{NoStop}%
	\bibitem [{\citenamefont {Gehring}\ \emph
		{et~al.}(2001{\natexlab{a}})\citenamefont {Gehring}, \citenamefont {Park},\
		and\ \citenamefont {Shirane}}]{Gehring01:63}%
	\BibitemOpen
	\bibfield  {author} {\bibinfo {author} {\bibfnamefont {P.~M.}\ \bibnamefont
			{Gehring}}, \bibinfo {author} {\bibfnamefont {S.-E.}\ \bibnamefont {Park}},\
		and\ \bibinfo {author} {\bibfnamefont {G.}~\bibnamefont {Shirane}},\ }\href
	{https://doi.org/10.1103/PhysRevB.63.224109} {\bibfield  {journal} {\bibinfo
			{journal} {Phys. Rev. B}\ }\textbf {\bibinfo {volume} {63}},\ \bibinfo
		{pages} {224109} (\bibinfo {year} {2001}{\natexlab{a}})}\BibitemShut
	{NoStop}%
	\bibitem [{\citenamefont {Hlinka}\ \emph {et~al.}(2003)\citenamefont {Hlinka},
		\citenamefont {Kamba}, \citenamefont {Petzelt}, \citenamefont {Kulda},
		\citenamefont {Randall},\ and\ \citenamefont {Zhang}}]{Hlinka03:91}%
	\BibitemOpen
	\bibfield  {author} {\bibinfo {author} {\bibfnamefont {J.}~\bibnamefont
			{Hlinka}}, \bibinfo {author} {\bibfnamefont {S.}~\bibnamefont {Kamba}},
		\bibinfo {author} {\bibfnamefont {J.}~\bibnamefont {Petzelt}}, \bibinfo
		{author} {\bibfnamefont {J.}~\bibnamefont {Kulda}}, \bibinfo {author}
		{\bibfnamefont {C.~A.}\ \bibnamefont {Randall}},\ and\ \bibinfo {author}
		{\bibfnamefont {S.~J.}\ \bibnamefont {Zhang}},\ }\href
	{https://doi.org/10.1103/PhysRevLett.91.107602} {\bibfield  {journal}
		{\bibinfo  {journal} {Phys. Rev. Lett.}\ }\textbf {\bibinfo {volume} {91}},\
		\bibinfo {pages} {107602} (\bibinfo {year} {2003})}\BibitemShut {NoStop}%
	\bibitem [{\citenamefont {Naberezhnov}\ \emph {et~al.}(1999)\citenamefont
		{Naberezhnov}, \citenamefont {Vakhrushev}, \citenamefont {Dorner},
		\citenamefont {Strauch},\ and\ \citenamefont {Moudden}}]{Nab99:11}%
	\BibitemOpen
	\bibfield  {author} {\bibinfo {author} {\bibfnamefont {A.}~\bibnamefont
			{Naberezhnov}}, \bibinfo {author} {\bibfnamefont {S.}~\bibnamefont
			{Vakhrushev}}, \bibinfo {author} {\bibfnamefont {B.}~\bibnamefont {Dorner}},
		\bibinfo {author} {\bibfnamefont {D.}~\bibnamefont {Strauch}},\ and\ \bibinfo
		{author} {\bibfnamefont {H.}~\bibnamefont {Moudden}},\ }\href
	{https://doi.org/10.1007/s100510050912} {\bibfield  {journal} {\bibinfo
			{journal} {Eur. Phys. J. B}\ }\textbf {\bibinfo {volume} {11}},\ \bibinfo
		{pages} {13} (\bibinfo {year} {1999})}\BibitemShut {NoStop}%
	\bibitem [{\citenamefont {Gehring}\ \emph {et~al.}(2000)\citenamefont
		{Gehring}, \citenamefont {Park},\ and\ \citenamefont
		{Shirane}}]{Gehring00:84}%
	\BibitemOpen
	\bibfield  {author} {\bibinfo {author} {\bibfnamefont {P.~M.}\ \bibnamefont
			{Gehring}}, \bibinfo {author} {\bibfnamefont {S.-E.}\ \bibnamefont {Park}},\
		and\ \bibinfo {author} {\bibfnamefont {G.}~\bibnamefont {Shirane}},\ }\href
	{https://doi.org/10.1103/PhysRevLett.84.5216} {\bibfield  {journal} {\bibinfo
			{journal} {Phys. Rev. Lett.}\ }\textbf {\bibinfo {volume} {84}},\ \bibinfo
		{pages} {5216} (\bibinfo {year} {2000})}\BibitemShut {NoStop}%
	\bibitem [{\citenamefont {Gehring}\ \emph
		{et~al.}(2001{\natexlab{b}})\citenamefont {Gehring}, \citenamefont
		{Wakimoto}, \citenamefont {Ye},\ and\ \citenamefont
		{Shirane}}]{Gehring01:87}%
	\BibitemOpen
	\bibfield  {author} {\bibinfo {author} {\bibfnamefont {P.~M.}\ \bibnamefont
			{Gehring}}, \bibinfo {author} {\bibfnamefont {S.}~\bibnamefont {Wakimoto}},
		\bibinfo {author} {\bibfnamefont {Z.~G.}\ \bibnamefont {Ye}},\ and\ \bibinfo
		{author} {\bibfnamefont {G.}~\bibnamefont {Shirane}},\ }\href
	{https://doi.org/10.1103/PhysRevLett.87.277601} {\bibfield  {journal}
		{\bibinfo  {journal} {Phys. Rev. Lett.}\ }\textbf {\bibinfo {volume} {87}},\
		\bibinfo {pages} {277601} (\bibinfo {year} {2001}{\natexlab{b}})}\BibitemShut
	{NoStop}%
	\bibitem [{\citenamefont {Wakimoto}\ \emph {et~al.}(2002)\citenamefont
		{Wakimoto}, \citenamefont {Stock}, \citenamefont {Birgeneau}, \citenamefont
		{Ye}, \citenamefont {Chen}, \citenamefont {Buyers}, \citenamefont {Gehring},\
		and\ \citenamefont {Shirane}}]{Waki02:65}%
	\BibitemOpen
	\bibfield  {author} {\bibinfo {author} {\bibfnamefont {S.}~\bibnamefont
			{Wakimoto}}, \bibinfo {author} {\bibfnamefont {C.}~\bibnamefont {Stock}},
		\bibinfo {author} {\bibfnamefont {R.~J.}\ \bibnamefont {Birgeneau}}, \bibinfo
		{author} {\bibfnamefont {Z.-G.}\ \bibnamefont {Ye}}, \bibinfo {author}
		{\bibfnamefont {W.}~\bibnamefont {Chen}}, \bibinfo {author} {\bibfnamefont
			{W.~J.~L.}\ \bibnamefont {Buyers}}, \bibinfo {author} {\bibfnamefont {P.~M.}\
			\bibnamefont {Gehring}},\ and\ \bibinfo {author} {\bibfnamefont
			{G.}~\bibnamefont {Shirane}},\ }\href
	{https://doi.org/10.1103/PhysRevB.65.172105} {\bibfield  {journal} {\bibinfo
			{journal} {Phys. Rev. B}\ }\textbf {\bibinfo {volume} {65}},\ \bibinfo
		{pages} {172105} (\bibinfo {year} {2002})}\BibitemShut {NoStop}%
	\bibitem [{\citenamefont {Cao}\ \emph {et~al.}(2008)\citenamefont {Cao},
		\citenamefont {Stock}, \citenamefont {Xu}, \citenamefont {Gehring},
		\citenamefont {Li},\ and\ \citenamefont {Viehland}}]{Cao08:78}%
	\BibitemOpen
	\bibfield  {author} {\bibinfo {author} {\bibfnamefont {H.}~\bibnamefont
			{Cao}}, \bibinfo {author} {\bibfnamefont {C.}~\bibnamefont {Stock}}, \bibinfo
		{author} {\bibfnamefont {G.}~\bibnamefont {Xu}}, \bibinfo {author}
		{\bibfnamefont {P.~M.}\ \bibnamefont {Gehring}}, \bibinfo {author}
		{\bibfnamefont {J.}~\bibnamefont {Li}},\ and\ \bibinfo {author}
		{\bibfnamefont {D.}~\bibnamefont {Viehland}},\ }\href
	{https://doi.org/10.1103/PhysRevB.78.104103} {\bibfield  {journal} {\bibinfo
			{journal} {Phys. Rev. B}\ }\textbf {\bibinfo {volume} {78}},\ \bibinfo
		{pages} {104103} (\bibinfo {year} {2008})}\BibitemShut {NoStop}%
	\bibitem [{\citenamefont {Vakhrushev}\ and\ \citenamefont
		{Shapiro}(2002)}]{Vak02:66}%
	\BibitemOpen
	\bibfield  {author} {\bibinfo {author} {\bibfnamefont {S.~B.}\ \bibnamefont
			{Vakhrushev}}\ and\ \bibinfo {author} {\bibfnamefont {S.~M.}\ \bibnamefont
			{Shapiro}},\ }\href {https://doi.org/10.1103/PhysRevB.66.214101} {\bibfield
		{journal} {\bibinfo  {journal} {Phys. Rev. B}\ }\textbf {\bibinfo {volume}
			{66}},\ \bibinfo {pages} {214101} (\bibinfo {year} {2002})}\BibitemShut
	{NoStop}%
	\bibitem [{\citenamefont {Kamba}\ \emph {et~al.}(2005)\citenamefont {Kamba},
		\citenamefont {Kempa}, \citenamefont {Bovtun}, \citenamefont {Petzelt},
		\citenamefont {Brinkman},\ and\ \citenamefont {Setter}}]{Kamba05:17}%
	\BibitemOpen
	\bibfield  {author} {\bibinfo {author} {\bibfnamefont {S.}~\bibnamefont
			{Kamba}}, \bibinfo {author} {\bibfnamefont {M.}~\bibnamefont {Kempa}},
		\bibinfo {author} {\bibfnamefont {V.}~\bibnamefont {Bovtun}}, \bibinfo
		{author} {\bibfnamefont {J.}~\bibnamefont {Petzelt}}, \bibinfo {author}
		{\bibfnamefont {K.}~\bibnamefont {Brinkman}},\ and\ \bibinfo {author}
		{\bibfnamefont {N.}~\bibnamefont {Setter}},\ }\href
	{https://doi.org/10.1088/0953-8984/17/25/022} {\bibfield  {journal} {\bibinfo
			{journal} {J. Phys. Condens. Matter}\ }\textbf {\bibinfo {volume} {17}},\
		\bibinfo {pages} {3965} (\bibinfo {year} {2005})}\BibitemShut {NoStop}%
	\bibitem [{\citenamefont {Nuzhnyy}\ \emph {et~al.}(2017)\citenamefont
		{Nuzhnyy}, \citenamefont {Petzelt}, \citenamefont {Bovtun}, \citenamefont
		{Kempa}, \citenamefont {Kamba}, \citenamefont {Hlinka},\ and\ \citenamefont
		{Hehlen}}]{Nuzhnyy17:96}%
	\BibitemOpen
	\bibfield  {author} {\bibinfo {author} {\bibfnamefont {D.}~\bibnamefont
			{Nuzhnyy}}, \bibinfo {author} {\bibfnamefont {J.}~\bibnamefont {Petzelt}},
		\bibinfo {author} {\bibfnamefont {V.}~\bibnamefont {Bovtun}}, \bibinfo
		{author} {\bibfnamefont {M.}~\bibnamefont {Kempa}}, \bibinfo {author}
		{\bibfnamefont {S.}~\bibnamefont {Kamba}}, \bibinfo {author} {\bibfnamefont
			{J.}~\bibnamefont {Hlinka}},\ and\ \bibinfo {author} {\bibfnamefont
			{B.}~\bibnamefont {Hehlen}},\ }\href
	{https://doi.org/10.1103/PhysRevB.96.174113} {\bibfield  {journal} {\bibinfo
			{journal} {Phys. Rev. B}\ }\textbf {\bibinfo {volume} {96}},\ \bibinfo
		{pages} {174113} (\bibinfo {year} {2017})}\BibitemShut {NoStop}%
	\bibitem [{\citenamefont {Bishop}\ \emph {et~al.}(2010)\citenamefont {Bishop},
		\citenamefont {Bussmann-Holder}, \citenamefont {Kamba},\ and\ \citenamefont
		{Maglione}}]{Bishop10:81}%
	\BibitemOpen
	\bibfield  {author} {\bibinfo {author} {\bibfnamefont {A.~R.}\ \bibnamefont
			{Bishop}}, \bibinfo {author} {\bibfnamefont {A.}~\bibnamefont
			{Bussmann-Holder}}, \bibinfo {author} {\bibfnamefont {S.}~\bibnamefont
			{Kamba}},\ and\ \bibinfo {author} {\bibfnamefont {M.}~\bibnamefont
			{Maglione}},\ }\href {https://doi.org/10.1103/PhysRevB.81.064106} {\bibfield
		{journal} {\bibinfo  {journal} {Phys. Rev. B}\ }\textbf {\bibinfo {volume}
			{81}},\ \bibinfo {pages} {064106} (\bibinfo {year} {2010})}\BibitemShut
	{NoStop}%
	\bibitem [{\citenamefont {Vakhrushev}\ \emph {et~al.}(1989)\citenamefont
		{Vakhrushev}, \citenamefont {Naberezhnov}, \citenamefont {Okuneva},\ and\
		\citenamefont {Toperverg}}]{Vak89:90}%
	\BibitemOpen
	\bibfield  {author} {\bibinfo {author} {\bibfnamefont {S.~B.}\ \bibnamefont
			{Vakhrushev}}, \bibinfo {author} {\bibfnamefont {A.~A.}\ \bibnamefont
			{Naberezhnov}}, \bibinfo {author} {\bibfnamefont {N.~M.}\ \bibnamefont
			{Okuneva}},\ and\ \bibinfo {author} {\bibfnamefont {B.~P.}\ \bibnamefont
			{Toperverg}},\ }\href {https://doi.org/10.1080/00150198908211287} {\bibfield
		{journal} {\bibinfo  {journal} {Ferroelectrics}\ }\textbf {\bibinfo {volume}
			{90}},\ \bibinfo {pages} {173} (\bibinfo {year} {1989})}\BibitemShut
	{NoStop}%
	\bibitem [{\citenamefont {You}\ and\ \citenamefont {Zhang}(1997)}]{You97:79}%
	\BibitemOpen
	\bibfield  {author} {\bibinfo {author} {\bibfnamefont {H.}~\bibnamefont
			{You}}\ and\ \bibinfo {author} {\bibfnamefont {Q.~M.}\ \bibnamefont
			{Zhang}},\ }\href {https://doi.org/10.1103/PhysRevLett.79.3950} {\bibfield
		{journal} {\bibinfo  {journal} {Phys. Rev. Lett.}\ }\textbf {\bibinfo
			{volume} {79}},\ \bibinfo {pages} {3950} (\bibinfo {year}
		{1997})}\BibitemShut {NoStop}%
	\bibitem [{\citenamefont {Dkhil}\ \emph {et~al.}(2001)\citenamefont {Dkhil},
		\citenamefont {Kiat}, \citenamefont {Calvarin}, \citenamefont {Baldinozzi},
		\citenamefont {Vakhrushev},\ and\ \citenamefont {Suard}}]{Dkhil01:65}%
	\BibitemOpen
	\bibfield  {author} {\bibinfo {author} {\bibfnamefont {B.}~\bibnamefont
			{Dkhil}}, \bibinfo {author} {\bibfnamefont {J.~M.}\ \bibnamefont {Kiat}},
		\bibinfo {author} {\bibfnamefont {G.}~\bibnamefont {Calvarin}}, \bibinfo
		{author} {\bibfnamefont {G.}~\bibnamefont {Baldinozzi}}, \bibinfo {author}
		{\bibfnamefont {S.~B.}\ \bibnamefont {Vakhrushev}},\ and\ \bibinfo {author}
		{\bibfnamefont {E.}~\bibnamefont {Suard}},\ }\href
	{https://doi.org/10.1103/PhysRevB.65.024104} {\bibfield  {journal} {\bibinfo
			{journal} {Phys. Rev. B}\ }\textbf {\bibinfo {volume} {65}},\ \bibinfo
		{pages} {024104} (\bibinfo {year} {2001})}\BibitemShut {NoStop}%
	\bibitem [{\citenamefont {Hirota}\ \emph {et~al.}(2002)\citenamefont {Hirota},
		\citenamefont {Ye}, \citenamefont {Wakimoto}, \citenamefont {Gehring},\ and\
		\citenamefont {Shirane}}]{Hirota02:65}%
	\BibitemOpen
	\bibfield  {author} {\bibinfo {author} {\bibfnamefont {K.}~\bibnamefont
			{Hirota}}, \bibinfo {author} {\bibfnamefont {Z.-G.}\ \bibnamefont {Ye}},
		\bibinfo {author} {\bibfnamefont {S.}~\bibnamefont {Wakimoto}}, \bibinfo
		{author} {\bibfnamefont {P.~M.}\ \bibnamefont {Gehring}},\ and\ \bibinfo
		{author} {\bibfnamefont {G.}~\bibnamefont {Shirane}},\ }\href
	{https://doi.org/10.1103/PhysRevB.65.104105} {\bibfield  {journal} {\bibinfo
			{journal} {Phys. Rev. B}\ }\textbf {\bibinfo {volume} {65}},\ \bibinfo
		{pages} {104105} (\bibinfo {year} {2002})}\BibitemShut {NoStop}%
	\bibitem [{\citenamefont {Xu}\ \emph {et~al.}(2004{\natexlab{a}})\citenamefont
		{Xu}, \citenamefont {Shirane}, \citenamefont {Copley},\ and\ \citenamefont
		{Gehring}}]{Xu04:69}%
	\BibitemOpen
	\bibfield  {author} {\bibinfo {author} {\bibfnamefont {G.}~\bibnamefont
			{Xu}}, \bibinfo {author} {\bibfnamefont {G.}~\bibnamefont {Shirane}},
		\bibinfo {author} {\bibfnamefont {J.~R.~D.}\ \bibnamefont {Copley}},\ and\
		\bibinfo {author} {\bibfnamefont {P.~M.}\ \bibnamefont {Gehring}},\ }\href
	{https://doi.org/10.1103/PhysRevB.69.064112} {\bibfield  {journal} {\bibinfo
			{journal} {Phys. Rev. B}\ }\textbf {\bibinfo {volume} {69}},\ \bibinfo
		{pages} {064112} (\bibinfo {year} {2004}{\natexlab{a}})}\BibitemShut
	{NoStop}%
	\bibitem [{\citenamefont {Matsuura}\ \emph {et~al.}(2006)\citenamefont
		{Matsuura}, \citenamefont {Hirota}, \citenamefont {Gehring}, \citenamefont
		{Ye}, \citenamefont {Chen},\ and\ \citenamefont {Shirane}}]{Mats06:74}%
	\BibitemOpen
	\bibfield  {author} {\bibinfo {author} {\bibfnamefont {M.}~\bibnamefont
			{Matsuura}}, \bibinfo {author} {\bibfnamefont {K.}~\bibnamefont {Hirota}},
		\bibinfo {author} {\bibfnamefont {P.~M.}\ \bibnamefont {Gehring}}, \bibinfo
		{author} {\bibfnamefont {Z.-G.}\ \bibnamefont {Ye}}, \bibinfo {author}
		{\bibfnamefont {W.}~\bibnamefont {Chen}},\ and\ \bibinfo {author}
		{\bibfnamefont {G.}~\bibnamefont {Shirane}},\ }\href
	{https://doi.org/10.1103/PhysRevB.74.144107} {\bibfield  {journal} {\bibinfo
			{journal} {Phys. Rev. B}\ }\textbf {\bibinfo {volume} {74}},\ \bibinfo
		{pages} {144107} (\bibinfo {year} {2006})}\BibitemShut {NoStop}%
	\bibitem [{\citenamefont {Stock}\ \emph
		{et~al.}(2010{\natexlab{a}})\citenamefont {Stock}, \citenamefont {Van~Eijck},
		\citenamefont {Fouquet}, \citenamefont {Maccarini}, \citenamefont {Gehring},
		\citenamefont {Xu}, \citenamefont {Luo}, \citenamefont {Zhao}, \citenamefont
		{Li},\ and\ \citenamefont {Viehland}}]{Stock10:81}%
	\BibitemOpen
	\bibfield  {author} {\bibinfo {author} {\bibfnamefont {C.}~\bibnamefont
			{Stock}}, \bibinfo {author} {\bibfnamefont {L.}~\bibnamefont {Van~Eijck}},
		\bibinfo {author} {\bibfnamefont {P.}~\bibnamefont {Fouquet}}, \bibinfo
		{author} {\bibfnamefont {M.}~\bibnamefont {Maccarini}}, \bibinfo {author}
		{\bibfnamefont {P.~M.}\ \bibnamefont {Gehring}}, \bibinfo {author}
		{\bibfnamefont {G.}~\bibnamefont {Xu}}, \bibinfo {author} {\bibfnamefont
			{H.}~\bibnamefont {Luo}}, \bibinfo {author} {\bibfnamefont {X.}~\bibnamefont
			{Zhao}}, \bibinfo {author} {\bibfnamefont {J.-F.}\ \bibnamefont {Li}},\ and\
		\bibinfo {author} {\bibfnamefont {D.}~\bibnamefont {Viehland}},\ }\href
	{https://doi.org/10.1103/PhysRevB.81.144127} {\bibfield  {journal} {\bibinfo
			{journal} {Phys. Rev. B}\ }\textbf {\bibinfo {volume} {81}},\ \bibinfo
		{pages} {144127} (\bibinfo {year} {2010}{\natexlab{a}})}\BibitemShut
	{NoStop}%
	\bibitem [{\citenamefont {Xu}\ \emph {et~al.}(2012)\citenamefont {Xu},
		\citenamefont {Wen}, \citenamefont {Mamontov}, \citenamefont {Stock},
		\citenamefont {Gehring},\ and\ \citenamefont {Xu}}]{Xu12:86}%
	\BibitemOpen
	\bibfield  {author} {\bibinfo {author} {\bibfnamefont {Z.}~\bibnamefont
			{Xu}}, \bibinfo {author} {\bibfnamefont {J.}~\bibnamefont {Wen}}, \bibinfo
		{author} {\bibfnamefont {E.}~\bibnamefont {Mamontov}}, \bibinfo {author}
		{\bibfnamefont {C.}~\bibnamefont {Stock}}, \bibinfo {author} {\bibfnamefont
			{P.~M.}\ \bibnamefont {Gehring}},\ and\ \bibinfo {author} {\bibfnamefont
			{G.}~\bibnamefont {Xu}},\ }\href {https://doi.org/10.1103/PhysRevB.86.144106}
	{\bibfield  {journal} {\bibinfo  {journal} {Phys. Rev. B}\ }\textbf {\bibinfo
			{volume} {86}},\ \bibinfo {pages} {144106} (\bibinfo {year}
		{2012})}\BibitemShut {NoStop}%
	\bibitem [{\citenamefont {Vakhrushev}\ \emph {et~al.}(2005)\citenamefont
		{Vakhrushev}, \citenamefont {Ivanov},\ and\ \citenamefont {Kulda}}]{Vak05:7}%
	\BibitemOpen
	\bibfield  {author} {\bibinfo {author} {\bibfnamefont {S.}~\bibnamefont
			{Vakhrushev}}, \bibinfo {author} {\bibfnamefont {A.}~\bibnamefont {Ivanov}},\
		and\ \bibinfo {author} {\bibfnamefont {J.}~\bibnamefont {Kulda}},\ }\href
	{https://doi.org/10.1039/b416454g} {\bibfield  {journal} {\bibinfo  {journal}
			{PCCP}\ }\textbf {\bibinfo {volume} {7}},\ \bibinfo {pages} {2340} (\bibinfo
		{year} {2005})}\BibitemShut {NoStop}%
	\bibitem [{\citenamefont {Vakhrushev}\ \emph {et~al.}(1998)\citenamefont
		{Vakhrushev}, \citenamefont {Naberezhnov}, \citenamefont {Okuneva},\ and\
		\citenamefont {Savenko}}]{Vak98:40}%
	\BibitemOpen
	\bibfield  {author} {\bibinfo {author} {\bibfnamefont {S.}~\bibnamefont
			{Vakhrushev}}, \bibinfo {author} {\bibfnamefont {A.}~\bibnamefont
			{Naberezhnov}}, \bibinfo {author} {\bibfnamefont {N.~M.}\ \bibnamefont
			{Okuneva}},\ and\ \bibinfo {author} {\bibfnamefont {B.~N.}\ \bibnamefont
			{Savenko}},\ }\href {https://doi.org/10.1134/1.1130645} {\bibfield  {journal}
		{\bibinfo  {journal} {Phys. Solid State}\ }\textbf {\bibinfo {volume} {40}},\
		\bibinfo {pages} {1728} (\bibinfo {year} {1998})}\BibitemShut {NoStop}%
	\bibitem [{\citenamefont {Vakhrushev}\ \emph {et~al.}(1995)\citenamefont
		{Vakhrushev}, \citenamefont {Naberezhnov}, \citenamefont {Okuneva},\ and\
		\citenamefont {Savenko}}]{Vak95:37}%
	\BibitemOpen
	\bibfield  {author} {\bibinfo {author} {\bibfnamefont {S.}~\bibnamefont
			{Vakhrushev}}, \bibinfo {author} {\bibfnamefont {A.}~\bibnamefont
			{Naberezhnov}}, \bibinfo {author} {\bibfnamefont {N.~M.}\ \bibnamefont
			{Okuneva}},\ and\ \bibinfo {author} {\bibfnamefont {B.~N.}\ \bibnamefont
			{Savenko}},\ }\href@noop {} {\bibfield  {journal} {\bibinfo  {journal}
			{Fizika Tverdogo Tela}\ }\textbf {\bibinfo {volume} {37}},\ \bibinfo {pages}
		{3621} (\bibinfo {year} {1995})}\BibitemShut {NoStop}%
	\bibitem [{\citenamefont {Gvasaliya}\ \emph {et~al.}(2003)\citenamefont
		{Gvasaliya}, \citenamefont {Roessli},\ and\ \citenamefont
		{Lushnikov}}]{Gvas03:63}%
	\BibitemOpen
	\bibfield  {author} {\bibinfo {author} {\bibfnamefont {S.~N.}\ \bibnamefont
			{Gvasaliya}}, \bibinfo {author} {\bibfnamefont {B.}~\bibnamefont {Roessli}},\
		and\ \bibinfo {author} {\bibfnamefont {S.~G.}\ \bibnamefont {Lushnikov}},\
	}\href {https://doi.org/10.1209/epl/i2003-00528-3} {\bibfield  {journal}
		{\bibinfo  {journal} {Europhys. Lett.}\ }\textbf {\bibinfo {volume} {63}},\
		\bibinfo {pages} {303} (\bibinfo {year} {2003})}\BibitemShut {NoStop}%
	\bibitem [{\citenamefont {Ye}\ and\ \citenamefont {Schmid}(1993)}]{Ye93:145}%
	\BibitemOpen
	\bibfield  {author} {\bibinfo {author} {\bibfnamefont {Z.-G.}\ \bibnamefont
			{Ye}}\ and\ \bibinfo {author} {\bibfnamefont {H.}~\bibnamefont {Schmid}},\
	}\href {https://doi.org/10.1080/00150199308222438} {\bibfield  {journal}
		{\bibinfo  {journal} {Ferroelectrics}\ }\textbf {\bibinfo {volume} {145}},\
		\bibinfo {pages} {83} (\bibinfo {year} {1993})}\BibitemShut {NoStop}%
	\bibitem [{\citenamefont {Stock}\ \emph {et~al.}(2007)\citenamefont {Stock},
		\citenamefont {Xu}, \citenamefont {Gehring}, \citenamefont {Luo},
		\citenamefont {Zhao}, \citenamefont {Cao}, \citenamefont {Li}, \citenamefont
		{Viehland},\ and\ \citenamefont {Shirane}}]{Stock07:76}%
	\BibitemOpen
	\bibfield  {author} {\bibinfo {author} {\bibfnamefont {C.}~\bibnamefont
			{Stock}}, \bibinfo {author} {\bibfnamefont {G.}~\bibnamefont {Xu}}, \bibinfo
		{author} {\bibfnamefont {P.~M.}\ \bibnamefont {Gehring}}, \bibinfo {author}
		{\bibfnamefont {H.}~\bibnamefont {Luo}}, \bibinfo {author} {\bibfnamefont
			{X.}~\bibnamefont {Zhao}}, \bibinfo {author} {\bibfnamefont {H.}~\bibnamefont
			{Cao}}, \bibinfo {author} {\bibfnamefont {J.~F.}\ \bibnamefont {Li}},
		\bibinfo {author} {\bibfnamefont {D.}~\bibnamefont {Viehland}},\ and\
		\bibinfo {author} {\bibfnamefont {G.}~\bibnamefont {Shirane}},\ }\href
	{https://doi.org/10.1103/PhysRevB.76.064122} {\bibfield  {journal} {\bibinfo
			{journal} {Phys. Rev. B}\ }\textbf {\bibinfo {volume} {76}},\ \bibinfo
		{pages} {064122} (\bibinfo {year} {2007})}\BibitemShut {NoStop}%
	\bibitem [{\citenamefont {Westphal}\ \emph {et~al.}(1992)\citenamefont
		{Westphal}, \citenamefont {Kleemann},\ and\ \citenamefont
		{Glinchuk}}]{Westphal92:68}%
	\BibitemOpen
	\bibfield  {author} {\bibinfo {author} {\bibfnamefont {V.}~\bibnamefont
			{Westphal}}, \bibinfo {author} {\bibfnamefont {W.}~\bibnamefont {Kleemann}},\
		and\ \bibinfo {author} {\bibfnamefont {M.~D.}\ \bibnamefont {Glinchuk}},\
	}\href {https://doi.org/10.1103/PhysRevLett.68.847} {\bibfield  {journal}
		{\bibinfo  {journal} {Phys. Rev. Lett.}\ }\textbf {\bibinfo {volume} {68}},\
		\bibinfo {pages} {847} (\bibinfo {year} {1992})}\BibitemShut {NoStop}%
	\bibitem [{\citenamefont {Fisch}(2003)}]{Fisch03:67}%
	\BibitemOpen
	\bibfield  {author} {\bibinfo {author} {\bibfnamefont {R.}~\bibnamefont
			{Fisch}},\ }\href {https://doi.org/10.1103/PhysRevB.67.094110} {\bibfield
		{journal} {\bibinfo  {journal} {Phys. Rev. B}\ }\textbf {\bibinfo {volume}
			{67}},\ \bibinfo {pages} {094110} (\bibinfo {year} {2003})}\BibitemShut
	{NoStop}%
	\bibitem [{\citenamefont {Stock}\ \emph {et~al.}(2004)\citenamefont {Stock},
		\citenamefont {Birgeneau}, \citenamefont {Wakimoto}, \citenamefont {Gardner},
		\citenamefont {Chen}, \citenamefont {Ye},\ and\ \citenamefont
		{Shirane}}]{Stock04:69}%
	\BibitemOpen
	\bibfield  {author} {\bibinfo {author} {\bibfnamefont {C.}~\bibnamefont
			{Stock}}, \bibinfo {author} {\bibfnamefont {R.~J.}\ \bibnamefont
			{Birgeneau}}, \bibinfo {author} {\bibfnamefont {S.}~\bibnamefont {Wakimoto}},
		\bibinfo {author} {\bibfnamefont {J.~S.}\ \bibnamefont {Gardner}}, \bibinfo
		{author} {\bibfnamefont {W.}~\bibnamefont {Chen}}, \bibinfo {author}
		{\bibfnamefont {Z.-G.}\ \bibnamefont {Ye}},\ and\ \bibinfo {author}
		{\bibfnamefont {G.}~\bibnamefont {Shirane}},\ }\href
	{https://doi.org/10.1103/PhysRevB.69.094104} {\bibfield  {journal} {\bibinfo
			{journal} {Phys. Rev. B}\ }\textbf {\bibinfo {volume} {69}},\ \bibinfo
		{pages} {094104} (\bibinfo {year} {2004})}\BibitemShut {NoStop}%
	\bibitem [{\citenamefont {Cowley}\ \emph {et~al.}(2009)\citenamefont {Cowley},
		\citenamefont {Gvasaliya},\ and\ \citenamefont {Roessli}}]{Cowley09:378}%
	\BibitemOpen
	\bibfield  {author} {\bibinfo {author} {\bibfnamefont {R.~A.}\ \bibnamefont
			{Cowley}}, \bibinfo {author} {\bibfnamefont {S.~N.}\ \bibnamefont
			{Gvasaliya}},\ and\ \bibinfo {author} {\bibfnamefont {B.}~\bibnamefont
			{Roessli}},\ }\href {https://doi.org/10.1080/00150190902845087} {\bibfield
		{journal} {\bibinfo  {journal} {Ferroelectrics}\ }\textbf {\bibinfo {volume}
			{378}},\ \bibinfo {pages} {53} (\bibinfo {year} {2009})}\BibitemShut
	{NoStop}%
	\bibitem [{\citenamefont {Imry}\ and\ \citenamefont {Ma}(1975)}]{Imry75:35}%
	\BibitemOpen
	\bibfield  {author} {\bibinfo {author} {\bibfnamefont {Y.}~\bibnamefont
			{Imry}}\ and\ \bibinfo {author} {\bibfnamefont {S.-k.}\ \bibnamefont {Ma}},\
	}\href {https://doi.org/10.1103/PhysRevLett.35.1399} {\bibfield  {journal}
		{\bibinfo  {journal} {Phys. Rev. Lett.}\ }\textbf {\bibinfo {volume} {35}},\
		\bibinfo {pages} {1399} (\bibinfo {year} {1975})}\BibitemShut {NoStop}%
	\bibitem [{\citenamefont {Fishman}\ and\ \citenamefont
		{Aharony}(1979)}]{Fishman79:12}%
	\BibitemOpen
	\bibfield  {author} {\bibinfo {author} {\bibfnamefont {R.}~\bibnamefont
			{Fishman}}\ and\ \bibinfo {author} {\bibfnamefont {A.}~\bibnamefont
			{Aharony}},\ }\href {https://doi.org/10.1088/0022-3719/12/18/006} {\bibfield
		{journal} {\bibinfo  {journal} {J. Phys. C: Solid State Phys.}\ }\textbf
		{\bibinfo {volume} {12}},\ \bibinfo {pages} {L729} (\bibinfo {year}
		{1979})}\BibitemShut {NoStop}%
	\bibitem [{\citenamefont {Birgeneau}\ \emph {et~al.}(1983)\citenamefont
		{Birgeneau}, \citenamefont {Yoshizawa}, \citenamefont {Cowley}, \citenamefont
		{Shirane},\ and\ \citenamefont {Ikeda}}]{Birgeneau83:28}%
	\BibitemOpen
	\bibfield  {author} {\bibinfo {author} {\bibfnamefont {R.~J.}\ \bibnamefont
			{Birgeneau}}, \bibinfo {author} {\bibfnamefont {H.}~\bibnamefont
			{Yoshizawa}}, \bibinfo {author} {\bibfnamefont {R.~A.}\ \bibnamefont
			{Cowley}}, \bibinfo {author} {\bibfnamefont {G.}~\bibnamefont {Shirane}},\
		and\ \bibinfo {author} {\bibfnamefont {H.}~\bibnamefont {Ikeda}},\ }\href
	{https://doi.org/10.1103/PhysRevB.28.1438} {\bibfield  {journal} {\bibinfo
			{journal} {Phys. Rev. B}\ }\textbf {\bibinfo {volume} {28}},\ \bibinfo
		{pages} {1438} (\bibinfo {year} {1983})}\BibitemShut {NoStop}%
	\bibitem [{\citenamefont {Bellini}\ \emph {et~al.}(2001)\citenamefont
		{Bellini}, \citenamefont {Radzihovsky}, \citenamefont {Toner},\ and\
		\citenamefont {Clark}}]{Bellini01:294}%
	\BibitemOpen
	\bibfield  {author} {\bibinfo {author} {\bibfnamefont {T.}~\bibnamefont
			{Bellini}}, \bibinfo {author} {\bibfnamefont {L.}~\bibnamefont
			{Radzihovsky}}, \bibinfo {author} {\bibfnamefont {J.}~\bibnamefont {Toner}},\
		and\ \bibinfo {author} {\bibfnamefont {N.~A.}\ \bibnamefont {Clark}},\ }\href
	{https://doi.org/10.1126/science.1057480} {\bibfield  {journal} {\bibinfo
			{journal} {Science}\ }\textbf {\bibinfo {volume} {294}},\ \bibinfo {pages}
		{1074} (\bibinfo {year} {2001})}\BibitemShut {NoStop}%
	\bibitem [{\citenamefont {Park}\ \emph {et~al.}(2002)\citenamefont {Park},
		\citenamefont {Leheny}, \citenamefont {Birgeneau}, \citenamefont {Gallani},
		\citenamefont {Garland},\ and\ \citenamefont {Iannacchione}}]{Park02:65}%
	\BibitemOpen
	\bibfield  {author} {\bibinfo {author} {\bibfnamefont {S.}~\bibnamefont
			{Park}}, \bibinfo {author} {\bibfnamefont {R.~L.}\ \bibnamefont {Leheny}},
		\bibinfo {author} {\bibfnamefont {R.~J.}\ \bibnamefont {Birgeneau}}, \bibinfo
		{author} {\bibfnamefont {J.-L.}\ \bibnamefont {Gallani}}, \bibinfo {author}
		{\bibfnamefont {C.~W.}\ \bibnamefont {Garland}},\ and\ \bibinfo {author}
		{\bibfnamefont {G.~S.}\ \bibnamefont {Iannacchione}},\ }\href
	{https://doi.org/10.1103/PhysRevE.65.050703} {\bibfield  {journal} {\bibinfo
			{journal} {Phys. Rev. E}\ }\textbf {\bibinfo {volume} {65}},\ \bibinfo
		{pages} {050703(R)} (\bibinfo {year} {2002})}\BibitemShut {NoStop}%
	\bibitem [{\citenamefont {Reppy}(1992)}]{Reppy92:87}%
	\BibitemOpen
	\bibfield  {author} {\bibinfo {author} {\bibfnamefont {J.~D.}\ \bibnamefont
			{Reppy}},\ }\href {https://doi.org/10.1007/BF00114905} {\bibfield  {journal}
		{\bibinfo  {journal} {J. Low Temp. Phys.}\ }\textbf {\bibinfo {volume}
			{87}},\ \bibinfo {pages} {205} (\bibinfo {year} {1992})}\BibitemShut
	{NoStop}%
	\bibitem [{\citenamefont {Gridnev}\ and\ \citenamefont
		{Kamynin}(2012)}]{Gridnev12:54}%
	\BibitemOpen
	\bibfield  {author} {\bibinfo {author} {\bibfnamefont {S.~A.}\ \bibnamefont
			{Gridnev}}\ and\ \bibinfo {author} {\bibfnamefont {A.~A.}\ \bibnamefont
			{Kamynin}},\ }\href {https://doi.org/10.1134/S1063783412050137} {\bibfield
		{journal} {\bibinfo  {journal} {Phys. Solid State}\ }\textbf {\bibinfo
			{volume} {54}},\ \bibinfo {pages} {1018} (\bibinfo {year}
		{2012})}\BibitemShut {NoStop}%
	\bibitem [{\citenamefont {Brzezinska}\ \emph {et~al.}(2018)\citenamefont
		{Brzezinska}, \citenamefont {Skulski}, \citenamefont {Bochenek},
		\citenamefont {Niemiec}, \citenamefont {Chroback}, \citenamefont
		{Fajfrowski},\ and\ \citenamefont {Matyjasik}}]{Brze18:737}%
	\BibitemOpen
	\bibfield  {author} {\bibinfo {author} {\bibfnamefont {D.}~\bibnamefont
			{Brzezinska}}, \bibinfo {author} {\bibfnamefont {R.}~\bibnamefont {Skulski}},
		\bibinfo {author} {\bibfnamefont {D.}~\bibnamefont {Bochenek}}, \bibinfo
		{author} {\bibfnamefont {P.}~\bibnamefont {Niemiec}}, \bibinfo {author}
		{\bibfnamefont {A.}~\bibnamefont {Chroback}}, \bibinfo {author}
		{\bibfnamefont {L.}~\bibnamefont {Fajfrowski}},\ and\ \bibinfo {author}
		{\bibfnamefont {S.}~\bibnamefont {Matyjasik}},\ }\href
	{https://doi.org/10.1016/j.jallcom.2017.12.055} {\bibfield  {journal}
		{\bibinfo  {journal} {J. Alloys Compd.}\ }\textbf {\bibinfo {volume} {727}},\
		\bibinfo {pages} {299} (\bibinfo {year} {2018})}\BibitemShut {NoStop}%
	\bibitem [{\citenamefont {Cheong}\ and\ \citenamefont
		{Mostovoy}(2007)}]{Cheong07:6}%
	\BibitemOpen
	\bibfield  {author} {\bibinfo {author} {\bibfnamefont {S.~W.}\ \bibnamefont
			{Cheong}}\ and\ \bibinfo {author} {\bibfnamefont {M.}~\bibnamefont
			{Mostovoy}},\ }\href {https://doi.org/10.1038/nmat1804} {\bibfield  {journal}
		{\bibinfo  {journal} {Nat. Mater.}\ }\textbf {\bibinfo {volume} {6}},\
		\bibinfo {pages} {13} (\bibinfo {year} {2007})}\BibitemShut {NoStop}%
	\bibitem [{\citenamefont {Kimura}\ \emph {et~al.}(2003)\citenamefont {Kimura},
		\citenamefont {Goto}, \citenamefont {Shintani}, \citenamefont {Ishizaka},
		\citenamefont {Arima},\ and\ \citenamefont {Tokura}}]{Kimura03:426}%
	\BibitemOpen
	\bibfield  {author} {\bibinfo {author} {\bibfnamefont {T.}~\bibnamefont
			{Kimura}}, \bibinfo {author} {\bibfnamefont {T.}~\bibnamefont {Goto}},
		\bibinfo {author} {\bibfnamefont {H.}~\bibnamefont {Shintani}}, \bibinfo
		{author} {\bibfnamefont {K.}~\bibnamefont {Ishizaka}}, \bibinfo {author}
		{\bibfnamefont {T.}~\bibnamefont {Arima}},\ and\ \bibinfo {author}
		{\bibfnamefont {Y.}~\bibnamefont {Tokura}},\ }\href
	{https://doi.org/10.1038/nature02018} {\bibfield  {journal} {\bibinfo
			{journal} {Nature}\ }\textbf {\bibinfo {volume} {426}},\ \bibinfo {pages}
		{55} (\bibinfo {year} {2003})}\BibitemShut {NoStop}%
	\bibitem [{\citenamefont {dos Santos}\ \emph {et~al.}(2002)\citenamefont {dos
			Santos}, \citenamefont {Parashar}, \citenamefont {Raju}, \citenamefont
		{Zhao}, \citenamefont {Cheetham},\ and\ \citenamefont {Rao}}]{Santos02:122}%
	\BibitemOpen
	\bibfield  {author} {\bibinfo {author} {\bibfnamefont {A.~M.}\ \bibnamefont
			{dos Santos}}, \bibinfo {author} {\bibfnamefont {S.}~\bibnamefont
			{Parashar}}, \bibinfo {author} {\bibfnamefont {A.~R.}\ \bibnamefont {Raju}},
		\bibinfo {author} {\bibfnamefont {Y.~S.}\ \bibnamefont {Zhao}}, \bibinfo
		{author} {\bibfnamefont {A.~K.}\ \bibnamefont {Cheetham}},\ and\ \bibinfo
		{author} {\bibfnamefont {C.~N.~R.}\ \bibnamefont {Rao}},\ }\href
	{https://doi.org/10.1016/S0038-1098(02)00087-X} {\bibfield  {journal}
		{\bibinfo  {journal} {Solid State Commun.}\ }\textbf {\bibinfo {volume}
			{122}},\ \bibinfo {pages} {49} (\bibinfo {year} {2002})}\BibitemShut
	{NoStop}%
	\bibitem [{\citenamefont {Laguta}\ \emph {et~al.}(2016)\citenamefont {Laguta},
		\citenamefont {Morozovska}, \citenamefont {Eliseev}, \citenamefont {Raevski},
		\citenamefont {Raevskaya}, \citenamefont {Sitalo}, \citenamefont
		{Prosandeev},\ and\ \citenamefont {Bellaiche}}]{Laguta16:51}%
	\BibitemOpen
	\bibfield  {author} {\bibinfo {author} {\bibfnamefont {V.~V.}\ \bibnamefont
			{Laguta}}, \bibinfo {author} {\bibfnamefont {A.~N.}\ \bibnamefont
			{Morozovska}}, \bibinfo {author} {\bibfnamefont {E.~A.}\ \bibnamefont
			{Eliseev}}, \bibinfo {author} {\bibfnamefont {I.~P.}\ \bibnamefont
			{Raevski}}, \bibinfo {author} {\bibfnamefont {S.~I.}\ \bibnamefont
			{Raevskaya}}, \bibinfo {author} {\bibfnamefont {E.~I.}\ \bibnamefont
			{Sitalo}}, \bibinfo {author} {\bibfnamefont {S.~A.}\ \bibnamefont
			{Prosandeev}},\ and\ \bibinfo {author} {\bibfnamefont {L.}~\bibnamefont
			{Bellaiche}},\ }\href {https://doi.org/10.1007/s10853-016-9836-4} {\bibfield
		{journal} {\bibinfo  {journal} {J. Mater. Sci.}\ }\textbf {\bibinfo {volume}
			{51}},\ \bibinfo {pages} {5330} (\bibinfo {year} {2016})}\BibitemShut
	{NoStop}%
	\bibitem [{\citenamefont {Laguta}\ \emph {et~al.}(2017)\citenamefont {Laguta},
		\citenamefont {Stephanovich}, \citenamefont {Raevski}, \citenamefont
		{Raevskaya}, \citenamefont {Titov}, \citenamefont {Smotrakov},\ and\
		\citenamefont {Eremkin}}]{Laguta17:95}%
	\BibitemOpen
	\bibfield  {author} {\bibinfo {author} {\bibfnamefont {V.~V.}\ \bibnamefont
			{Laguta}}, \bibinfo {author} {\bibfnamefont {V.~A.}\ \bibnamefont
			{Stephanovich}}, \bibinfo {author} {\bibfnamefont {I.~P.}\ \bibnamefont
			{Raevski}}, \bibinfo {author} {\bibfnamefont {S.~I.}\ \bibnamefont
			{Raevskaya}}, \bibinfo {author} {\bibfnamefont {V.~V.}\ \bibnamefont
			{Titov}}, \bibinfo {author} {\bibfnamefont {V.~G.}\ \bibnamefont
			{Smotrakov}},\ and\ \bibinfo {author} {\bibfnamefont {V.~V.}\ \bibnamefont
			{Eremkin}},\ }\href {https://doi.org/10.1103/PhysRevB.95.014207} {\bibfield
		{journal} {\bibinfo  {journal} {Phys. Rev. B}\ }\textbf {\bibinfo {volume}
			{95}},\ \bibinfo {pages} {014207} (\bibinfo {year} {2017})}\BibitemShut
	{NoStop}%
	\bibitem [{\citenamefont {Yang}\ \emph {et~al.}(2004)\citenamefont {Yang},
		\citenamefont {Liu}, \citenamefont {Huang}, \citenamefont {Zou},
		\citenamefont {Bao},\ and\ \citenamefont {Liu}}]{Yang04:70}%
	\BibitemOpen
	\bibfield  {author} {\bibinfo {author} {\bibfnamefont {Y.}~\bibnamefont
			{Yang}}, \bibinfo {author} {\bibfnamefont {J.-M.}\ \bibnamefont {Liu}},
		\bibinfo {author} {\bibfnamefont {H.~B.}\ \bibnamefont {Huang}}, \bibinfo
		{author} {\bibfnamefont {W.~Q.}\ \bibnamefont {Zou}}, \bibinfo {author}
		{\bibfnamefont {P.}~\bibnamefont {Bao}},\ and\ \bibinfo {author}
		{\bibfnamefont {Z.~G.}\ \bibnamefont {Liu}},\ }\href
	{https://doi.org/10.1103/PhysRevB.70.132101} {\bibfield  {journal} {\bibinfo
			{journal} {Phys. Rev. B}\ }\textbf {\bibinfo {volume} {70}},\ \bibinfo
		{pages} {132101} (\bibinfo {year} {2004})}\BibitemShut {NoStop}%
	\bibitem [{\citenamefont {Correa}\ \emph {et~al.}(2011)\citenamefont {Correa},
		\citenamefont {Kumar}, \citenamefont {Priya}, \citenamefont {Katiyar},\ and\
		\citenamefont {Scott}}]{Correa11:83}%
	\BibitemOpen
	\bibfield  {author} {\bibinfo {author} {\bibfnamefont {M.}~\bibnamefont
			{Correa}}, \bibinfo {author} {\bibfnamefont {A.}~\bibnamefont {Kumar}},
		\bibinfo {author} {\bibfnamefont {S.}~\bibnamefont {Priya}}, \bibinfo
		{author} {\bibfnamefont {R.~S.}\ \bibnamefont {Katiyar}},\ and\ \bibinfo
		{author} {\bibfnamefont {J.~F.}\ \bibnamefont {Scott}},\ }\href
	{https://doi.org/10.1103/PhysRevB.83.014302} {\bibfield  {journal} {\bibinfo
			{journal} {Phys. Rev. B}\ }\textbf {\bibinfo {volume} {83}},\ \bibinfo
		{pages} {014302} (\bibinfo {year} {2011})}\BibitemShut {NoStop}%
	\bibitem [{\citenamefont {Garcia-Flores}\ \emph {et~al.}(2011)\citenamefont
		{Garcia-Flores}, \citenamefont {Tenne}, \citenamefont {Choi}, \citenamefont
		{Ren}, \citenamefont {Xi},\ and\ \citenamefont {Cheong}}]{Garcia11:23}%
	\BibitemOpen
	\bibfield  {author} {\bibinfo {author} {\bibfnamefont {A.~F.}\ \bibnamefont
			{Garcia-Flores}}, \bibinfo {author} {\bibfnamefont {D.~A.}\ \bibnamefont
			{Tenne}}, \bibinfo {author} {\bibfnamefont {Y.~J.}\ \bibnamefont {Choi}},
		\bibinfo {author} {\bibfnamefont {W.~J.}\ \bibnamefont {Ren}}, \bibinfo
		{author} {\bibfnamefont {X.~X.}\ \bibnamefont {Xi}},\ and\ \bibinfo {author}
		{\bibfnamefont {S.~W.}\ \bibnamefont {Cheong}},\ }\href
	{https://doi.org/10.1088/0953-8984/23/1/015401} {\bibfield  {journal}
		{\bibinfo  {journal} {J. Phys.: Condens. Matter}\ }\textbf {\bibinfo {volume}
			{23}},\ \bibinfo {pages} {015401} (\bibinfo {year} {2011})}\BibitemShut
	{NoStop}%
	\bibitem [{\citenamefont {Turik}\ \emph {et~al.}(2012)\citenamefont {Turik},
		\citenamefont {Pavlenko}, \citenamefont {Andryushin}, \citenamefont
		{Shevtsova}, \citenamefont {Reznichenko},\ and\ \citenamefont
		{Chernobabov}}]{Turik13:88}%
	\BibitemOpen
	\bibfield  {author} {\bibinfo {author} {\bibfnamefont {A.~V.}\ \bibnamefont
			{Turik}}, \bibinfo {author} {\bibfnamefont {A.~V.}\ \bibnamefont {Pavlenko}},
		\bibinfo {author} {\bibfnamefont {K.~P.}\ \bibnamefont {Andryushin}},
		\bibinfo {author} {\bibfnamefont {S.~I.}\ \bibnamefont {Shevtsova}}, \bibinfo
		{author} {\bibfnamefont {L.~A.}\ \bibnamefont {Reznichenko}},\ and\ \bibinfo
		{author} {\bibfnamefont {A.~I.}\ \bibnamefont {Chernobabov}},\ }\href
	{https://doi.org/10.1134/S1063783412050447} {\bibfield  {journal} {\bibinfo
			{journal} {Phys. Solid State}\ }\textbf {\bibinfo {volume} {54}},\ \bibinfo
		{pages} {947} (\bibinfo {year} {2012})}\BibitemShut {NoStop}%
	\bibitem [{\citenamefont {Chillal}\ \emph {et~al.}(2014)\citenamefont
		{Chillal}, \citenamefont {Gvasaliya}, \citenamefont {Zheludev}, \citenamefont
		{Schroeter}, \citenamefont {Kraken}, \citenamefont {Litterst}, \citenamefont
		{Shaplygina},\ and\ \citenamefont {Lushnikov}}]{Chillal14:89}%
	\BibitemOpen
	\bibfield  {author} {\bibinfo {author} {\bibfnamefont {S.}~\bibnamefont
			{Chillal}}, \bibinfo {author} {\bibfnamefont {S.~N.}\ \bibnamefont
			{Gvasaliya}}, \bibinfo {author} {\bibfnamefont {A.}~\bibnamefont {Zheludev}},
		\bibinfo {author} {\bibfnamefont {D.}~\bibnamefont {Schroeter}}, \bibinfo
		{author} {\bibfnamefont {M.}~\bibnamefont {Kraken}}, \bibinfo {author}
		{\bibfnamefont {F.~J.}\ \bibnamefont {Litterst}}, \bibinfo {author}
		{\bibfnamefont {T.~A.}\ \bibnamefont {Shaplygina}},\ and\ \bibinfo {author}
		{\bibfnamefont {S.~G.}\ \bibnamefont {Lushnikov}},\ }\href
	{https://doi.org/10.1103/PhysRevB.89.174418} {\bibfield  {journal} {\bibinfo
			{journal} {Phys. Rev. B}\ }\textbf {\bibinfo {volume} {89}},\ \bibinfo
		{pages} {174418(R)} (\bibinfo {year} {2014})}\BibitemShut {NoStop}%
	\bibitem [{\citenamefont {Chen}\ \emph {et~al.}(2016)\citenamefont {Chen},
		\citenamefont {Bokov}, \citenamefont {Zhu}, \citenamefont {Zhuang},
		\citenamefont {Zhang}, \citenamefont {Tailor}, \citenamefont {Ren},\ and\
		\citenamefont {Ye}}]{Chen16:6}%
	\BibitemOpen
	\bibfield  {author} {\bibinfo {author} {\bibfnamefont {L.}~\bibnamefont
			{Chen}}, \bibinfo {author} {\bibfnamefont {A.~A.}\ \bibnamefont {Bokov}},
		\bibinfo {author} {\bibfnamefont {W.}~\bibnamefont {Zhu}}, \bibinfo {author}
		{\bibfnamefont {J.}~\bibnamefont {Zhuang}}, \bibinfo {author} {\bibfnamefont
			{N.}~\bibnamefont {Zhang}}, \bibinfo {author} {\bibfnamefont {H.~N.}\
			\bibnamefont {Tailor}}, \bibinfo {author} {\bibfnamefont {W.}~\bibnamefont
			{Ren}},\ and\ \bibinfo {author} {\bibfnamefont {Z.~G.}\ \bibnamefont {Ye}},\
	}\href {https://doi.org/10.1038/srep22327} {\bibfield  {journal} {\bibinfo
			{journal} {Sci. Rep.}\ }\textbf {\bibinfo {volume} {6}},\ \bibinfo {pages}
		{22327} (\bibinfo {year} {2016})}\BibitemShut {NoStop}%
	\bibitem [{\citenamefont {Pavlenko}\ \emph {et~al.}(2012)\citenamefont
		{Pavlenko}, \citenamefont {Kozakov}, \citenamefont {Kubrin}, \citenamefont
		{Pavelko}, \citenamefont {Guglev}, \citenamefont {Shilkina}, \citenamefont
		{Verbenko}, \citenamefont {Sarichev},\ and\ \citenamefont
		{Reznichenko}}]{Pavlenko12:38}%
	\BibitemOpen
	\bibfield  {author} {\bibinfo {author} {\bibfnamefont {A.~V.}\ \bibnamefont
			{Pavlenko}}, \bibinfo {author} {\bibfnamefont {A.~T.}\ \bibnamefont
			{Kozakov}}, \bibinfo {author} {\bibfnamefont {S.~P.}\ \bibnamefont {Kubrin}},
		\bibinfo {author} {\bibfnamefont {A.~A.}\ \bibnamefont {Pavelko}}, \bibinfo
		{author} {\bibfnamefont {K.~A.}\ \bibnamefont {Guglev}}, \bibinfo {author}
		{\bibfnamefont {L.~A.}\ \bibnamefont {Shilkina}}, \bibinfo {author}
		{\bibfnamefont {I.~A.}\ \bibnamefont {Verbenko}}, \bibinfo {author}
		{\bibfnamefont {D.~A.}\ \bibnamefont {Sarichev}},\ and\ \bibinfo {author}
		{\bibfnamefont {L.~A.}\ \bibnamefont {Reznichenko}},\ }\href
	{https://doi.org/10.1016/j.ceramint.2012.04.066} {\bibfield  {journal}
		{\bibinfo  {journal} {Ceram. Int.}\ }\textbf {\bibinfo {volume} {38}},\
		\bibinfo {pages} {6157} (\bibinfo {year} {2012})}\BibitemShut {NoStop}%
	\bibitem [{\citenamefont {Raevski}\ \emph {et~al.}(2019)\citenamefont
		{Raevski}, \citenamefont {Kubrin}, \citenamefont {Pushkarev}, \citenamefont
		{Olekhnovich}, \citenamefont {Radyush}, \citenamefont {Li}, \citenamefont
		{Chou}, \citenamefont {Raevskaya}, \citenamefont {Titov},\ and\ \citenamefont
		{Malitskaya}}]{Raevski19:542}%
	\BibitemOpen
	\bibfield  {author} {\bibinfo {author} {\bibfnamefont {I.~P.}\ \bibnamefont
			{Raevski}}, \bibinfo {author} {\bibfnamefont {S.~P.}\ \bibnamefont {Kubrin}},
		\bibinfo {author} {\bibfnamefont {A.~V.}\ \bibnamefont {Pushkarev}}, \bibinfo
		{author} {\bibfnamefont {N.~M.}\ \bibnamefont {Olekhnovich}}, \bibinfo
		{author} {\bibfnamefont {Y.~V.}\ \bibnamefont {Radyush}}, \bibinfo {author}
		{\bibfnamefont {G.~R.}\ \bibnamefont {Li}}, \bibinfo {author} {\bibfnamefont
			{C.~C.}\ \bibnamefont {Chou}}, \bibinfo {author} {\bibfnamefont {S.~I.}\
			\bibnamefont {Raevskaya}}, \bibinfo {author} {\bibfnamefont {V.~V.}\
			\bibnamefont {Titov}},\ and\ \bibinfo {author} {\bibfnamefont {M.~A.}\
			\bibnamefont {Malitskaya}},\ }\href
	{https://doi.org/10.1080/00150193.2019.1574660} {\bibfield  {journal}
		{\bibinfo  {journal} {Ferroelectrics}\ }\textbf {\bibinfo {volume} {542}},\
		\bibinfo {pages} {36} (\bibinfo {year} {2019})}\BibitemShut {NoStop}%
	\bibitem [{\citenamefont {Pietrzak}\ \emph {et~al.}(1981)\citenamefont
		{Pietrzak}, \citenamefont {Maryanowska},\ and\ \citenamefont
		{Leciejewicz}}]{Pietrzak81:65}%
	\BibitemOpen
	\bibfield  {author} {\bibinfo {author} {\bibfnamefont {J.}~\bibnamefont
			{Pietrzak}}, \bibinfo {author} {\bibfnamefont {A.}~\bibnamefont
			{Maryanowska}},\ and\ \bibinfo {author} {\bibfnamefont {J.}~\bibnamefont
			{Leciejewicz}},\ }\href {https://doi.org/10.1002/pssa.2210650164} {\bibfield
		{journal} {\bibinfo  {journal} {Phys. Stat. Sol.}\ }\textbf {\bibinfo
			{volume} {65}},\ \bibinfo {pages} {K79} (\bibinfo {year} {1981})}\BibitemShut
	{NoStop}%
	\bibitem [{\citenamefont {Kleemann}\ \emph {et~al.}(2010)\citenamefont
		{Kleemann}, \citenamefont {Shvartsman}, \citenamefont {Borisov},\ and\
		\citenamefont {Kania}}]{Kleeman10:105}%
	\BibitemOpen
	\bibfield  {author} {\bibinfo {author} {\bibfnamefont {W.}~\bibnamefont
			{Kleemann}}, \bibinfo {author} {\bibfnamefont {V.~V.}\ \bibnamefont
			{Shvartsman}}, \bibinfo {author} {\bibfnamefont {P.}~\bibnamefont
			{Borisov}},\ and\ \bibinfo {author} {\bibfnamefont {A.}~\bibnamefont
			{Kania}},\ }\href {https://doi.org/10.1103/PhysRevLett.105.257202} {\bibfield
		{journal} {\bibinfo  {journal} {Phys. Rev. Lett.}\ }\textbf {\bibinfo
			{volume} {105}},\ \bibinfo {pages} {257202} (\bibinfo {year}
		{2010})}\BibitemShut {NoStop}%
	\bibitem [{\citenamefont {Luo}\ \emph {et~al.}(2000)\citenamefont {Luo},
		\citenamefont {Xu}, \citenamefont {Xu}, \citenamefont {Wang},\ and\
		\citenamefont {Yin}}]{Luo00:39}%
	\BibitemOpen
	\bibfield  {author} {\bibinfo {author} {\bibfnamefont {H.}~\bibnamefont
			{Luo}}, \bibinfo {author} {\bibfnamefont {G.}~\bibnamefont {Xu}}, \bibinfo
		{author} {\bibfnamefont {H.}~\bibnamefont {Xu}}, \bibinfo {author}
		{\bibfnamefont {P.}~\bibnamefont {Wang}},\ and\ \bibinfo {author}
		{\bibfnamefont {Z.}~\bibnamefont {Yin}},\ }\href
	{https://doi.org/10.1143/JJAP.39.5581} {\bibfield  {journal} {\bibinfo
			{journal} {Jpn. J. Appl. Phys.}\ }\textbf {\bibinfo {volume} {39}},\ \bibinfo
		{pages} {5581} (\bibinfo {year} {2000})}\BibitemShut {NoStop}%
	\bibitem [{\citenamefont {Kozlenko}\ \emph {et~al.}(2014)\citenamefont
		{Kozlenko}, \citenamefont {Kichanov}, \citenamefont {Lukin}, \citenamefont
		{Dang}, \citenamefont {Dubrovinsky}, \citenamefont {Liermann}, \citenamefont
		{Morgenroth}, \citenamefont {Kamynin}, \citenamefont {Gridnev},\ and\
		\citenamefont {Savenko}}]{Kozlenko14:89}%
	\BibitemOpen
	\bibfield  {author} {\bibinfo {author} {\bibfnamefont {D.~P.}\ \bibnamefont
			{Kozlenko}}, \bibinfo {author} {\bibfnamefont {S.~E.}\ \bibnamefont
			{Kichanov}}, \bibinfo {author} {\bibfnamefont {E.~V.}\ \bibnamefont {Lukin}},
		\bibinfo {author} {\bibfnamefont {N.~T.}\ \bibnamefont {Dang}}, \bibinfo
		{author} {\bibfnamefont {L.~S.}\ \bibnamefont {Dubrovinsky}}, \bibinfo
		{author} {\bibfnamefont {H.-P.}\ \bibnamefont {Liermann}}, \bibinfo {author}
		{\bibfnamefont {W.}~\bibnamefont {Morgenroth}}, \bibinfo {author}
		{\bibfnamefont {A.~A.}\ \bibnamefont {Kamynin}}, \bibinfo {author}
		{\bibfnamefont {S.~A.}\ \bibnamefont {Gridnev}},\ and\ \bibinfo {author}
		{\bibfnamefont {B.~N.}\ \bibnamefont {Savenko}},\ }\href
	{https://doi.org/10.1103/PhysRevB.89.174107} {\bibfield  {journal} {\bibinfo
			{journal} {Phys. Rev. B}\ }\textbf {\bibinfo {volume} {89}},\ \bibinfo
		{pages} {174107} (\bibinfo {year} {2014})}\BibitemShut {NoStop}%
	\bibitem [{\citenamefont {Stock}\ \emph {et~al.}(2013)\citenamefont {Stock},
		\citenamefont {Dunsiger}, \citenamefont {Mole}, \citenamefont {Li},\ and\
		\citenamefont {Luo}}]{Stock13:88}%
	\BibitemOpen
	\bibfield  {author} {\bibinfo {author} {\bibfnamefont {C.}~\bibnamefont
			{Stock}}, \bibinfo {author} {\bibfnamefont {S.~R.}\ \bibnamefont {Dunsiger}},
		\bibinfo {author} {\bibfnamefont {R.~A.}\ \bibnamefont {Mole}}, \bibinfo
		{author} {\bibfnamefont {X.}~\bibnamefont {Li}},\ and\ \bibinfo {author}
		{\bibfnamefont {H.}~\bibnamefont {Luo}},\ }\href
	{https://doi.org/10.1103/PhysRevB.88.094105} {\bibfield  {journal} {\bibinfo
			{journal} {Phys. Rev. B}\ }\textbf {\bibinfo {volume} {88}},\ \bibinfo
		{pages} {094105} (\bibinfo {year} {2013})}\BibitemShut {NoStop}%
	\bibitem [{\citenamefont {Wilfong}\ \emph
		{et~al.}(2016{\natexlab{a}})\citenamefont {Wilfong}, \citenamefont {Ahart},
		\citenamefont {Gramsch}, \citenamefont {Stock}, \citenamefont {Li},
		\citenamefont {Luo},\ and\ \citenamefont {Hemley}}]{Wilfong16:47}%
	\BibitemOpen
	\bibfield  {author} {\bibinfo {author} {\bibfnamefont {B.}~\bibnamefont
			{Wilfong}}, \bibinfo {author} {\bibfnamefont {M.}~\bibnamefont {Ahart}},
		\bibinfo {author} {\bibfnamefont {S.~A.}\ \bibnamefont {Gramsch}}, \bibinfo
		{author} {\bibfnamefont {C.}~\bibnamefont {Stock}}, \bibinfo {author}
		{\bibfnamefont {Z.}~\bibnamefont {Li}}, \bibinfo {author} {\bibfnamefont
			{H.}~\bibnamefont {Luo}},\ and\ \bibinfo {author} {\bibfnamefont {R.~J.}\
			\bibnamefont {Hemley}},\ }\href {https://doi.org/10.1002/jrs.4779} {\bibfield
		{journal} {\bibinfo  {journal} {J. Raman Spectrosc.}\ }\textbf {\bibinfo
			{volume} {47}},\ \bibinfo {pages} {227} (\bibinfo {year}
		{2016}{\natexlab{a}})}\BibitemShut {NoStop}%
	\bibitem [{\citenamefont {Bonny}\ \emph {et~al.}(1997)\citenamefont {Bonny},
		\citenamefont {Bonin}, \citenamefont {Sciau}, \citenamefont {Schenk},\ and\
		\citenamefont {Chapuis}}]{Bonny97:102}%
	\BibitemOpen
	\bibfield  {author} {\bibinfo {author} {\bibfnamefont {V.}~\bibnamefont
			{Bonny}}, \bibinfo {author} {\bibfnamefont {M.}~\bibnamefont {Bonin}},
		\bibinfo {author} {\bibfnamefont {P.}~\bibnamefont {Sciau}}, \bibinfo
		{author} {\bibfnamefont {K.~J.}\ \bibnamefont {Schenk}},\ and\ \bibinfo
		{author} {\bibfnamefont {G.}~\bibnamefont {Chapuis}},\ }\href
	{https://doi.org/10.1016/S0038-1098(97)00022-7} {\bibfield  {journal}
		{\bibinfo  {journal} {Solid State Commun.}\ }\textbf {\bibinfo {volume}
			{102}},\ \bibinfo {pages} {347} (\bibinfo {year} {1997})}\BibitemShut
	{NoStop}%
	\bibitem [{\citenamefont {Zagorodniy}\ \emph {et~al.}(2018)\citenamefont
		{Zagorodniy}, \citenamefont {Kuzian}, \citenamefont {Kondakova},
		\citenamefont {Mary\ifmmode~\check{s}\else \v{s}\fi{}ko}, \citenamefont
		{Chlan}, \citenamefont {\ifmmode \check{S}\else
			\v{S}\fi{}t\ifmmode~\check{e}\else \v{e}\fi{}p\'ankov\'a}, \citenamefont
		{Olekhnovich}, \citenamefont {Pushkarev}, \citenamefont {Radyush},
		\citenamefont {Raevski}, \citenamefont {Zalar}, \citenamefont {Laguta},\ and\
		\citenamefont {Stephanovich}}]{Zakorodniy18:2}%
	\BibitemOpen
	\bibfield  {author} {\bibinfo {author} {\bibfnamefont {Y.~O.}\ \bibnamefont
			{Zagorodniy}}, \bibinfo {author} {\bibfnamefont {R.~O.}\ \bibnamefont
			{Kuzian}}, \bibinfo {author} {\bibfnamefont {I.~V.}\ \bibnamefont
			{Kondakova}}, \bibinfo {author} {\bibfnamefont {M.}~\bibnamefont
			{Mary\ifmmode~\check{s}\else \v{s}\fi{}ko}}, \bibinfo {author} {\bibfnamefont
			{V.}~\bibnamefont {Chlan}}, \bibinfo {author} {\bibfnamefont
			{H.}~\bibnamefont {\ifmmode \check{S}\else \v{S}\fi{}t\ifmmode~\check{e}\else
				\v{e}\fi{}p\'ankov\'a}}, \bibinfo {author} {\bibfnamefont {N.~M.}\
			\bibnamefont {Olekhnovich}}, \bibinfo {author} {\bibfnamefont {A.~V.}\
			\bibnamefont {Pushkarev}}, \bibinfo {author} {\bibfnamefont {Y.~V.}\
			\bibnamefont {Radyush}}, \bibinfo {author} {\bibfnamefont {I.~P.}\
			\bibnamefont {Raevski}}, \bibinfo {author} {\bibfnamefont {B.}~\bibnamefont
			{Zalar}}, \bibinfo {author} {\bibfnamefont {V.~V.}\ \bibnamefont {Laguta}},\
		and\ \bibinfo {author} {\bibfnamefont {V.~A.}\ \bibnamefont {Stephanovich}},\
	}\href {https://doi.org/10.1103/PhysRevMaterials.2.014401} {\bibfield
		{journal} {\bibinfo  {journal} {Phys. Rev. Materials}\ }\textbf {\bibinfo
			{volume} {2}},\ \bibinfo {pages} {014401} (\bibinfo {year}
		{2018})}\BibitemShut {NoStop}%
	\bibitem [{\citenamefont {Bochenek}\ and\ \citenamefont
		{Niemiec}(2018)}]{Bochenek18:11}%
	\BibitemOpen
	\bibfield  {author} {\bibinfo {author} {\bibfnamefont {D.}~\bibnamefont
			{Bochenek}}\ and\ \bibinfo {author} {\bibfnamefont {P.}~\bibnamefont
			{Niemiec}},\ }\href {https://doi.org/10.3390/ma11122504} {\bibfield
		{journal} {\bibinfo  {journal} {Materials}\ }\textbf {\bibinfo {volume}
			{11}},\ \bibinfo {pages} {2504} (\bibinfo {year} {2018})}\BibitemShut
	{NoStop}%
	\bibitem [{\citenamefont {Hiraka}\ \emph {et~al.}(2004)\citenamefont {Hiraka},
		\citenamefont {Lee}, \citenamefont {Gehring}, \citenamefont {Xu},\ and\
		\citenamefont {Shirane}}]{Hiraka04:70}%
	\BibitemOpen
	\bibfield  {author} {\bibinfo {author} {\bibfnamefont {H.}~\bibnamefont
			{Hiraka}}, \bibinfo {author} {\bibfnamefont {S.-H.}\ \bibnamefont {Lee}},
		\bibinfo {author} {\bibfnamefont {P.~M.}\ \bibnamefont {Gehring}}, \bibinfo
		{author} {\bibfnamefont {G.}~\bibnamefont {Xu}},\ and\ \bibinfo {author}
		{\bibfnamefont {G.}~\bibnamefont {Shirane}},\ }\href
	{https://doi.org/10.1103/PhysRevB.70.184105} {\bibfield  {journal} {\bibinfo
			{journal} {Phys. Rev. B}\ }\textbf {\bibinfo {volume} {70}},\ \bibinfo
		{pages} {184105} (\bibinfo {year} {2004})}\BibitemShut {NoStop}%
	\bibitem [{\citenamefont {Zinkin}\ \emph {et~al.}(1997)\citenamefont {Zinkin},
		\citenamefont {Harris},\ and\ \citenamefont {Zeiske}}]{Zinkin97:56}%
	\BibitemOpen
	\bibfield  {author} {\bibinfo {author} {\bibfnamefont {M.~P.}\ \bibnamefont
			{Zinkin}}, \bibinfo {author} {\bibfnamefont {M.~J.}\ \bibnamefont {Harris}},\
		and\ \bibinfo {author} {\bibfnamefont {T.}~\bibnamefont {Zeiske}},\ }\href
	{https://doi.org/10.1103/PhysRevB.56.11786} {\bibfield  {journal} {\bibinfo
			{journal} {Phys. Rev. B}\ }\textbf {\bibinfo {volume} {56}},\ \bibinfo
		{pages} {11786} (\bibinfo {year} {1997})}\BibitemShut {NoStop}%
	\bibitem [{\citenamefont {Schweika}\ and\ \citenamefont
		{Boni}(2001)}]{Schwieka01:297}%
	\BibitemOpen
	\bibfield  {author} {\bibinfo {author} {\bibfnamefont {W.}~\bibnamefont
			{Schweika}}\ and\ \bibinfo {author} {\bibfnamefont {P.}~\bibnamefont
			{Boni}},\ }\href {https://doi.org/10.1016/S0921-4526(00)00858-9} {\bibfield
		{journal} {\bibinfo  {journal} {Physica B}\ }\textbf {\bibinfo {volume}
			{297}},\ \bibinfo {pages} {155} (\bibinfo {year} {2001})}\BibitemShut
	{NoStop}%
	\bibitem [{\citenamefont {Stewart}\ \emph {et~al.}(2009)\citenamefont
		{Stewart}, \citenamefont {Deen}, \citenamefont {Andersen}, \citenamefont
		{Schober}, \citenamefont {Barthelemy}, \citenamefont {Hillier}, \citenamefont
		{Murani}, \citenamefont {Hayes},\ and\ \citenamefont
		{Lindenau}}]{Stewart09:42}%
	\BibitemOpen
	\bibfield  {author} {\bibinfo {author} {\bibfnamefont {J.~R.}\ \bibnamefont
			{Stewart}}, \bibinfo {author} {\bibfnamefont {P.~P.}\ \bibnamefont {Deen}},
		\bibinfo {author} {\bibfnamefont {K.~H.}\ \bibnamefont {Andersen}}, \bibinfo
		{author} {\bibfnamefont {H.}~\bibnamefont {Schober}}, \bibinfo {author}
		{\bibfnamefont {J.~F.}\ \bibnamefont {Barthelemy}}, \bibinfo {author}
		{\bibfnamefont {J.~M.}\ \bibnamefont {Hillier}}, \bibinfo {author}
		{\bibfnamefont {A.~P.}\ \bibnamefont {Murani}}, \bibinfo {author}
		{\bibfnamefont {T.}~\bibnamefont {Hayes}},\ and\ \bibinfo {author}
		{\bibfnamefont {B.}~\bibnamefont {Lindenau}},\ }\href
	{https://doi.org/10.1107/S0021889808039162} {\bibfield  {journal} {\bibinfo
			{journal} {J. Appl. Cryst.}\ }\textbf {\bibinfo {volume} {42}},\ \bibinfo
		{pages} {69} (\bibinfo {year} {2009})}\BibitemShut {NoStop}%
	\bibitem [{\citenamefont {Wilfong}\ \emph
		{et~al.}(2016{\natexlab{b}})\citenamefont {Wilfong}, \citenamefont {Ahart},
		\citenamefont {Gramsch}, \citenamefont {Stock}, \citenamefont {Li},
		\citenamefont {Luo},\ and\ \citenamefont {Hemley}}]{Wilfong15:70}%
	\BibitemOpen
	\bibfield  {author} {\bibinfo {author} {\bibfnamefont {B.}~\bibnamefont
			{Wilfong}}, \bibinfo {author} {\bibfnamefont {M.}~\bibnamefont {Ahart}},
		\bibinfo {author} {\bibfnamefont {S.~A.}\ \bibnamefont {Gramsch}}, \bibinfo
		{author} {\bibfnamefont {C.}~\bibnamefont {Stock}}, \bibinfo {author}
		{\bibfnamefont {X.}~\bibnamefont {Li}}, \bibinfo {author} {\bibfnamefont
			{H.}~\bibnamefont {Luo}},\ and\ \bibinfo {author} {\bibfnamefont {R.~J.}\
			\bibnamefont {Hemley}},\ }\href {https://doi.org/10.1002/jrs.4779} {\bibfield
		{journal} {\bibinfo  {journal} {J. Raman Spectrosc.}\ }\textbf {\bibinfo
			{volume} {47}},\ \bibinfo {pages} {227} (\bibinfo {year}
		{2016}{\natexlab{b}})}\BibitemShut {NoStop}%
	\bibitem [{\citenamefont {Swainson}\ \emph {et~al.}(2009)\citenamefont
		{Swainson}, \citenamefont {Stock}, \citenamefont {Gehring}, \citenamefont
		{Xu}, \citenamefont {Hirota}, \citenamefont {Qiu}, \citenamefont {Luo},
		\citenamefont {Zhao}, \citenamefont {Li},\ and\ \citenamefont
		{Viehland}}]{Swainson09:79}%
	\BibitemOpen
	\bibfield  {author} {\bibinfo {author} {\bibfnamefont {I.~P.}\ \bibnamefont
			{Swainson}}, \bibinfo {author} {\bibfnamefont {C.}~\bibnamefont {Stock}},
		\bibinfo {author} {\bibfnamefont {P.~M.}\ \bibnamefont {Gehring}}, \bibinfo
		{author} {\bibfnamefont {G.}~\bibnamefont {Xu}}, \bibinfo {author}
		{\bibfnamefont {K.}~\bibnamefont {Hirota}}, \bibinfo {author} {\bibfnamefont
			{Y.}~\bibnamefont {Qiu}}, \bibinfo {author} {\bibfnamefont {H.}~\bibnamefont
			{Luo}}, \bibinfo {author} {\bibfnamefont {X.}~\bibnamefont {Zhao}}, \bibinfo
		{author} {\bibfnamefont {J.-F.}\ \bibnamefont {Li}},\ and\ \bibinfo {author}
		{\bibfnamefont {D.}~\bibnamefont {Viehland}},\ }\href
	{https://doi.org/10.1103/PhysRevB.79.224301} {\bibfield  {journal} {\bibinfo
			{journal} {Phys. Rev. B}\ }\textbf {\bibinfo {volume} {79}},\ \bibinfo
		{pages} {224301} (\bibinfo {year} {2009})}\BibitemShut {NoStop}%
	\bibitem [{\citenamefont {Lampis}\ \emph {et~al.}(1999)\citenamefont {Lampis},
		\citenamefont {Sciau},\ and\ \citenamefont {Lehmann}}]{Lampis99:11}%
	\BibitemOpen
	\bibfield  {author} {\bibinfo {author} {\bibfnamefont {N.}~\bibnamefont
			{Lampis}}, \bibinfo {author} {\bibfnamefont {P.}~\bibnamefont {Sciau}},\ and\
		\bibinfo {author} {\bibfnamefont {A.~G.}\ \bibnamefont {Lehmann}},\ }\href
	{https://doi.org/10.1088/0953-8984/11/17/307} {\bibfield  {journal} {\bibinfo
			{journal} {J. Phys.: Condens. Matter}\ }\textbf {\bibinfo {volume} {11}},\
		\bibinfo {pages} {3489} (\bibinfo {year} {1999})}\BibitemShut {NoStop}%
	\bibitem [{\citenamefont {Majumder}\ \emph {et~al.}(2006)\citenamefont
		{Majumder}, \citenamefont {Bhattacharyya}, \citenamefont {Katiyar},
		\citenamefont {Mannivannan}, \citenamefont {Dutta},\ and\ \citenamefont
		{Seehra}}]{Maj06:99}%
	\BibitemOpen
	\bibfield  {author} {\bibinfo {author} {\bibfnamefont {S.~B.}\ \bibnamefont
			{Majumder}}, \bibinfo {author} {\bibfnamefont {S.}~\bibnamefont
			{Bhattacharyya}}, \bibinfo {author} {\bibfnamefont {R.~S.}\ \bibnamefont
			{Katiyar}}, \bibinfo {author} {\bibfnamefont {A.}~\bibnamefont
			{Mannivannan}}, \bibinfo {author} {\bibfnamefont {P.}~\bibnamefont {Dutta}},\
		and\ \bibinfo {author} {\bibfnamefont {M.~S.}\ \bibnamefont {Seehra}},\
	}\href {https://doi.org/10.1063/1.2158131} {\bibfield  {journal} {\bibinfo
			{journal} {J. Appl. Phys.}\ }\textbf {\bibinfo {volume} {99}},\ \bibinfo
		{pages} {024108} (\bibinfo {year} {2006})}\BibitemShut {NoStop}%
	\bibitem [{\citenamefont {Xu}\ \emph {et~al.}(2004{\natexlab{b}})\citenamefont
		{Xu}, \citenamefont {Zhong}, \citenamefont {Hiraka},\ and\ \citenamefont
		{Shirane}}]{Xu04:70}%
	\BibitemOpen
	\bibfield  {author} {\bibinfo {author} {\bibfnamefont {G.}~\bibnamefont
			{Xu}}, \bibinfo {author} {\bibfnamefont {Z.}~\bibnamefont {Zhong}}, \bibinfo
		{author} {\bibfnamefont {H.}~\bibnamefont {Hiraka}},\ and\ \bibinfo {author}
		{\bibfnamefont {G.}~\bibnamefont {Shirane}},\ }\href
	{https://doi.org/10.1103/PhysRevB.70.174109} {\bibfield  {journal} {\bibinfo
			{journal} {Phys. Rev. B}\ }\textbf {\bibinfo {volume} {70}},\ \bibinfo
		{pages} {174109} (\bibinfo {year} {2004}{\natexlab{b}})}\BibitemShut
	{NoStop}%
	\bibitem [{\citenamefont {Stock}\ \emph {et~al.}(2018)\citenamefont {Stock},
		\citenamefont {Gehring}, \citenamefont {Ewings}, \citenamefont {Xu},
		\citenamefont {Li}, \citenamefont {Viehland},\ and\ \citenamefont
		{Luo}}]{Stock18:2}%
	\BibitemOpen
	\bibfield  {author} {\bibinfo {author} {\bibfnamefont {C.}~\bibnamefont
			{Stock}}, \bibinfo {author} {\bibfnamefont {P.~M.}\ \bibnamefont {Gehring}},
		\bibinfo {author} {\bibfnamefont {R.~A.}\ \bibnamefont {Ewings}}, \bibinfo
		{author} {\bibfnamefont {G.}~\bibnamefont {Xu}}, \bibinfo {author}
		{\bibfnamefont {J.}~\bibnamefont {Li}}, \bibinfo {author} {\bibfnamefont
			{D.}~\bibnamefont {Viehland}},\ and\ \bibinfo {author} {\bibfnamefont
			{H.}~\bibnamefont {Luo}},\ }\href
	{https://doi.org/10.1103/PhysRevMaterials.2.024404} {\bibfield  {journal}
		{\bibinfo  {journal} {Phys. Rev. Materials}\ }\textbf {\bibinfo {volume}
			{2}},\ \bibinfo {pages} {024404} (\bibinfo {year} {2018})}\BibitemShut
	{NoStop}%
	\bibitem [{\citenamefont {Stock}\ \emph {et~al.}(2012)\citenamefont {Stock},
		\citenamefont {Gehring}, \citenamefont {Hiraka}, \citenamefont {Swainson},
		\citenamefont {Xu}, \citenamefont {Ye}, \citenamefont {Luo}, \citenamefont
		{Li},\ and\ \citenamefont {Viehland}}]{Stock12:86}%
	\BibitemOpen
	\bibfield  {author} {\bibinfo {author} {\bibfnamefont {C.}~\bibnamefont
			{Stock}}, \bibinfo {author} {\bibfnamefont {P.~M.}\ \bibnamefont {Gehring}},
		\bibinfo {author} {\bibfnamefont {H.}~\bibnamefont {Hiraka}}, \bibinfo
		{author} {\bibfnamefont {I.}~\bibnamefont {Swainson}}, \bibinfo {author}
		{\bibfnamefont {G.}~\bibnamefont {Xu}}, \bibinfo {author} {\bibfnamefont
			{Z.-G.}\ \bibnamefont {Ye}}, \bibinfo {author} {\bibfnamefont
			{H.}~\bibnamefont {Luo}}, \bibinfo {author} {\bibfnamefont {J.-F.}\
			\bibnamefont {Li}},\ and\ \bibinfo {author} {\bibfnamefont {D.}~\bibnamefont
			{Viehland}},\ }\href {https://doi.org/10.1103/PhysRevB.86.104108} {\bibfield
		{journal} {\bibinfo  {journal} {Phys. Rev. B}\ }\textbf {\bibinfo {volume}
			{86}},\ \bibinfo {pages} {104108} (\bibinfo {year} {2012})}\BibitemShut
	{NoStop}%
	\bibitem [{\citenamefont {Stock}\ \emph {et~al.}(2005)\citenamefont {Stock},
		\citenamefont {Luo}, \citenamefont {Viehland}, \citenamefont {Li},
		\citenamefont {Swainson}, \citenamefont {Birgeneau},\ and\ \citenamefont
		{Shirane}}]{Stock05:74}%
	\BibitemOpen
	\bibfield  {author} {\bibinfo {author} {\bibfnamefont {C.}~\bibnamefont
			{Stock}}, \bibinfo {author} {\bibfnamefont {H.}~\bibnamefont {Luo}}, \bibinfo
		{author} {\bibfnamefont {D.}~\bibnamefont {Viehland}}, \bibinfo {author}
		{\bibfnamefont {J.~F.}\ \bibnamefont {Li}}, \bibinfo {author} {\bibfnamefont
			{I.~P.}\ \bibnamefont {Swainson}}, \bibinfo {author} {\bibfnamefont {R.~J.}\
			\bibnamefont {Birgeneau}},\ and\ \bibinfo {author} {\bibfnamefont
			{G.}~\bibnamefont {Shirane}},\ }\href {https://doi.org/10.1143/JPSJ.74.3002}
	{\bibfield  {journal} {\bibinfo  {journal} {J. Phys. Soc. Jpn.}\ }\textbf
		{\bibinfo {volume} {74}},\ \bibinfo {pages} {3002} (\bibinfo {year}
		{2005})}\BibitemShut {NoStop}%
	\bibitem [{\citenamefont {Gvasaliay}\ \emph {et~al.}(2005)\citenamefont
		{Gvasaliay}, \citenamefont {Roessli}, \citenamefont {Cowley}, \citenamefont
		{Huber},\ and\ \citenamefont {Lusnikov}}]{Gvas05:17}%
	\BibitemOpen
	\bibfield  {author} {\bibinfo {author} {\bibfnamefont {S.~N.}\ \bibnamefont
			{Gvasaliay}}, \bibinfo {author} {\bibfnamefont {B.}~\bibnamefont {Roessli}},
		\bibinfo {author} {\bibfnamefont {R.~A.}\ \bibnamefont {Cowley}}, \bibinfo
		{author} {\bibfnamefont {P.}~\bibnamefont {Huber}},\ and\ \bibinfo {author}
		{\bibfnamefont {S.~G.}\ \bibnamefont {Lusnikov}},\ }\href
	{https://doi.org/10.1088/0953-8984/17/27/010} {\bibfield  {journal} {\bibinfo
			{journal} {J. Phys. Condens. Matter}\ }\textbf {\bibinfo {volume} {17}},\
		\bibinfo {pages} {4343} (\bibinfo {year} {2005})}\BibitemShut {NoStop}%
	\bibitem [{\citenamefont {Gvasaliya}\ \emph
		{et~al.}(2004{\natexlab{a}})\citenamefont {Gvasaliya}, \citenamefont
		{Lushnikov},\ and\ \citenamefont {Roessli}}]{Gvas04:69}%
	\BibitemOpen
	\bibfield  {author} {\bibinfo {author} {\bibfnamefont {S.~N.}\ \bibnamefont
			{Gvasaliya}}, \bibinfo {author} {\bibfnamefont {S.~G.}\ \bibnamefont
			{Lushnikov}},\ and\ \bibinfo {author} {\bibfnamefont {B.}~\bibnamefont
			{Roessli}},\ }\href {https://doi.org/10.1103/PhysRevB.69.092105} {\bibfield
		{journal} {\bibinfo  {journal} {Phys. Rev. B}\ }\textbf {\bibinfo {volume}
			{69}},\ \bibinfo {pages} {092105} (\bibinfo {year}
		{2004}{\natexlab{a}})}\BibitemShut {NoStop}%
	\bibitem [{\citenamefont {Gvasaliya}\ \emph
		{et~al.}(2004{\natexlab{b}})\citenamefont {Gvasaliya}, \citenamefont
		{Lushnikov},\ and\ \citenamefont {Roessli}}]{Gvas04:49}%
	\BibitemOpen
	\bibfield  {author} {\bibinfo {author} {\bibfnamefont {S.~N.}\ \bibnamefont
			{Gvasaliya}}, \bibinfo {author} {\bibfnamefont {S.~G.}\ \bibnamefont
			{Lushnikov}},\ and\ \bibinfo {author} {\bibfnamefont {B.}~\bibnamefont
			{Roessli}},\ }\href {https://doi.org/10.1134/1.1643970} {\bibfield  {journal}
		{\bibinfo  {journal} {Crystallogr. Rep.}\ }\textbf {\bibinfo {volume} {49}},\
		\bibinfo {pages} {108} (\bibinfo {year} {2004}{\natexlab{b}})}\BibitemShut
	{NoStop}%
	\bibitem [{\citenamefont {Murani}\ and\ \citenamefont
		{Heidemann}(1978)}]{Murani78:41}%
	\BibitemOpen
	\bibfield  {author} {\bibinfo {author} {\bibfnamefont {A.~P.}\ \bibnamefont
			{Murani}}\ and\ \bibinfo {author} {\bibfnamefont {A.}~\bibnamefont
			{Heidemann}},\ }\href {https://doi.org/10.1103/PhysRevLett.41.1402}
	{\bibfield  {journal} {\bibinfo  {journal} {Phys. Rev. Lett.}\ }\textbf
		{\bibinfo {volume} {41}},\ \bibinfo {pages} {1402} (\bibinfo {year}
		{1978})}\BibitemShut {NoStop}%
	\bibitem [{\citenamefont {Welberry}\ \emph {et~al.}(2003)\citenamefont
		{Welberry}, \citenamefont {Goossens}, \citenamefont {David}, \citenamefont
		{Gutmann}, \citenamefont {Bull},\ and\ \citenamefont
		{Heerdegen}}]{Welberry03:36}%
	\BibitemOpen
	\bibfield  {author} {\bibinfo {author} {\bibfnamefont {T.~R.}\ \bibnamefont
			{Welberry}}, \bibinfo {author} {\bibfnamefont {D.~J.}\ \bibnamefont
			{Goossens}}, \bibinfo {author} {\bibfnamefont {W.~I.~F.}\ \bibnamefont
			{David}}, \bibinfo {author} {\bibfnamefont {M.~J.}\ \bibnamefont {Gutmann}},
		\bibinfo {author} {\bibfnamefont {M.~J.}\ \bibnamefont {Bull}},\ and\
		\bibinfo {author} {\bibfnamefont {A.~P.}\ \bibnamefont {Heerdegen}},\ }\href
	{https://doi.org/10.1107/S002188980302209X} {\bibfield  {journal} {\bibinfo
			{journal} {J. Appl. Cryst.}\ }\textbf {\bibinfo {volume} {36}},\ \bibinfo
		{pages} {1440} (\bibinfo {year} {2003})}\BibitemShut {NoStop}%
	\bibitem [{\citenamefont {Hohlwein}\ \emph {et~al.}(2003)\citenamefont
		{Hohlwein}, \citenamefont {Hoffmann},\ and\ \citenamefont
		{Schneider}}]{Hohlwein03:68}%
	\BibitemOpen
	\bibfield  {author} {\bibinfo {author} {\bibfnamefont {D.}~\bibnamefont
			{Hohlwein}}, \bibinfo {author} {\bibfnamefont {J.-U.}\ \bibnamefont
			{Hoffmann}},\ and\ \bibinfo {author} {\bibfnamefont {R.}~\bibnamefont
			{Schneider}},\ }\href {https://doi.org/10.1103/PhysRevB.68.140408} {\bibfield
		{journal} {\bibinfo  {journal} {Phys. Rev. B}\ }\textbf {\bibinfo {volume}
			{68}},\ \bibinfo {pages} {140408(R)} (\bibinfo {year} {2003})}\BibitemShut
	{NoStop}%
	\bibitem [{\citenamefont {Zeiske}\ \emph {et~al.}(1997)\citenamefont {Zeiske},
		\citenamefont {Hohlwein},\ and\ \citenamefont {Prandl}}]{Zeiske97:241}%
	\BibitemOpen
	\bibfield  {author} {\bibinfo {author} {\bibfnamefont {T.}~\bibnamefont
			{Zeiske}}, \bibinfo {author} {\bibfnamefont {D.}~\bibnamefont {Hohlwein}},\
		and\ \bibinfo {author} {\bibfnamefont {W.}~\bibnamefont {Prandl}},\ }\href
	{https://doi.org/S0921-4526(97)00664-9} {\bibfield  {journal} {\bibinfo
			{journal} {Physica B}\ }\textbf {\bibinfo {volume} {241}},\ \bibinfo {pages}
		{628} (\bibinfo {year} {1997})}\BibitemShut {NoStop}%
	\bibitem [{\citenamefont {Hohlwein}\ \emph {et~al.}(1997)\citenamefont
		{Hohlwein}, \citenamefont {Mayer},\ and\ \citenamefont
		{Zeiske}}]{Hohlwein97:234}%
	\BibitemOpen
	\bibfield  {author} {\bibinfo {author} {\bibfnamefont {D.}~\bibnamefont
			{Hohlwein}}, \bibinfo {author} {\bibfnamefont {H.~M.}\ \bibnamefont
			{Mayer}},\ and\ \bibinfo {author} {\bibfnamefont {T.}~\bibnamefont
			{Zeiske}},\ }\href {https://doi.org/10.1016/S0921-4526(97)89268-X} {\bibfield
		{journal} {\bibinfo  {journal} {Physica B}\ }\textbf {\bibinfo {volume}
			{234}},\ \bibinfo {pages} {1109} (\bibinfo {year} {1997})}\BibitemShut
	{NoStop}%
	\bibitem [{\citenamefont {Falqui}\ \emph {et~al.}(2005)\citenamefont {Falqui},
		\citenamefont {Lampis}, \citenamefont {Geddo-Lehmann},\ and\ \citenamefont
		{Pinna}}]{Falqui05:109}%
	\BibitemOpen
	\bibfield  {author} {\bibinfo {author} {\bibfnamefont {A.}~\bibnamefont
			{Falqui}}, \bibinfo {author} {\bibfnamefont {N.}~\bibnamefont {Lampis}},
		\bibinfo {author} {\bibfnamefont {A.}~\bibnamefont {Geddo-Lehmann}},\ and\
		\bibinfo {author} {\bibfnamefont {G.}~\bibnamefont {Pinna}},\ }\href
	{https://doi.org/10.1021/jp0551014} {\bibfield  {journal} {\bibinfo
			{journal} {J. Phys. Chem. B}\ }\textbf {\bibinfo {volume} {109}},\ \bibinfo
		{pages} {22967} (\bibinfo {year} {2005})}\BibitemShut {NoStop}%
	\bibitem [{\citenamefont {Kumar}\ \emph
		{et~al.}(2008{\natexlab{a}})\citenamefont {Kumar}, \citenamefont {Katiyar},
		\citenamefont {Rinaldi}, \citenamefont {Lushnikov},\ and\ \citenamefont
		{Shaplygina}}]{Kuma08:93}%
	\BibitemOpen
	\bibfield  {author} {\bibinfo {author} {\bibfnamefont {A.}~\bibnamefont
			{Kumar}}, \bibinfo {author} {\bibfnamefont {R.~S.}\ \bibnamefont {Katiyar}},
		\bibinfo {author} {\bibfnamefont {C.}~\bibnamefont {Rinaldi}}, \bibinfo
		{author} {\bibfnamefont {S.~G.}\ \bibnamefont {Lushnikov}},\ and\ \bibinfo
		{author} {\bibfnamefont {T.~A.}\ \bibnamefont {Shaplygina}},\ }\href
	{https://doi.org/10.1063/1.3043686} {\bibfield  {journal} {\bibinfo
			{journal} {Appl. Phys. Lett.}\ }\textbf {\bibinfo {volume} {93}},\ \bibinfo
		{pages} {232902} (\bibinfo {year} {2008}{\natexlab{a}})}\BibitemShut
	{NoStop}%
	\bibitem [{\citenamefont {Aeppli}\ \emph
		{et~al.}(1984{\natexlab{a}})\citenamefont {Aeppli}, \citenamefont {Shapiro},
		\citenamefont {Maletta}, \citenamefont {Birgeneau},\ and\ \citenamefont
		{Chen}}]{Aeppli84:55}%
	\BibitemOpen
	\bibfield  {author} {\bibinfo {author} {\bibfnamefont {G.}~\bibnamefont
			{Aeppli}}, \bibinfo {author} {\bibfnamefont {S.~M.}\ \bibnamefont {Shapiro}},
		\bibinfo {author} {\bibfnamefont {H.}~\bibnamefont {Maletta}}, \bibinfo
		{author} {\bibfnamefont {R.~J.}\ \bibnamefont {Birgeneau}},\ and\ \bibinfo
		{author} {\bibfnamefont {H.~S.}\ \bibnamefont {Chen}},\ }\href
	{https://doi.org/10.1063/1.333426} {\bibfield  {journal} {\bibinfo  {journal}
			{J. Appl. Phys.}\ }\textbf {\bibinfo {volume} {55}},\ \bibinfo {pages} {1628}
		(\bibinfo {year} {1984}{\natexlab{a}})}\BibitemShut {NoStop}%
	\bibitem [{\citenamefont {Chillal}\ \emph {et~al.}(2013)\citenamefont
		{Chillal}, \citenamefont {Thede}, \citenamefont {Litterst}, \citenamefont
		{Gvasaliya}, \citenamefont {Shaplygina}, \citenamefont {Lushnikov},\ and\
		\citenamefont {Zheludev}}]{Chillal13:87}%
	\BibitemOpen
	\bibfield  {author} {\bibinfo {author} {\bibfnamefont {S.}~\bibnamefont
			{Chillal}}, \bibinfo {author} {\bibfnamefont {M.}~\bibnamefont {Thede}},
		\bibinfo {author} {\bibfnamefont {F.~J.}\ \bibnamefont {Litterst}}, \bibinfo
		{author} {\bibfnamefont {S.~N.}\ \bibnamefont {Gvasaliya}}, \bibinfo {author}
		{\bibfnamefont {T.~A.}\ \bibnamefont {Shaplygina}}, \bibinfo {author}
		{\bibfnamefont {S.~G.}\ \bibnamefont {Lushnikov}},\ and\ \bibinfo {author}
		{\bibfnamefont {A.}~\bibnamefont {Zheludev}},\ }\href
	{https://doi.org/10.1103/PhysRevB.87.220403} {\bibfield  {journal} {\bibinfo
			{journal} {Phys. Rev. B}\ }\textbf {\bibinfo {volume} {87}},\ \bibinfo
		{pages} {220403(R)} (\bibinfo {year} {2013})}\BibitemShut {NoStop}%
	\bibitem [{\citenamefont {Nagata}\ \emph {et~al.}(1979)\citenamefont {Nagata},
		\citenamefont {Keesom},\ and\ \citenamefont {Harrison}}]{Nagata79:19}%
	\BibitemOpen
	\bibfield  {author} {\bibinfo {author} {\bibfnamefont {S.}~\bibnamefont
			{Nagata}}, \bibinfo {author} {\bibfnamefont {P.~H.}\ \bibnamefont {Keesom}},\
		and\ \bibinfo {author} {\bibfnamefont {H.~R.}\ \bibnamefont {Harrison}},\
	}\href {https://doi.org/10.1103/PhysRevB.19.1633} {\bibfield  {journal}
		{\bibinfo  {journal} {Phys. Rev. B}\ }\textbf {\bibinfo {volume} {19}},\
		\bibinfo {pages} {1633} (\bibinfo {year} {1979})}\BibitemShut {NoStop}%
	\bibitem [{\citenamefont {Sherington}(1975)}]{Sherrington75:8}%
	\BibitemOpen
	\bibfield  {author} {\bibinfo {author} {\bibfnamefont {D.}~\bibnamefont
			{Sherington}},\ }\href {https://doi.org/10.1088/0022-3719/8/10/021}
	{\bibfield  {journal} {\bibinfo  {journal} {J. Phys. C: Solid State Phys.}\
		}\textbf {\bibinfo {volume} {8}},\ \bibinfo {pages} {L208} (\bibinfo {year}
		{1975})}\BibitemShut {NoStop}%
	\bibitem [{\citenamefont {Yamani}\ \emph {et~al.}(2015)\citenamefont {Yamani},
		\citenamefont {Buyers}, \citenamefont {Wang}, \citenamefont {Kim},
		\citenamefont {Chung}, \citenamefont {Chang}, \citenamefont {Gehring},
		\citenamefont {Gasparovic}, \citenamefont {Stock}, \citenamefont {Broholm},
		\citenamefont {Baglo}, \citenamefont {Liang}, \citenamefont {Bonn},\ and\
		\citenamefont {Hardy}}]{Yamani15:91}%
	\BibitemOpen
	\bibfield  {author} {\bibinfo {author} {\bibfnamefont {Z.}~\bibnamefont
			{Yamani}}, \bibinfo {author} {\bibfnamefont {W.~J.~L.}\ \bibnamefont
			{Buyers}}, \bibinfo {author} {\bibfnamefont {F.}~\bibnamefont {Wang}},
		\bibinfo {author} {\bibfnamefont {Y.-J.}\ \bibnamefont {Kim}}, \bibinfo
		{author} {\bibfnamefont {J.-H.}\ \bibnamefont {Chung}}, \bibinfo {author}
		{\bibfnamefont {S.}~\bibnamefont {Chang}}, \bibinfo {author} {\bibfnamefont
			{P.~M.}\ \bibnamefont {Gehring}}, \bibinfo {author} {\bibfnamefont
			{G.}~\bibnamefont {Gasparovic}}, \bibinfo {author} {\bibfnamefont
			{C.}~\bibnamefont {Stock}}, \bibinfo {author} {\bibfnamefont {C.~L.}\
			\bibnamefont {Broholm}}, \bibinfo {author} {\bibfnamefont {J.~C.}\
			\bibnamefont {Baglo}}, \bibinfo {author} {\bibfnamefont {R.}~\bibnamefont
			{Liang}}, \bibinfo {author} {\bibfnamefont {D.~A.}\ \bibnamefont {Bonn}},\
		and\ \bibinfo {author} {\bibfnamefont {W.~N.}\ \bibnamefont {Hardy}},\ }\href
	{https://doi.org/10.1103/PhysRevB.91.134427} {\bibfield  {journal} {\bibinfo
			{journal} {Phys. Rev. B}\ }\textbf {\bibinfo {volume} {91}},\ \bibinfo
		{pages} {134427} (\bibinfo {year} {2015})}\BibitemShut {NoStop}%
	\bibitem [{\citenamefont {Stock}\ \emph {et~al.}(2008)\citenamefont {Stock},
		\citenamefont {Buyers}, \citenamefont {Yamani}, \citenamefont {Tun},
		\citenamefont {Birgeneau}, \citenamefont {Liang}, \citenamefont {Bonn},\ and\
		\citenamefont {Hardy}}]{Stock08:77}%
	\BibitemOpen
	\bibfield  {author} {\bibinfo {author} {\bibfnamefont {C.}~\bibnamefont
			{Stock}}, \bibinfo {author} {\bibfnamefont {W.~J.~L.}\ \bibnamefont
			{Buyers}}, \bibinfo {author} {\bibfnamefont {Z.}~\bibnamefont {Yamani}},
		\bibinfo {author} {\bibfnamefont {Z.}~\bibnamefont {Tun}}, \bibinfo {author}
		{\bibfnamefont {R.~J.}\ \bibnamefont {Birgeneau}}, \bibinfo {author}
		{\bibfnamefont {R.}~\bibnamefont {Liang}}, \bibinfo {author} {\bibfnamefont
			{D.}~\bibnamefont {Bonn}},\ and\ \bibinfo {author} {\bibfnamefont {W.~N.}\
			\bibnamefont {Hardy}},\ }\href {https://doi.org/10.1103/PhysRevB.77.104513}
	{\bibfield  {journal} {\bibinfo  {journal} {Phys. Rev. B}\ }\textbf {\bibinfo
			{volume} {77}},\ \bibinfo {pages} {104513} (\bibinfo {year}
		{2008})}\BibitemShut {NoStop}%
	\bibitem [{\citenamefont {Bernhoeft}(2001)}]{Bernhoeft01:13}%
	\BibitemOpen
	\bibfield  {author} {\bibinfo {author} {\bibfnamefont {N.}~\bibnamefont
			{Bernhoeft}},\ }\href {https://doi.org/10.1088/0953-8984/13/39/201}
	{\bibfield  {journal} {\bibinfo  {journal} {J. Phys.: Condens. Matter}\
		}\textbf {\bibinfo {volume} {13}},\ \bibinfo {pages} {R771} (\bibinfo {year}
		{2001})}\BibitemShut {NoStop}%
	\bibitem [{\citenamefont {Stock}\ \emph {et~al.}(2020)\citenamefont {Stock},
		\citenamefont {Songvilay}, \citenamefont {Gehring}, \citenamefont {Xu},\ and\
		\citenamefont {Roessli}}]{Stock20:32}%
	\BibitemOpen
	\bibfield  {author} {\bibinfo {author} {\bibfnamefont {C.}~\bibnamefont
			{Stock}}, \bibinfo {author} {\bibfnamefont {M.}~\bibnamefont {Songvilay}},
		\bibinfo {author} {\bibfnamefont {P.~M.}\ \bibnamefont {Gehring}}, \bibinfo
		{author} {\bibfnamefont {G.}~\bibnamefont {Xu}},\ and\ \bibinfo {author}
		{\bibfnamefont {B.}~\bibnamefont {Roessli}},\ }\href
	{https://doi.org/10.1088/1361-648X/ab86ee} {\bibfield  {journal} {\bibinfo
			{journal} {J. Phys. Condens. Matter}\ }\textbf {\bibinfo {volume} {32}},\
		\bibinfo {pages} {374012} (\bibinfo {year} {2020})}\BibitemShut {NoStop}%
	\bibitem [{\citenamefont {Sherrington}(2019)}]{Sherrington19:52}%
	\BibitemOpen
	\bibfield  {author} {\bibinfo {author} {\bibfnamefont {D.}~\bibnamefont
			{Sherrington}},\ }\href {https://doi.org/10.1088/1751-8121/ab2166} {\bibfield
		{journal} {\bibinfo  {journal} {J. Phys. A: Math. Theor.}\ }\textbf
		{\bibinfo {volume} {52}},\ \bibinfo {pages} {264001} (\bibinfo {year}
		{2019})}\BibitemShut {NoStop}%
	\bibitem [{\citenamefont {Kumar}\ \emph
		{et~al.}(2008{\natexlab{b}})\citenamefont {Kumar}, \citenamefont {Katiyar},
		\citenamefont {Rinaldi}, \citenamefont {Lushnikov},\ and\ \citenamefont
		{Shaplygina}}]{Kumar08:93}%
	\BibitemOpen
	\bibfield  {author} {\bibinfo {author} {\bibfnamefont {A.}~\bibnamefont
			{Kumar}}, \bibinfo {author} {\bibfnamefont {R.}~\bibnamefont {Katiyar}},
		\bibinfo {author} {\bibfnamefont {C.}~\bibnamefont {Rinaldi}}, \bibinfo
		{author} {\bibfnamefont {S.}~\bibnamefont {Lushnikov}},\ and\ \bibinfo
		{author} {\bibfnamefont {T.}~\bibnamefont {Shaplygina}},\ }\href
	{https://doi.org/10.1063/1.3043686} {\bibfield  {journal} {\bibinfo
			{journal} {Appl. Phys. Lett.}\ }\textbf {\bibinfo {volume} {3}},\ \bibinfo
		{pages} {232902} (\bibinfo {year} {2008}{\natexlab{b}})}\BibitemShut
	{NoStop}%
	\bibitem [{\citenamefont {Stock}\ \emph
		{et~al.}(2010{\natexlab{b}})\citenamefont {Stock}, \citenamefont {Jonas},
		\citenamefont {Broholm}, \citenamefont {Nakatsuji}, \citenamefont {Nambu},
		\citenamefont {Onuma}, \citenamefont {Maeno},\ and\ \citenamefont
		{Chung}}]{Stock10:105}%
	\BibitemOpen
	\bibfield  {author} {\bibinfo {author} {\bibfnamefont {C.}~\bibnamefont
			{Stock}}, \bibinfo {author} {\bibfnamefont {S.}~\bibnamefont {Jonas}},
		\bibinfo {author} {\bibfnamefont {C.}~\bibnamefont {Broholm}}, \bibinfo
		{author} {\bibfnamefont {S.}~\bibnamefont {Nakatsuji}}, \bibinfo {author}
		{\bibfnamefont {Y.}~\bibnamefont {Nambu}}, \bibinfo {author} {\bibfnamefont
			{K.}~\bibnamefont {Onuma}}, \bibinfo {author} {\bibfnamefont
			{Y.}~\bibnamefont {Maeno}},\ and\ \bibinfo {author} {\bibfnamefont {J.-H.}\
			\bibnamefont {Chung}},\ }\href
	{https://doi.org/10.1103/PhysRevLett.105.037402} {\bibfield  {journal}
		{\bibinfo  {journal} {Phys. Rev. Lett.}\ }\textbf {\bibinfo {volume} {105}},\
		\bibinfo {pages} {037402} (\bibinfo {year} {2010}{\natexlab{b}})}\BibitemShut
	{NoStop}%
	\bibitem [{\citenamefont {Aeppli}\ \emph {et~al.}(1983)\citenamefont {Aeppli},
		\citenamefont {Shapiro}, \citenamefont {Birgeneau},\ and\ \citenamefont
		{Chen}}]{Aeppli83:28}%
	\BibitemOpen
	\bibfield  {author} {\bibinfo {author} {\bibfnamefont {G.}~\bibnamefont
			{Aeppli}}, \bibinfo {author} {\bibfnamefont {S.~M.}\ \bibnamefont {Shapiro}},
		\bibinfo {author} {\bibfnamefont {R.~J.}\ \bibnamefont {Birgeneau}},\ and\
		\bibinfo {author} {\bibfnamefont {H.~S.}\ \bibnamefont {Chen}},\ }\href
	{https://doi.org/10.1103/PhysRevB.28.5160} {\bibfield  {journal} {\bibinfo
			{journal} {Phys. Rev. B}\ }\textbf {\bibinfo {volume} {28}},\ \bibinfo
		{pages} {5160} (\bibinfo {year} {1983})}\BibitemShut {NoStop}%
	\bibitem [{\citenamefont {Aeppli}\ \emph
		{et~al.}(1984{\natexlab{b}})\citenamefont {Aeppli}, \citenamefont {Shapiro},
		\citenamefont {Birgeneau},\ and\ \citenamefont {Chen}}]{Aeepli84:29}%
	\BibitemOpen
	\bibfield  {author} {\bibinfo {author} {\bibfnamefont {G.}~\bibnamefont
			{Aeppli}}, \bibinfo {author} {\bibfnamefont {S.~M.}\ \bibnamefont {Shapiro}},
		\bibinfo {author} {\bibfnamefont {R.~J.}\ \bibnamefont {Birgeneau}},\ and\
		\bibinfo {author} {\bibfnamefont {H.~S.}\ \bibnamefont {Chen}},\ }\href
	{https://doi.org/10.1103/PhysRevB.29.2589} {\bibfield  {journal} {\bibinfo
			{journal} {Phys. Rev. B}\ }\textbf {\bibinfo {volume} {29}},\ \bibinfo
		{pages} {2589} (\bibinfo {year} {1984}{\natexlab{b}})}\BibitemShut {NoStop}%
	\bibitem [{\citenamefont {Shvartsman}\ \emph {et~al.}(2008)\citenamefont
		{Shvartsman}, \citenamefont {Bedanta}, \citenamefont {Borisov}, \citenamefont
		{Kleemann}, \citenamefont {Tkach},\ and\ \citenamefont
		{Vilarinho}}]{Shvartsman08:101}%
	\BibitemOpen
	\bibfield  {author} {\bibinfo {author} {\bibfnamefont {V.~V.}\ \bibnamefont
			{Shvartsman}}, \bibinfo {author} {\bibfnamefont {S.}~\bibnamefont {Bedanta}},
		\bibinfo {author} {\bibfnamefont {P.}~\bibnamefont {Borisov}}, \bibinfo
		{author} {\bibfnamefont {W.}~\bibnamefont {Kleemann}}, \bibinfo {author}
		{\bibfnamefont {A.}~\bibnamefont {Tkach}},\ and\ \bibinfo {author}
		{\bibfnamefont {P.~M.}\ \bibnamefont {Vilarinho}},\ }\href
	{https://doi.org/10.1103/PhysRevLett.101.165704} {\bibfield  {journal}
		{\bibinfo  {journal} {Phys. Rev. Lett.}\ }\textbf {\bibinfo {volume} {101}},\
		\bibinfo {pages} {165704} (\bibinfo {year} {2008})}\BibitemShut {NoStop}%
	\bibitem [{\citenamefont {Aeppli}\ \emph {et~al.}(1982)\citenamefont {Aeppli},
		\citenamefont {Shapiro}, \citenamefont {Birgeneau},\ and\ \citenamefont
		{Chen}}]{Aeppli82:25}%
	\BibitemOpen
	\bibfield  {author} {\bibinfo {author} {\bibfnamefont {G.}~\bibnamefont
			{Aeppli}}, \bibinfo {author} {\bibfnamefont {S.~M.}\ \bibnamefont {Shapiro}},
		\bibinfo {author} {\bibfnamefont {R.~J.}\ \bibnamefont {Birgeneau}},\ and\
		\bibinfo {author} {\bibfnamefont {H.~S.}\ \bibnamefont {Chen}},\ }\href
	{https://doi.org/10.1103/PhysRevB.25.4882} {\bibfield  {journal} {\bibinfo
			{journal} {Phys. Rev. B}\ }\textbf {\bibinfo {volume} {25}},\ \bibinfo
		{pages} {4882} (\bibinfo {year} {1982})}\BibitemShut {NoStop}%
	\bibitem [{\citenamefont {Stephanovich}\ and\ \citenamefont
		{Laguta}(2016)}]{Stephanovich16:18}%
	\BibitemOpen
	\bibfield  {author} {\bibinfo {author} {\bibfnamefont {V.~A.}\ \bibnamefont
			{Stephanovich}}\ and\ \bibinfo {author} {\bibfnamefont {V.~V.}\ \bibnamefont
			{Laguta}},\ }\href {https://doi.org/10.1039/c6cp00054a} {\bibfield  {journal}
		{\bibinfo  {journal} {Phys. Chem. Chem. Phys.}\ }\textbf {\bibinfo {volume}
			{18}},\ \bibinfo {pages} {7229} (\bibinfo {year} {2016})}\BibitemShut
	{NoStop}%
	\bibitem [{\citenamefont {Kuzian}\ \emph {et~al.}(2014)\citenamefont {Kuzian},
		\citenamefont {Kondakova}, \citenamefont {Dar\'e},\ and\ \citenamefont
		{Laguta}}]{Kuzian14:89}%
	\BibitemOpen
	\bibfield  {author} {\bibinfo {author} {\bibfnamefont {R.~O.}\ \bibnamefont
			{Kuzian}}, \bibinfo {author} {\bibfnamefont {I.~V.}\ \bibnamefont
			{Kondakova}}, \bibinfo {author} {\bibfnamefont {A.~M.}\ \bibnamefont
			{Dar\'e}},\ and\ \bibinfo {author} {\bibfnamefont {V.~V.}\ \bibnamefont
			{Laguta}},\ }\href {https://doi.org/10.1103/PhysRevB.89.024402} {\bibfield
		{journal} {\bibinfo  {journal} {Phys. Rev. B}\ }\textbf {\bibinfo {volume}
			{89}},\ \bibinfo {pages} {024402} (\bibinfo {year} {2014})}\BibitemShut
	{NoStop}%
	\bibitem [{\citenamefont {Birgeneau}(1998)}]{Birgeneau98:177}%
	\BibitemOpen
	\bibfield  {author} {\bibinfo {author} {\bibfnamefont {R.~J.}\ \bibnamefont
			{Birgeneau}},\ }\href {https://doi.org/10.1016/S0304-8853(97)00998-0}
	{\bibfield  {journal} {\bibinfo  {journal} {J. Magn. Magn. Mater.}\ }\textbf
		{\bibinfo {volume} {177-181}},\ \bibinfo {pages} {1} (\bibinfo {year}
		{1998})}\BibitemShut {NoStop}%
	\bibitem [{\citenamefont {Rotaru}\ \emph {et~al.}(2009)\citenamefont {Rotaru},
		\citenamefont {Roessli}, \citenamefont {Amato}, \citenamefont {Gvasaliya},
		\citenamefont {Mudry}, \citenamefont {Lushnikov},\ and\ \citenamefont
		{Shaplygina}}]{Rotaru09:79}%
	\BibitemOpen
	\bibfield  {author} {\bibinfo {author} {\bibfnamefont {G.~M.}\ \bibnamefont
			{Rotaru}}, \bibinfo {author} {\bibfnamefont {B.}~\bibnamefont {Roessli}},
		\bibinfo {author} {\bibfnamefont {A.}~\bibnamefont {Amato}}, \bibinfo
		{author} {\bibfnamefont {S.~N.}\ \bibnamefont {Gvasaliya}}, \bibinfo {author}
		{\bibfnamefont {C.}~\bibnamefont {Mudry}}, \bibinfo {author} {\bibfnamefont
			{S.~G.}\ \bibnamefont {Lushnikov}},\ and\ \bibinfo {author} {\bibfnamefont
			{T.~A.}\ \bibnamefont {Shaplygina}},\ }\href
	{https://doi.org/10.1103/PhysRevB.79.184430} {\bibfield  {journal} {\bibinfo
			{journal} {Phys. Rev. B}\ }\textbf {\bibinfo {volume} {79}},\ \bibinfo
		{pages} {184430} (\bibinfo {year} {2009})}\BibitemShut {NoStop}%
	\bibitem [{\citenamefont {Blinc}\ \emph {et~al.}(2008)\citenamefont {Blinc},
		\citenamefont {Laguta}, \citenamefont {Zalar}, \citenamefont {Zupancic},\
		and\ \citenamefont {Itoh}}]{Blinc08:104}%
	\BibitemOpen
	\bibfield  {author} {\bibinfo {author} {\bibfnamefont {R.}~\bibnamefont
			{Blinc}}, \bibinfo {author} {\bibfnamefont {V.~V.}\ \bibnamefont {Laguta}},
		\bibinfo {author} {\bibfnamefont {B.}~\bibnamefont {Zalar}}, \bibinfo
		{author} {\bibfnamefont {B.}~\bibnamefont {Zupancic}},\ and\ \bibinfo
		{author} {\bibfnamefont {M.}~\bibnamefont {Itoh}},\ }\href
	{https://doi.org/10.1063/1.2957077} {\bibfield  {journal} {\bibinfo
			{journal} {J. Appl. Phys.}\ }\textbf {\bibinfo {volume} {104}},\ \bibinfo
		{pages} {084105} (\bibinfo {year} {2008})}\BibitemShut {NoStop}%
	\bibitem [{\citenamefont {Bhat}\ \emph {et~al.}(2004)\citenamefont {Bhat},
		\citenamefont {Ramunujachary}, \citenamefont {Lofland},\ and\ \citenamefont
		{Umarji}}]{Bhat04:280}%
	\BibitemOpen
	\bibfield  {author} {\bibinfo {author} {\bibfnamefont {V.~V.}\ \bibnamefont
			{Bhat}}, \bibinfo {author} {\bibfnamefont {K.~V.}\ \bibnamefont
			{Ramunujachary}}, \bibinfo {author} {\bibfnamefont {S.~E.}\ \bibnamefont
			{Lofland}},\ and\ \bibinfo {author} {\bibfnamefont {A.~M.}\ \bibnamefont
			{Umarji}},\ }\href {https://doi.org/10.1016/j.jmmm.2004.03.016} {\bibfield
		{journal} {\bibinfo  {journal} {J. Magn. Magn. Mater.}\ }\textbf {\bibinfo
			{volume} {280}},\ \bibinfo {pages} {221} (\bibinfo {year}
		{2004})}\BibitemShut {NoStop}%
	\bibitem [{\citenamefont {Bhat}\ \emph {et~al.}(2005)\citenamefont {Bhat},
		\citenamefont {Umarji}, \citenamefont {Shenoy},\ and\ \citenamefont
		{Waghmare}}]{Bhat05:72}%
	\BibitemOpen
	\bibfield  {author} {\bibinfo {author} {\bibfnamefont {V.~V.}\ \bibnamefont
			{Bhat}}, \bibinfo {author} {\bibfnamefont {A.~M.}\ \bibnamefont {Umarji}},
		\bibinfo {author} {\bibfnamefont {V.~B.}\ \bibnamefont {Shenoy}},\ and\
		\bibinfo {author} {\bibfnamefont {U.~V.}\ \bibnamefont {Waghmare}},\ }\href
	{https://doi.org/10.1103/PhysRevB.72.014104} {\bibfield  {journal} {\bibinfo
			{journal} {Phys. Rev. B}\ }\textbf {\bibinfo {volume} {72}},\ \bibinfo
		{pages} {014104} (\bibinfo {year} {2005})}\BibitemShut {NoStop}%
	\bibitem [{\citenamefont {Xu}\ \emph {et~al.}(2003)\citenamefont {Xu},
		\citenamefont {Zhong}, \citenamefont {Bing}, \citenamefont {Ye},
		\citenamefont {Stock},\ and\ \citenamefont {Shirane}}]{Xu03:67}%
	\BibitemOpen
	\bibfield  {author} {\bibinfo {author} {\bibfnamefont {G.}~\bibnamefont
			{Xu}}, \bibinfo {author} {\bibfnamefont {Z.}~\bibnamefont {Zhong}}, \bibinfo
		{author} {\bibfnamefont {Y.}~\bibnamefont {Bing}}, \bibinfo {author}
		{\bibfnamefont {Z.-G.}\ \bibnamefont {Ye}}, \bibinfo {author} {\bibfnamefont
			{C.}~\bibnamefont {Stock}},\ and\ \bibinfo {author} {\bibfnamefont
			{G.}~\bibnamefont {Shirane}},\ }\href
	{https://doi.org/10.1103/PhysRevB.67.104102} {\bibfield  {journal} {\bibinfo
			{journal} {Phys. Rev. B}\ }\textbf {\bibinfo {volume} {67}},\ \bibinfo
		{pages} {104102} (\bibinfo {year} {2003})}\BibitemShut {NoStop}%
	\bibitem [{\citenamefont {Conlon}\ \emph {et~al.}(2004)\citenamefont {Conlon},
		\citenamefont {Luo}, \citenamefont {Viehland}, \citenamefont {Li},
		\citenamefont {Whan}, \citenamefont {Fox}, \citenamefont {Stock},\ and\
		\citenamefont {Shirane}}]{Conlon04:70}%
	\BibitemOpen
	\bibfield  {author} {\bibinfo {author} {\bibfnamefont {K.~H.}\ \bibnamefont
			{Conlon}}, \bibinfo {author} {\bibfnamefont {H.}~\bibnamefont {Luo}},
		\bibinfo {author} {\bibfnamefont {D.}~\bibnamefont {Viehland}}, \bibinfo
		{author} {\bibfnamefont {J.~F.}\ \bibnamefont {Li}}, \bibinfo {author}
		{\bibfnamefont {T.}~\bibnamefont {Whan}}, \bibinfo {author} {\bibfnamefont
			{J.~H.}\ \bibnamefont {Fox}}, \bibinfo {author} {\bibfnamefont
			{C.}~\bibnamefont {Stock}},\ and\ \bibinfo {author} {\bibfnamefont
			{G.}~\bibnamefont {Shirane}},\ }\href
	{https://doi.org/10.1103/PhysRevB.70.172204} {\bibfield  {journal} {\bibinfo
			{journal} {Phys. Rev. B}\ }\textbf {\bibinfo {volume} {70}},\ \bibinfo
		{pages} {172204} (\bibinfo {year} {2004})}\BibitemShut {NoStop}%
	\bibitem [{\citenamefont {Kisi}\ and\ \citenamefont
		{Forrester}(2005)}]{Kisi05:17}%
	\BibitemOpen
	\bibfield  {author} {\bibinfo {author} {\bibfnamefont {E.~H.}\ \bibnamefont
			{Kisi}}\ and\ \bibinfo {author} {\bibfnamefont {J.~S.}\ \bibnamefont
			{Forrester}},\ }\href {https://doi.org/10.1088/0953-8984/17/36/l01}
	{\bibfield  {journal} {\bibinfo  {journal} {J. Phys. Condens. Matter}\
		}\textbf {\bibinfo {volume} {17}},\ \bibinfo {pages} {L381} (\bibinfo {year}
		{2005})}\BibitemShut {NoStop}%
	\bibitem [{\citenamefont {Xu}\ \emph {et~al.}(2006)\citenamefont {Xu},
		\citenamefont {Gehring}, \citenamefont {Stock},\ and\ \citenamefont
		{Conlon}}]{Xu06:79}%
	\BibitemOpen
	\bibfield  {author} {\bibinfo {author} {\bibfnamefont {G.}~\bibnamefont
			{Xu}}, \bibinfo {author} {\bibfnamefont {P.~M.}\ \bibnamefont {Gehring}},
		\bibinfo {author} {\bibfnamefont {C.}~\bibnamefont {Stock}},\ and\ \bibinfo
		{author} {\bibfnamefont {K.}~\bibnamefont {Conlon}},\ }\href
	{https://doi.org/10.1080/02652030600558682} {\bibfield  {journal} {\bibinfo
			{journal} {Phase Transit.}\ }\textbf {\bibinfo {volume} {79}},\ \bibinfo
		{pages} {135} (\bibinfo {year} {2006})}\BibitemShut {NoStop}%
	\bibitem [{\citenamefont {Phelan}\ \emph {et~al.}(2015)\citenamefont {Phelan},
		\citenamefont {Rodriguez}, \citenamefont {Gao}, \citenamefont {Bing},
		\citenamefont {Ye}, \citenamefont {Huang}, \citenamefont {Wen}, \citenamefont
		{Xu}, \citenamefont {Stock}, \citenamefont {Matsuura},\ and\ \citenamefont
		{Gehring}}]{Phelan15:88}%
	\BibitemOpen
	\bibfield  {author} {\bibinfo {author} {\bibfnamefont {D.}~\bibnamefont
			{Phelan}}, \bibinfo {author} {\bibfnamefont {E.~E.}\ \bibnamefont
			{Rodriguez}}, \bibinfo {author} {\bibfnamefont {J.}~\bibnamefont {Gao}},
		\bibinfo {author} {\bibfnamefont {Y.}~\bibnamefont {Bing}}, \bibinfo {author}
		{\bibfnamefont {Z.~G.}\ \bibnamefont {Ye}}, \bibinfo {author} {\bibfnamefont
			{Q.}~\bibnamefont {Huang}}, \bibinfo {author} {\bibfnamefont {J.~S.}\
			\bibnamefont {Wen}}, \bibinfo {author} {\bibfnamefont {G.}~\bibnamefont
			{Xu}}, \bibinfo {author} {\bibfnamefont {C.}~\bibnamefont {Stock}}, \bibinfo
		{author} {\bibfnamefont {M.}~\bibnamefont {Matsuura}},\ and\ \bibinfo
		{author} {\bibfnamefont {P.~M.}\ \bibnamefont {Gehring}},\ }\href
	{https://doi.org/10.1080/01411594.2014.989226} {\bibfield  {journal}
		{\bibinfo  {journal} {Phase Trans.}\ }\textbf {\bibinfo {volume} {88}},\
		\bibinfo {pages} {283} (\bibinfo {year} {2015})}\BibitemShut {NoStop}%
	\bibitem [{\citenamefont {Gehring}\ \emph {et~al.}(2004)\citenamefont
		{Gehring}, \citenamefont {Chen}, \citenamefont {Ye},\ and\ \citenamefont
		{Shirane}}]{Gehring04:16}%
	\BibitemOpen
	\bibfield  {author} {\bibinfo {author} {\bibfnamefont {P.~M.}\ \bibnamefont
			{Gehring}}, \bibinfo {author} {\bibfnamefont {W.}~\bibnamefont {Chen}},
		\bibinfo {author} {\bibfnamefont {Z.~G.}\ \bibnamefont {Ye}},\ and\ \bibinfo
		{author} {\bibfnamefont {G.}~\bibnamefont {Shirane}},\ }\href
	{https://doi.org/10.1088/0953-8984/16/39/042} {\bibfield  {journal} {\bibinfo
			{journal} {J. Phys. Condens. Matter}\ }\textbf {\bibinfo {volume} {16}},\
		\bibinfo {pages} {7113} (\bibinfo {year} {2004})}\BibitemShut {NoStop}%
	\bibitem [{\citenamefont {Brown}\ \emph {et~al.}(2018)\citenamefont {Brown},
		\citenamefont {Stockdale}, \citenamefont {Luo}, \citenamefont {Zhao},
		\citenamefont {Li}, \citenamefont {Viehland}, \citenamefont {Xu},
		\citenamefont {Gehring}, \citenamefont {Isida}, \citenamefont {Hillier},\
		and\ \citenamefont {Stock}}]{Brown18:30}%
	\BibitemOpen
	\bibfield  {author} {\bibinfo {author} {\bibfnamefont {K.~L.}\ \bibnamefont
			{Brown}}, \bibinfo {author} {\bibfnamefont {C.~P.~J.}\ \bibnamefont
			{Stockdale}}, \bibinfo {author} {\bibfnamefont {H.}~\bibnamefont {Luo}},
		\bibinfo {author} {\bibfnamefont {X.}~\bibnamefont {Zhao}}, \bibinfo {author}
		{\bibfnamefont {J.~F.}\ \bibnamefont {Li}}, \bibinfo {author} {\bibfnamefont
			{D.}~\bibnamefont {Viehland}}, \bibinfo {author} {\bibfnamefont
			{G.}~\bibnamefont {Xu}}, \bibinfo {author} {\bibfnamefont {P.~M.}\
			\bibnamefont {Gehring}}, \bibinfo {author} {\bibfnamefont {K.}~\bibnamefont
			{Isida}}, \bibinfo {author} {\bibfnamefont {A.~D.}\ \bibnamefont {Hillier}},\
		and\ \bibinfo {author} {\bibfnamefont {C.}~\bibnamefont {Stock}},\ }\href
	{https://doi.org/10.1088/1361-648X/aaade3} {\bibfield  {journal} {\bibinfo
			{journal} {J. Phys. Condens. Matter}\ }\textbf {\bibinfo {volume} {30}},\
		\bibinfo {pages} {125703} (\bibinfo {year} {2018})}\BibitemShut {NoStop}%
	\bibitem [{\citenamefont {Hill}\ \emph {et~al.}(1991)\citenamefont {Hill},
		\citenamefont {Thurston}, \citenamefont {Erwin}, \citenamefont {Ramstad},\
		and\ \citenamefont {Birgeneau}}]{Hill91:66}%
	\BibitemOpen
	\bibfield  {author} {\bibinfo {author} {\bibfnamefont {J.~P.}\ \bibnamefont
			{Hill}}, \bibinfo {author} {\bibfnamefont {T.~R.}\ \bibnamefont {Thurston}},
		\bibinfo {author} {\bibfnamefont {R.~W.}\ \bibnamefont {Erwin}}, \bibinfo
		{author} {\bibfnamefont {M.~J.}\ \bibnamefont {Ramstad}},\ and\ \bibinfo
		{author} {\bibfnamefont {R.~J.}\ \bibnamefont {Birgeneau}},\ }\href
	{https://doi.org/10.1103/PhysRevLett.66.3281} {\bibfield  {journal} {\bibinfo
			{journal} {Phys. Rev. Lett.}\ }\textbf {\bibinfo {volume} {66}},\ \bibinfo
		{pages} {3281} (\bibinfo {year} {1991})}\BibitemShut {NoStop}%
	\bibitem [{\citenamefont {Shirane}\ \emph {et~al.}(2004)\citenamefont
		{Shirane}, \citenamefont {Shapiro},\ and\ \citenamefont
		{Tranquada}}]{Shirane_book}%
	\BibitemOpen
	\bibfield  {author} {\bibinfo {author} {\bibfnamefont {G.}~\bibnamefont
			{Shirane}}, \bibinfo {author} {\bibfnamefont {S.~M.}\ \bibnamefont
			{Shapiro}},\ and\ \bibinfo {author} {\bibfnamefont {J.~M.}\ \bibnamefont
			{Tranquada}},\ }\href@noop {} {\emph {\bibinfo {title} {Neutron Scattering
				with a Triple-Axis Spectrometer}}}\ (\bibinfo  {publisher} {Cambridge
		University Press, Cambridge},\ \bibinfo {year} {2004})\BibitemShut {NoStop}%
	\bibitem [{\citenamefont {Bacon}(1975)}]{Bacon_book}%
	\BibitemOpen
	\bibfield  {author} {\bibinfo {author} {\bibfnamefont {G.~E.}\ \bibnamefont
			{Bacon}},\ }\href@noop {} {\emph {\bibinfo {title} {Neutron Diffraction}}}\
	(\bibinfo  {publisher} {Oxford University Press, London},\ \bibinfo {year}
	{1975})\BibitemShut {NoStop}%
	\bibitem [{\citenamefont {Moon}\ \emph {et~al.}(1969)\citenamefont {Moon},
		\citenamefont {Riste},\ and\ \citenamefont {Koehler}}]{Moon69:181}%
	\BibitemOpen
	\bibfield  {author} {\bibinfo {author} {\bibfnamefont {R.~M.}\ \bibnamefont
			{Moon}}, \bibinfo {author} {\bibfnamefont {T.}~\bibnamefont {Riste}},\ and\
		\bibinfo {author} {\bibfnamefont {W.~C.}\ \bibnamefont {Koehler}},\ }\href
	{https://doi.org/10.1103/PhysRev.181.920} {\bibfield  {journal} {\bibinfo
			{journal} {Phys. Rev.}\ }\textbf {\bibinfo {volume} {181}},\ \bibinfo {pages}
		{920} (\bibinfo {year} {1969})}\BibitemShut {NoStop}%
	\bibitem [{\citenamefont {Stewart}(2000)}]{Stewart00:87}%
	\BibitemOpen
	\bibfield  {author} {\bibinfo {author} {\bibfnamefont {J.~R.}\ \bibnamefont
			{Stewart}},\ }\href {https://doi.org/10.1063/1.373364} {\bibfield  {journal}
		{\bibinfo  {journal} {J. Appl. Phys.}\ }\textbf {\bibinfo {volume} {87}},\
		\bibinfo {pages} {5420} (\bibinfo {year} {2000})}\BibitemShut {NoStop}%
	\bibitem [{\citenamefont {Benkhaled}\ \emph {et~al.}(2020)\citenamefont
		{Benkhaled}, \citenamefont {Djermouni}, \citenamefont {Zaoui}, \citenamefont
		{Kondakova}, \citenamefont {Kuzian},\ and\ \citenamefont
		{Hayn}}]{Benkhaled20:525}%
	\BibitemOpen
	\bibfield  {author} {\bibinfo {author} {\bibfnamefont {N.}~\bibnamefont
			{Benkhaled}}, \bibinfo {author} {\bibfnamefont {M.}~\bibnamefont
			{Djermouni}}, \bibinfo {author} {\bibfnamefont {A.}~\bibnamefont {Zaoui}},
		\bibinfo {author} {\bibfnamefont {I.}~\bibnamefont {Kondakova}}, \bibinfo
		{author} {\bibfnamefont {R.}~\bibnamefont {Kuzian}},\ and\ \bibinfo {author}
		{\bibfnamefont {R.}~\bibnamefont {Hayn}},\ }\href
	{https://doi.org/10.1016/j.jmmm.2020.167309} {\bibfield  {journal} {\bibinfo
			{journal} {J. Mag. Mag. Mater.}\ }\textbf {\bibinfo {volume} {515}},\
		\bibinfo {pages} {167309} (\bibinfo {year} {2020})}\BibitemShut {NoStop}%
	\bibitem [{\citenamefont {Liu}\ \emph {et~al.}(2021)\citenamefont {Liu},
		\citenamefont {Lane}, \citenamefont {Frost}, \citenamefont {Ewings},
		\citenamefont {Attfield},\ and\ \citenamefont {Stock}}]{Liu21:54}%
	\BibitemOpen
	\bibfield  {author} {\bibinfo {author} {\bibfnamefont {Z.~H.}\ \bibnamefont
			{Liu}}, \bibinfo {author} {\bibfnamefont {H.}~\bibnamefont {Lane}}, \bibinfo
		{author} {\bibfnamefont {C.~D.}\ \bibnamefont {Frost}}, \bibinfo {author}
		{\bibfnamefont {R.~A.}\ \bibnamefont {Ewings}}, \bibinfo {author}
		{\bibfnamefont {J.~P.}\ \bibnamefont {Attfield}},\ and\ \bibinfo {author}
		{\bibfnamefont {C.}~\bibnamefont {Stock}},\ }\href
	{https://doi.org/10.1107/S1600576721004234} {\bibfield  {journal} {\bibinfo
			{journal} {J. Appl. Crystallogr.}\ }\textbf {\bibinfo {volume} {54}},\
		\bibinfo {pages} {957} (\bibinfo {year} {2021})}\BibitemShut {NoStop}%
\end{thebibliography}
\end{document}